\documentclass[usenatbib]{mnras}

\usepackage{newtxtext,newtxmath}
\usepackage[T1]{fontenc}
\usepackage{ae,aecompl}

\usepackage{color}

\usepackage{mathtools} 
\usepackage{amsmath} 
\usepackage{subfigure}
\usepackage{amsfonts} 
\usepackage{mathrsfs} 
\usepackage{graphicx} 
\usepackage{float} 
\usepackage{caption} 
\usepackage{bigstrut} 
\usepackage{tabularx} 
\usepackage{cancel} 
\usepackage{soul}
\usepackage{upgreek}

\usepackage{dcolumn}
\usepackage{bm}
\usepackage{xcolor}
\usepackage{ulem}

\usepackage{slashed} 
\usepackage{epstopdf} 

\title[Universal relations for hybrid stars]{Universal relations for rapidly rotating cold and hot hybrid stars}

\author[Largani et al.]{
Noshad Khosravi Largani,$^{1}$\thanks{E-mail: noshad.khosravilargani@uwr.edu.pl}
Tobias~Fischer,$^{1}$\thanks{E-mail: tobias.fischer@uwr.edu.pl}
Armen~Sedrakian,$^{1,2}$\thanks{E-mail: armen.sedrakian@uwr.edu.pl}
Mateusz~Cierniak,$^{1}$
\newauthor
David~E.~Alvarez-Castillo$^{3}$
and
David~B.~Blaschke,$^{1}$
\\
$^{1}$Institute of Theoretical Physics, University of Wroclaw,  Plac Maksa Borna 9, 50-204 Wroclaw, Poland
\\
$^{2}$Frankfurt Institute for Advanced Studies, Ruth-Moufang str. 1, 60438 Frankfurt am Main, Germany
\\
$^{3}$Institute of Nuclear Physics, Polish Academy of Sciences, Radzikowskiego 152, 31-342 Cracow, Poland
\\
}

\date{Accepted 2022 June 26. Received 2022 June 3; in original form 2021 December 20}

\pubyear{2021}

\begin{document}
\label{firstpage}
\pagerange{\pageref{firstpage}--\pageref{lastpage}}
\maketitle

\begin{abstract}

{Several global parameters of compact stars are related via empirical relations, which are (nearly) independent of the underlying equation of state (EoS) of dense matter and, therefore, are said to be universal. We investigate the universality of relations that express the maximum mass and the radius of non-rotating and maximally rapidly rotating configurations, as well as their moment of inertia, in terms of the compactness of the star. For this, we first utilize a collection of cold (zero-temperature) and hot (isentropic) nucleonic EoS and confirm that the universal relations are holding for our collection of EoS. 
We then go on, to add to our collection and test for the same universality models of  EoS that admit a strong first-order phase transition from nucleonic to deconfined quark matter. Also in this case we find that the universal relations hold, in particular for hot, isentropic hybrid stars. By fitting the universal relations to our computed data, we determine the coefficients entering these relations and the accuracy to which they hold.}

\end{abstract}

\section{Introduction}
\label{sec:intro}
The global properties of a compact star such as the maximum mass, the corresponding radius, moment of inertia, etc. sensitively depend on the microscopic equation of state (EoS). This applies to the cold (zero-temperature) and hot (finite-temperature) regimes and is the case for static and rotating stars. At the same time, certain relations among global parameters of compact stars were established~\citep{HaenselZdunik:1989,Friedman:1989,Shapiro:1989,Haensel:1995,Haensel:1996,Haensel:2009} and intensively studied in recent years that are highly insensitive to the input EoS and, therefore, are named {\it universal}~\citep{Yagi2017x}. The universal relations have been derived for a variety of systems under different conditions, for example, both for non-rotating~\citep{Yagi2013Sci,Sotani2013MNRAS, Majumder2015PhRvD,Steiner2016EPJA,Lenka2017,Wei2019JPhG,Kumar2019PhRvD,Raduta2020MNRAS}, slowly rotating~\citep{Silva2016MNRAS} and rapidly rotating stars~\citep{Breu:2016ufb,Paschalidis:2017qmb,Riahi2019,Bozzola:2019tit,Khadkikar2021,Koliogiannis:2020}, magnetized stars~\citep{Haskell2014MNRAS}, finite temperature stars~\citep{Raduta2020MNRAS,Khadkikar2021} in alternative theories of gravity~\citep{Doneva2014PhRvD,Pappas2019,Popchev2019EPJC,Yagi2021PhRvD} and in binaries~\citep{Manoharan2021}. Moreover, the level of validity of these universal relations with respect to the specific approach used in attaching the compact star crust with its core EoS has been reported in~\cite{Suleiman2021PhRvC}. For a review, see~\cite{Yagi2017x}.

The first observation of gravitational waves from a binary neutron star merger event GW170817~\citep{TheLIGOScientific:2017qsa} and its electromagnetic counterparts offers a wide range of insights into the physics of compact stars. In particular, several authors pointed out that the maximum mass of compact stars can be constrained under a reasonable and sufficiently general scenario of the formation of a hypermassive compact star and its subsequent collapse into a black hole~\citep{Bauswein:2017vtn,Shibata:2017xdx,Rezzolla:2017aly,Shibata:2019ctb,Khadkikar2021}. This scenario uses a universal relation between the maximum mass of a neutron star, rotating at the mass-shedding (Keplerian) limit and its non-rotating counterpart in the case of cold compact stars~\citep{Breu:2016ufb,Rezzolla:2017aly,Shibata:2019ctb} and isentropic hot stars~\citep{Khadkikar2021}. Earlier, \cite{Bauswein:2013} demonstrated that the maximum mass of cold and non-rotating neutron stars can be extracted from the gravitational wave signal of binary neutron star mergers using a tight correlation between threshold mass for a prompt collapse to a black hole and the compactness of non-rotating maximum-mass configuration.  Clearly, setting a limit on the maximum mass of non-rotating neutron star from gravitational wave observations offers a new astrophysical constraint on the EoS of matter at suprasaturation density, which is complementary to the astrophysical information coming from alternative sources, such as the X-ray sources associated with low-mass binary systems analysed by \citet{Steiner:2010} and by the \textit{Neutron star Interior Composition ExploreR (NICER)} experiment~\citep{Riley2021ApJ,Miller2021ApJ} as well as radio emission from neutron stars in binaries with white dwarfs~\citep{Cromartie:2020NatAs}.

Hypermassive compact stars formed in binary mergers are hot with temperatures ranging up to 100 MeV, therefore the above scenario needs to include the finite temperature EoS. The universal relations for isentropic stars were studied in recent work by \cite{Khadkikar2021}, who established universal relations for static and rapidly rotating hot configurations. They found that the universal relations hold for a variety of hadronic EoS at selected constant values of the entropy per particle and lepton fraction. Such thermodynamics conditions are important not only in the context of the particular scenario above, but also for other finite temperature astrophysics applications, such as core-collapse supernovae, which contain hot, dense, and lepton-rich protoneutron stars as central compact objects, as well as the pre- and post-merger phases of binary neutron star mergers \citep[cf.][and references therein]{Rosswog_15}. For reviews of the role of the EoS in simulations of core-collapse supernovae, see \citet{Janka:2007PhR} and \citet{Fischer:2017}. The role of the EoS in binary neutron star mergers is reviewed in \citet{Baiotti2019} and \citet{Chatziioannou2020GReGr}.

The deconfinement phase transition from nuclear (or more generally hadronic) matter to the quark-gluon plasma in the interiors of compact stars has been studied for a long time, but remains in the focus of current research. Multimessenger astronomy offers a new avenue to address the problem of relevant degrees of freedom in high density, cold matter, and possible phase transition from nucleonic to quark degrees of freedom. If a deconfinement phase transition occurs in neutron star interiors, it will affect the maximum mass of hybrid stars, i.e. stars containing quark cores surrounded by nuclear envelopes.

The underlying theory for strong interactions, quantum chromodynamics (QCD), with quarks and gluons as the degrees of freedom can be solved from first principles within the strong coupling regime via lattice simulations~\citep[cf.][]{Bazavov:2014,Borsanyi:2014,Philipsen2021}. However, lattice studies are still limited to the small baryon density sector of the phase diagram, where a smooth crossover transition between the quark and nucleonic phases is predicted. The recent results of \citet{Bazavov:2019} predict a pseudo-critical temperature of about 156~MeV at zero baryon chemical potential. At high baryon density, relevant for compact stars, phenomenological models have been used such as the MIT bag models~\citep{Farhi:1984qu} or the class of Lagrangian-based Nambu--Jona-Lasinio (NJL) models~\citep{NJL:1961}. There have been first-principle computations of the EoS of cold quark matter from perturbative QCD, which is valid only at extremely high densities far beyond those achieved in compact stars~\citep[for a recent study of perturbative QCD, see][and references therein]{Kurkela:2020NatPh}.  Furthermore, models that describe on the same footing the quark and hadronic phases for astrophysics applications are lacking, therefore a first-order phase transition is often constructed between models of hadronic and quark matter with different underlying theories. A first-order phase transition will affect not only the static properties of compact stars, such as tidal deformabilities measured in gravitational wave events~\citep{Paschalidis:2017qmb,Montana2019,Sieniawska2019,Essick2020,Otto2020,Lijj_2020a,Lijj_2021,Ferreira2021,Ayriyan:2021prr}, but also their dynamics~\citep{Fischer:2018, OConnor:2020, Most:2019PhRvL,Bauswein:2019,Prakash:2021,Kuroda:2022}. Specifically, \citet{Fischer:2018} and \citet{OConnor:2020} proposed that neutrino and gravitational wave signals from supernova explosions may contain observable signatures of a phase transition. Several authors \citep{Most:2019PhRvL,Bauswein:2019,Prakash:2021} performed binary neutron star mergers simulations with hot hybrid EoS, which revealed the possible signatures of the phase transition in the gravitational wave signals from those events.

The purpose of this paper is to address the universal relations for hybrid star configurations. We focus on universal relations for the maximum masses and the corresponding radii, as well as the moment of inertia, of non-rotating and rapidly rotating compact stars. The moment of inertia is particularly interesting since it might become possible to measure this quantity at high precision \citep[for details, cf.][and references therein]{Kramer:2021}. Specifically, universal relations for the moment of inertia have been studied by \citet{RavenhallPethick:1994}, \citet{LattimerSchutz:2005} and \citet{Breu:2016ufb}, also for hot EoS by \citet{Khadkikar2021}. Here, we use a large set of hadronic EoS, available in the literature in combination with phenomenological EoS for quark matter. Our work complements the earlier study by \citet{Khadkikar2021} that focused on hadronic EoS in that we include in our study a large set of finite-temperature EoS that contains a phase transition from nucleonic or hadronic matter to quark matter. Furthermore, we expand the set of zero-temperature EoS for hybrid stars by extending the previously investigated sets of multipolytrope and constantspeed of sound (CSS) EoS to new ones that are based on NJL and string-flip models (see below).

Rotating hybrid stars have received relatively little attention so far, with some works focusing on the effect of phase transition on the rotationally induced phase transition~\citep{Glendenning1997,Ayvazyan:2013,Haensel2016EPJA}, others on the tidal deformabilities and universal relations~\citep{Paschalidis:2017qmb,Bozzola:2017qbu,Bozzola:2019tit}. The key extension of this work over the previous ones is the use of the finite temperature EoS for the hadronic and quark phases of hybrid stars. We aim to confirm that (some of) the universal relations found for isentropic stars \citep{Khadkikar2021} hold also for finite-temperature isentropic hybrid stars. Furthermore, we will extend the sets of EoS used previously \citep{Paschalidis:2017qmb,Bozzola:2017qbu,Bozzola:2019tit} which were based on
multipolytropic EoS \citep{Alvarez-Castillo:2017qki} and CSS parametrization~\citep{Alford:2017qgh} to alternative phenomenological models.

This paper is organized as follows. In sec.~\ref{sec:RNS} we briefly review the {\tt{RNS}} code that is the foundation of this study to construct equilibrium configurations of hadronic and hybrid stars. The sets of hadronic and hybrid EoS are introduced in sec.~\ref{sec:EoS} and the universal relations obtained are discussed in sec.~\ref{sec:results}. We close with a summary in sec.~\ref{sec:summary}.

\begin{table*}
\centering
\caption{Abbreviations of the hadronic EoS and the corresponding references. 
The EoS of the crust for all models, except of the last four, is due to 
\citet{Hempel:2010}.}
\begin{tabular}{l c l}
\textsc{EoS} & & References \\     
\hline
    DD2           & & \citet{Typel:2010}      \\
    DD2ev      & & \citet{Typel:2010,Typel:2016}      \\
    DD2F         & & \citet{Alvarez-Castillo:2016oln}      \\
    DD2Fev      & & \citet{Alvarez-Castillo:2016oln,Typel:2016}      \\
    DD2esym    &  & \citet{Typel:2010,Typel:2014a,Typel:2014b}      \\
    
    NL3           &  & \citet{Lalazissis:1997} \\
    SFHo       &  & \citet{Steiner:2013}   \\
    SFHx       &   & \citet{Steiner:2013}    \\
    TM1         &   & \citet{Sugahara:1994}  \\
    TMA         &   & \citet{Toki:1995}  \\
    IUFSU      &   & \citet{RocaMaza:2008}  \\
     BHB$\Lambda \phi$ & &  \citet{Banik:2014}   \\
     GS      & &  \citet{Shen:2011}\\
    APR       &  & \citet{AP:1997,APR:1998}              \\
    LS       &  & \citet{Lattimer:1991}              \\
    BSk       &  & \citet{Goriely:2010}              \\

\hline
\end{tabular}
\label{tab:EoS-hadron}
\end{table*}

\begin{table*}
\centering
\caption{Abbreviations of the hybrid EoS and the corresponding references. 
The EoS of the crust for all models, except of the last four, is due to 
\citet{Hempel:2010}.}
\begin{tabular}{l c l}
\textsc{EoS} & & References                        \\     
\hline
DD2F--VBAG & & \citet{Klaehn:2015,Alvarez-Castillo:2016oln,Klaehn:2017} \\
MITBAG & & \cite{Sagert:2009,Fischer:2011} \\
RDF & & \citet{Kaltenborn:2017,Fischer:2018,Bastian:2021}  \\
DD2F--CSS & & \citet{Alford:2013,Alvarez-Castillo:2016oln,Alford:2017qgh} \\
APR--nlNJL & & \citet{AP:1997,APR:1998,Baym:2018,Ayriyan:2021} \\
DD2p40--nlNJL & & \citet{Typel:2010,Typel:2016,Alvarez:2019,Blaschke:2020qqj} \\
DDQuark$G_{\rm V}08$ & & \citet{Ayvazyan:2013} \\
\hline
\end{tabular}
\label{tab:EoS-hybrid}
\end{table*}

\section{{\tt{RNS}} -- Numerical model for rotating neutron stars}
\label{sec:RNS}
For the numerical simulations of rotating compact stars in hydrostatic equilibrium, we employ the {\tt{RNS}} open-source code developed by \cite{Stergioulas:1995}.  This code is based on the formalism of \cite{Komatsu:1989} and modifications introduced by \cite{Cook:1994}, see also complementary and alternative implementations~\citep{Bonazzola1993,Cook:1994,Nozawa1998}. It is built on solutions of Einstein's field equations for axially symmetric and stationary space-time in spherical coordinates,
\begin{eqnarray}
  ds^2 = -e^{\gamma+\rho}dt^2 + e^{2\alpha}\left(dr^2+r^2d\theta^2\right) 
  + e^{\gamma-\rho}\,r^2\sin^2\theta\left(d\phi-\omega dt\right)^2,
\end{eqnarray}
where $\gamma(r,\theta)$, $\rho(r,\theta)$ and $\alpha(r,\theta)$ are the metric functions, which depend on the radial coordinate $r$ and on the lateral angle $\theta$, and $\omega$ is the angular velocity of the local frame.  The equations that are solved through an iterative procedure comprise the differential equations for the metric functions and the equation of hydrostatic equilibrium, which involves the entropy for barotropic fluids \citep[see Appendix~A of][] {Cook:1994}. Given an input tabulated EoS the {\tt{RNS}} outputs global parameters of rotating stars, such as the gravitational and baryonic masses, the circumferential radius at the equator, the moment of inertia, angular momentum, the rotational kinetic and gravitational potential energies \citep[for explicit expressions, see Appendix~B of][]{Cook:1994}. The conserved stress-energy tensor is taken to be that of the perfect fluid, whereas the matter four-velocity is proportional to the proper and angular velocity, directed along the $z$-axis.
%

Of particular interest for our discussion are the limiting cases of static spherical symmetric stars and the mass-shedding (Keplerian) limit where the surface rotational energy exceeds the gravitational binding. We consider exclusively uniformly rotating stars, i.e. we employ a uniform rotation law.  In the mass-shedding limit, the equilibrium solution is marginally stable against quasi-radial perturbations. We note that the stability of rapidly rotating stars may be determined by non-axisymmetric modes, which may onset at rotational rates below those corresponding to the Keplerian one \citep[see for instance][where the authors also consider the Chandrasekhar-Friedman-Schutz (CFS) instability of the f-mode and present universal relations for it]{Kruger:2019zuz}\footnote{For the Chandrasekhar-Friedman-Schutz (CFS) instability, see \citep[][]{Chandrasekhar:1970pjp,1975ApJ...199L.157F,1978ApJ...221L..99F}.}. Therefore, the Keplerian limit provides only an upper bound to the stability of the star. We also note that neutron star sequences are limited from below by the minimum mass of a star, which we will not consider here \citep[for further details, see][]{Stergioulas:1995,Nozawa1998}.

\section{Equations of state}
\label{sec:EoS}
We consider purely hadronic and hybrid EoS, the latter being constructed assuming a first-order phase transition from nucleonic to quark matter. Tables~\ref{tab:EoS-hadron} and \ref{tab:EoS-hybrid} list the hadronic and hybrid EoS, respectively, with the corresponding references. The selected EoS all fulfil the maximum mass constraint $M_{\rm max}\ge 2M_\odot$ set by the observations of pulsars in binaries with white dwarfs with masses of about 2~M$_\odot$~\citep{Fonseca:2016,Cromartie:2020NatAs}.
Tables~\ref{tab:EoS-properties_hadron_T0} -- \ref{tab:EoS-properties_hybrid_s3} 
list the values of $M_{\rm max}$ for our collection of EoS of hadronic and hybrid EoS. In addition, cold equations of state that do not fulfil the various radius constraints, derived from the tidal deformability constraints for a 1.4~M$_\odot$ neutron star, in connection with GW170817 \citep[cf.][]{TheLIGOScientific:2018}, $\Lambda_{1.4}=70-580$, are marked with a superscript $\dagger$ symbol, and those which are in conflict with the NICER measurements, for the massive pulsar J0030+0451 of about 1.4~M$_\odot$ 
\citep[][]{Riley:2019,Miller:2019}, 
$R\simeq11-14$~km,
and for the massive pulsar J0740+6620 of about 2~M$_\odot$ 
\citep[][]{Riley2021ApJ,Miller2021ApJ}, 
$R\simeq11.5-16$~km,
are marked with a superscript $\ddagger$ symbol accordingly.
These constraint regions are also shown in Figs.~\ref{fig:MR_hadronic_cold} and ~\ref{fig:MR_hybrid_cold}.

\subsection{Hadronic EoS}
\label{sec:EoS_hadronic}
The set of hadronic EoS considered here is partly the same as in \citet{Bauswein:2017vtn} and \citet{Bauswein:2019},  with several additions. The results here provided for the EoS are complementary to those for hybrid EoS and are designed to serve as a reference.

\subsubsection{DD-class of relativistic mean-field models}

We employ several EoS derived from the relativistic mean-field (RMF) theory of nuclear matter with density-dependent nucleon-meson couplings. The DD2F, DD2Fev, and DD2esym models belong to this class with only nucleonic degrees of freedom. Apart from astrophysical constraints, these models fulfil the constraint derived from the elliptic flow of heavy-ion collisions by \citet{Danielewicz:2002Sci}. In addition to density-dependent coupling, DD2Fev EoS includes a geometric excluded volume. Its parametrizations that are explored here range from extreme stiffening (p35) to extreme softening (m15) relative to the reference case DD2F. In the DD2esym version of these models the nuclear symmetry energy and its slope are varied instead; we use here two cases of this type denoted as DD2esym(++) and DD2esym(-- --), see Table~\ref{tab:EoS-properties_hadron_T0}. Further details of these models are given in \citet{Typel:2014b}.

Within the same class of models is the BHB$\Lambda\phi$ model, which however includes the $\Lambda$ hyperon. In this model a repulsive $\Lambda$--$\Lambda$ interaction is accounted for through the inclusion of a strange $\phi$ meson interaction channel. This provides additional stiffness to the EoS at high density, which is necessary to fulfil the maximum neutron star mass constraint.

The EoS of the curst for these models is taken from \citet{Hempel:2010}, they obtained the EoS
from a nuclear statistical equilibrium model that includes 1000 nuclear species with tabulated and partly calculated nuclear masses, at densities below nuclear saturation density.

\subsubsection{Non-linear relativistic mean-field models}

The GS, NL3, SHFo/x, TM1, TMA, and IUFSU models belong to the mean-field models that include non-linear mesonic terms in the Lagrangian of the model and density and temperature-independent meson-nucleon coupling constants. As with all mean-field modes, they have been fitted to provide good nuclear matter properties at nuclear saturation density, see table~1 in \cite{Fischer:2014} for a summary. The FSUGold model was excluded from the collection of \citet{Bauswein:2019} here
as it has a maximum mass 1.74~$M_\odot$. As above, the crust EoS is taken from \citet{Hempel:2010}.

\begin{figure}
\includegraphics[width=0.475\textwidth]{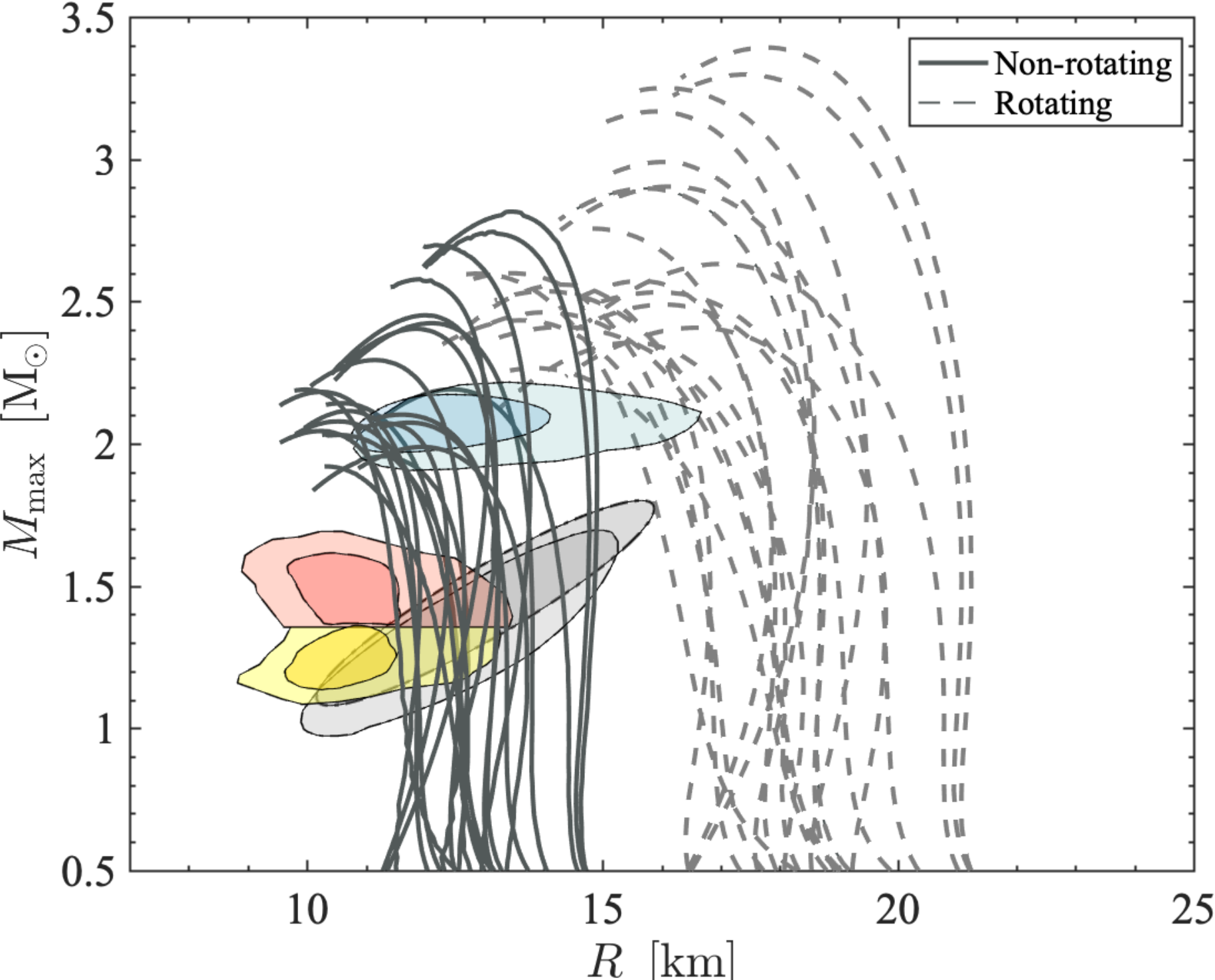}
\caption{(colour online) Mass-radius relations for the different sets of hadronic EoS, for non-rotating (solid lines) and rotating configurations (dashed lines) for $T=0$ and $\beta$-equilibrium, including the neutron star radii constraints marked by the elliptical regions due to GW170817 \citep[][]{TheLIGOScientific:2018} for the 68\% and 95\% confidence levels for both low-mass (yellow regions) and high-mass priors (red regions), as well as the low-mass (grey regions) and high-mass (light blue regions) NICER measurements \citep[][]{Miller:2019,Riley:2019,Miller2021ApJ,Riley2021ApJ,Raaijmakers:2019} at the 68\% confidence level.
The radii for the rapidly rotating configurations correspond to the equatorial radii.}
\label{fig:MR_hadronic_cold}
\end{figure}

\begin{figure}
\includegraphics[width=0.475\textwidth]{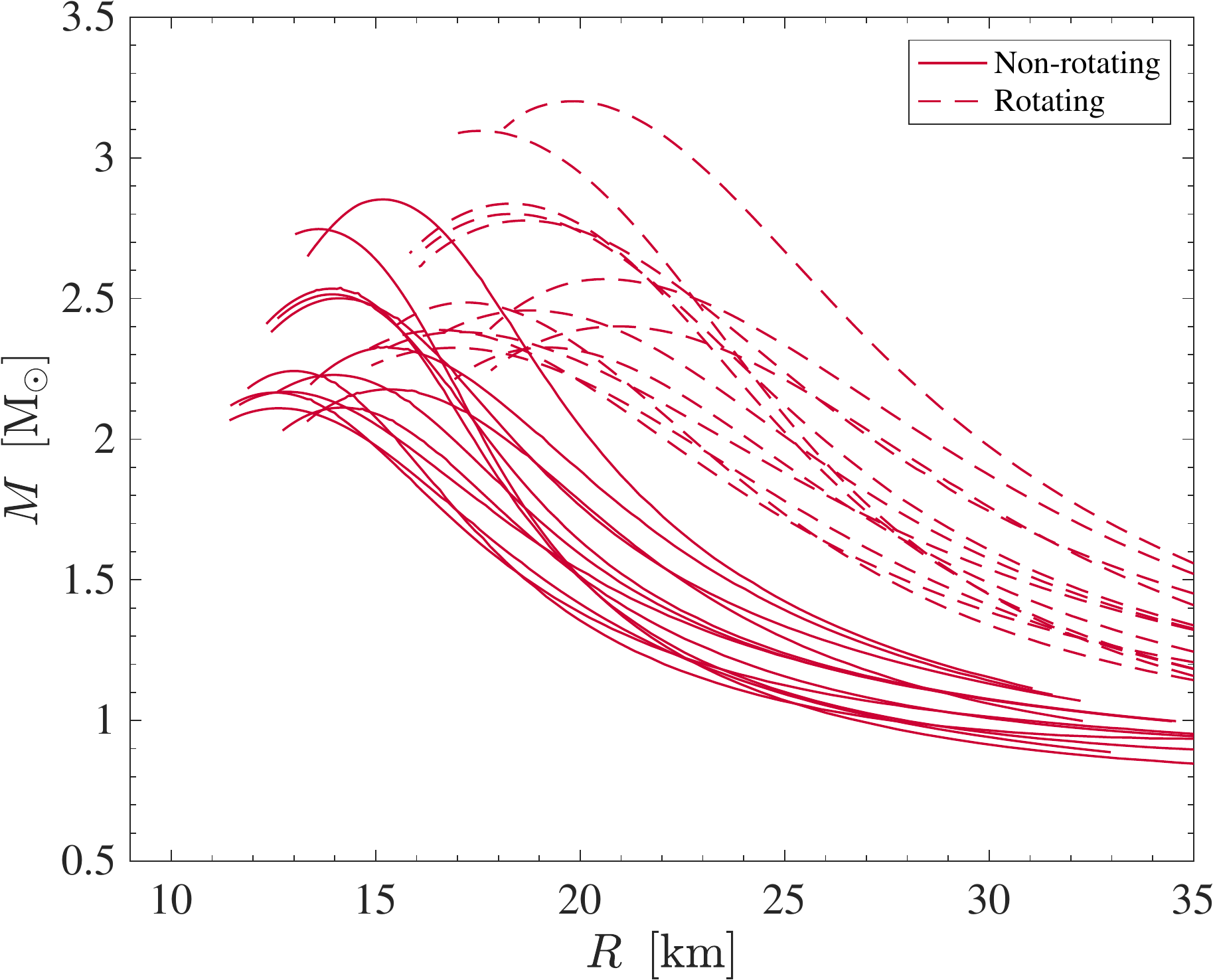}
\caption{(colour online) The same as Fig.~\ref{fig:MR_hadronic_cold} but for constant entropy per particle of $s=3~k_{\rm B}$ and fixed electron lepton fraction of $Y_{\rm L}=0.3$.}
\label{fig:MR_hadronic_hot}
\end{figure}

\begin{table*}
\centering
\caption{ Properties of the hadronic EoS used in this study for $T=0$ and $\beta$-equilibrium. The columns  show from left to right:  the maximum mass of non-rotating star $M_{\rm max}$, its radius $R_{\rm max}$,
the dimensionless tidal deformability $\Lambda_{1.4}$ of a 1.4~M$_\odot$ neutron star \citep[][]{Shibata:2016PhRvD}
, the radius of $M=2.0$~M$_\odot$ and $M = 1.4M_{\odot}$ stars, the central density, the speed of sound of the maximum-mass configuration, and the moment of inertia of the maximum mass of the rigidly rotating configuration $\left.\rho_{\rm central}\right\vert_{M_{\rm max}}$, $\left.c^2_{s\rm ,central}\right\vert_{M_{\rm max}}$ and $\left.I^*\right\vert_{M^*_{\rm max}}$, respectively, as well as the ratio of the masses of Keplerian and static configurations $M^*_{\rm max}/M_{\rm max}$, and the ratio radii of the same configurations $R^*_{\rm max}/R_{\rm max}$.
}
\begin{tabular}{l c c c c c c c c c c c}
\hline
EoS & $M_{\rm max}$ & $R_{\rm max}$ & $R_{2.0}$ &  $R_{1.4}$ & $\Lambda_{1.4}$ & $\left.\rho_{\rm central}\right\vert_{M_{\rm max}}$ & $\left.c^2_{s\rm ,central}\right\vert_{M_{\rm max}}$  &  $\left.I^*\right\vert_{M^*_{\rm max}}$ & $M^*_{\rm max}/M_{\rm max}$ & $R^*_{\rm max}/R_{\rm max}$   \\ 
 &  $[$M$_\odot]$ & $[$km$]$ & $[$km$]$ & $[$km$]$ & & $[10^{14}\,\,\,$g~cm$^{-3}]$  & $[$c$^2]$ &$[10^{45}\,\,\,$g~cm$^{2}]$ & & \\
\hline
DD2$^{\dagger}$ &   2.42  & 11.90 & 13.11 &  13.21 & 673 & 19.73  & 0.75 & 5.31 & 1.2051 & 1.3282 \\
DD2F &   2.07 & 10.46 & 13.35 &  12.42 & 425 & 25.18  & 0.75 & 3.37 & 1.1845 & 1.3350 \\
DD2ev--p15$^{\dagger,\diamond}$ &  2.58 & 11.91 &  13.05 &  12.76 & 739 & 24.20  & 1.14 & 6.55 & 1.2288 & 1.3341 \\
DD2ev--m20 &   2.07 & 11.33 & 11.97 &  12.34 &  563 & 20.99 & 0.45 & 4.44 & 1.2418 & 1.3798 \\
DD2Fev--p35&   2.14 & 10.57 & 11.65 &  12.49 &  433 & 25.75  & 0.87 & 3.54 & 1.1837 & 1.3256 \\
DD2Fev--m10${^\ddagger}$ &  1.92 & 10.33 &  -- -- &  12.12 &  347 & 27.56  & 0.57 & 2.82 & 1.1800 & 1.3444 \\
DD2esym$^{+ +,\dagger}$ &   2.40 & 11.99 & 13.57  &  13.81 &  880 & 19.19  & 0.80 & 5.43 & 1.1981 & 1.3324 \\
DD2esym$^{- -,\dagger}$ &   2.45 & 12.11 &  13.18 &  13.06 &  702 & 18.94  & 0.69 & 5.85 & 1.2204 & 1.3226 \\
GS1$^{\dagger,\ddagger}$ &   2.73 & 13.17 &   14.73 &  14.74 &  1346 & 17.89 & 0.83 & 7.58 & 1.2015 & 1.3210 \\
GS2$^\dagger$ &   2.07 & 11.73 &  12.56 &  13.36 &  691 & 20.72  & 0.47 & 4.11 & 1.2001 & 1.3647 \\
NL3$^{\dagger,\ddagger}$ &   2.79 & 13.43 & 14.90 &  14.98 &  1417 & 15.25  & 0.77 & 8.26 & 1.2049 & 1.3237 \\
SFHo &   2.06 & 10.32 &   10.82 &  11.80 &  318 & 27.23  & 0.79 & 3.19 & 1.1903 & 1.3308 \\
SHFx &   2.13 & 10.76 &  11.46  &  11.91 &  379 & 24.11  & 0.63 & 3.82 & 1.2101 & 1.3386 \\
TM1$^{\dagger,\ddagger}$ &  2.21 & 12.57 &  13.68  &  14.36 &  1098 & 18.58 & 0.44 & 4.83 & 1.2010 & 1.3641 \\
TMA${^\dagger}$ &   2.02 & 12.09 &   12.11  &  13.70 &  890 & 20.01 & 0.38 & 4.06 & 1.2108 & 1.3862 \\
IUFSU &   1.95 & 11.31 &  -- -- &  12.68 &  522 & 20.40 & 0.39 & 3.58 & 1.2018 & 1.3591 \\
APR $^\diamond$&   2.19 & 10.00 &  10.96 &  11.52 &  256 & 27.57 & 1.31 & 3.39 & 1.1861 & 1.2851 \\
LS220 &   2.04 & 10.62 & 11.13 &  12.59 &  512 &  29.35 & 0.90 & 3.33 & 1.1870 & 1.3606 \\
LS375$^{\dagger,\diamond}$ &   2.71 & 12.14 & 13.76 &  13.65 &  921 & 20.27  & 1.29 & 6.78 & 1.2060 & 1.3414 \\
BSk20 $^\diamond$ &  2.18 & 10.22 & 11.29 &  11.85 &  333 & 26.60 & 1.13 & 3.59 & 1.1912 & 1.3121 \\
BSk21 &  2.29 & 11.14 & 12.39 &  12.66 &  533 & 22.36  & 0.91 & 4.51 & 1.2015 & 1.3316 \\
BHB$\Lambda\phi^{\dagger}$ & 2.10 & 11.51 & 12.62 &  13.23 &  700 & 21.19  & 0.55 & 4.24 & 1.2046 & 1.3800 \\
\hline
\end{tabular}
\label{tab:EoS-properties_hadron_T0} 
\\
\begin{flushleft}
{$^\dagger$:~Excluded due to the tidal deformability constraint \citep[see][and references therein]{TheLIGOScientific:2018} \\
$^\ddagger$:~Excluded due to the NICER constraints \citep[][]{Riley:2019,Miller:2019,Riley2021ApJ,Miller2021ApJ}\\
$\diamond$:~Excluded due to breaking the causality limit when $c_s$ exceeds the speed of light.}
\end{flushleft}
\end{table*}

\begin{table*}
\centering
\caption{The same as Table~\ref{tab:EoS-properties_hadron_T0}, 
but for constant entropy per particle of $s=3~k_{\rm B}$ and a constant electron lepton fraction of $Y_{\rm L}=0.3$.}
\begin{tabular}{l c c c c c c c c}
\hline
EoS & $M_{\rm max}$ & $R_{\rm max}$ & $\left.\rho_{\rm central}\right\vert_{M_{\rm max}}$  & $\left.c^2_{s\rm ,central}\right\vert_{M_{\rm max}}$ &$\left.I^*\right\vert_{M^*_{\rm max}}$& $M^*_{\rm max}/M_{\rm max}$ & $R^*_{\rm max}/R_{\rm max}$   \\ 
 &  $[$M$_\odot]$ & $[$km$]$ & $[10^{14}\,\,\,$g~cm$^{-3}]$ & $[$c$^2]$ &$[10^{45}\,\,\,$g~cm$^{2}]$ & \\
\hline
DD2 &  2.51 & 13.94 & 17.38 & 0.70 & 5.01 & 1.1136 & 1.3168 \\
DD2F &  2.17 & 12.83 & 22.15 & 0.68 & 3.23 & 1.0979 & 1.3292 \\
DD2esym$^{+ +}$ &  2.50 & 14.13 & 17.23 & 0.71 & 4.96 & 1.1106 & 1.3175 \\
DD2esym$^{- -}$ &  2.53 & 13.91 & 15.93 & 0.73 & 5.21 & 1.1193 & 1.3195 \\
NL3 &   2.85 & 15.19 & 14.12  & 0.79 & 7.21 & 1.1224 & 1.3055 \\
SFHo &  2.17 & 12.54 & 22.63  & 0.65 & 3.20 & 1.1028 &  1.3214 \\
SHFx &  2.24 & 13.00 & 20.22 & 0.62 & 3.66 & 1.1086 &  1.3225 \\
TM1 &  2.33 & 15.27 & 15.60 & 0.42 & 4.72 & 1.1033 &  1.3513 \\
TMA &  2.18 & 15.29 & 15.84 & 0.37 & 4.21 & 1.1028 &  1.3681 \\
IUFSU &  2.11 & 14.20 & 18.18 & 0.40 & 3.57 & 1.1012 & 1.3530 \\
LS220 &  2.11 & 12.62 & 22.71 & 0.69 & 3.10 & 1.1018  & 1.3404 \\
LS375 $^\diamond$ &   2.75 & 13.62 & 16.86 & 1.21 & 6.04 & 1.1270  & 1.2908 \\
BHB$\Lambda\phi$ &  2.23 & 13.99 & 18.45 & 0.55 & 3.88 & 1.1028 & 1.3426 \\
\hline
\end{tabular}
\label{tab:EoS-properties_hadron_s3}
\\
\begin{flushleft}
{$\diamond$ EOS excluded due to breaking the causality limit.}
\end{flushleft}
\end{table*}

\begin{table*}
\centering
\caption{The same as Table~\ref{tab:EoS-properties_hadron_T0} but for the hybrid EoS considered at $T=0$ and $\beta$-equilibrium.}
\begin{tabular}{l  c c c c c c c c c c}
\hline
EoS & $M_{\rm max}$ & $R_{\rm max}$ &   $R_{2.0}$ &   $R_{1.4}$ &  $\Lambda_{1.4}$ & $\left.\rho_{\rm central}\right\vert_{M_{\rm max}}$  & $\left.c^2_{s\rm ,central}\right\vert_{M_{\rm max}}$ &$\left.I^*\right\vert_{M^*_{\rm max}}$ & $M^*_{\rm max}/M_{\rm max}$ & $R^*_{\rm max}/R_{\rm max}$   \\ 
& $[$M$_\odot]$ & $[$km$]$ & $[$km$]$ & $[$km$]$ &  & $[10^{14}\,\,\,$g~cm$^{-3}]$ & $[$c$^2]$ &$[10^{45}\,\,\,$g~cm$^{2}]$& & \\

\hline

APR--nlNJL $\eta_D=0.71$: \\
$\eta_V=0.06$  & 1.99 & 10.87 &  -- -- &  11.90 & 337 & 24.07 & 0.44 & 3.48 & 1.2077 & 1.3454 \\
$\eta_V=0.08$  & 2.07 & 11.13 &   11.66 &  12.00 &  362 & 21.81 & 0.47 & 3.93 & 1.2181 & 1.3458 \\
$\eta_V=0.10$  & 2.15 & 11.37 &  12.02 &  12.08 &  385 & 21.76 & 0.52 & 4.41 & 1.2239 & 1.3431 \\
$\eta_V=0.12$ & 2.25 & 11.66 &   12.28 &  12.18 &  405 & 19.31 & 0.57 & 5.00 & 1.2995 & 1.3419 \\
$\eta_V=0.14$  & 2.34 & 11.91 &  12.47 &  12.24 &  424 & 17.99 & 0.66 & 5.65 & 1.2440 & 1.3383 \\
$\eta_V=0.16$  & 2.45 & 12.11 &  12.63 &  12.27 &  440 & 17.36 & 0.99 & 6.35 & 1.2467 & 1.3448 \\
$\eta_V=0.18$  & 2.56 & 12.34 &  12.76 &  12.34 &  455 & 16.89 & 0.99 & 7.11 & 1.2520 & 1.3378 \\
$\eta_V=0.20$  & 2.69 & 12.59 &  12.85 &  12.36 &  466 & 15.67 & 0.99 & 7.94 & 1.2539 & 1.3277 \\
\hline
APR--nlNJL $\eta_D=0.79$: \\
$\eta_V=0.06 ^{\ \ddagger}$  & 1.92 & 10.36 &  -- -- &  11.72 &  296 & 26.39 & 0.45 & 2.94 & 1.1974 & 1.3376 \\
$\eta_V=0.08$  & 2.00 & 10.62 &  10.62 &  11.85 &  324 & 24.90 & 0.49 & 3.29 & 1.1994 & 1.3415 \\
$\eta_V=0.10$  & 2.07 & 10.89 &  11.44 & 11.96 &  350 & 23.49 & 0.55 & 3.70 & 1.2062 & 1.3392 \\
$\eta_V=0.12$  & 2.15 & 11.09 &  11.87 &  12.04 &  374 & 21.97 & 0.66 & 4.13 & 1.2126 & 1.3337 \\
$\eta_V=0.14$  & 2.25 & 11.20 &  12.16 &  12.12 &  396 & 21.79 & 0.99 & 4.68 & 1.2177 & 1.3510 \\
$\eta_V=0.16$  & 2.36 & 11.39 & 12.39 &  12.21 &  416 & 20.53 & 0.99 & 5.25 & 1.2185 & 1.3636 \\
$\eta_V=0.18$  & 2.47 & 11.62 &  12.57 &  12.26 &  433 & 19.59 & 0.99 & 5.88 & 1.2221 & 1.3536 \\
$\eta_V=0.20$  & 2.58 & 11.94 &  12.70 &  12.30 &  449 & 18.29 & 0.99 & 6.62 & 1.2301 & 1.3359 \\
\hline
DD2p40--nlNJL: \\
$\mu_< = 950$   & 2.14 & 10.46 &  11.13 &  11.37 &  227 & 25.64 & 0.70 & 3.58 & 1.2028 & 1.3037 \\
$\mu_< = 1000$ & 2.12 & 10.39 &  11.03 &  11.46 &  223 & 26.29 & 0.68 & 3.41 & 1.1879 & 1.3046 \\
$\mu_< = 1050$ & 2.11 & 10.38 &  10.99 &  11.72 &  237 & 26.58 & 0.71 & 3.25 & 1.1804 & 1.2979 \\
$\mu_< = 1100$ & 2.14 & 10.13 &  10.93 &  11.98 &  283 & 27.94 & 0.86 & 3.15 & 1.1702 & 1.3003 \\
$\mu_< = 1150$ & 2.12 & 10.06 &  10.88 &  12.51 &  400 & 28.05 & 0.90 & 3.05 & 1.1667 & 1.3019 \\
$\mu_< = 1200 ^\dagger$ & 2.13 & 10.05 &  10.81 &  13.80 &  913 & 29.63 & 0.99 & 2.88 & 1.1600 & 1.2870\\
$\mu_< = 1250 ^\dagger$ & 2.08 & 9.94 &  10.65 &  13.86 &  969 & 29.88 & 0.99 & 2.77 & 1.1560 & 1.2898 \\
$\mu_< = 1300 ^\dagger$ & 2.04 & 9.93 &  10.50 &  13.90 &  982 & 30.38 & 0.99 & 2.67 & 1.1519 & 1.3034 \\
$\mu_< = 1350 ^{\dagger,\ddagger}$ & 2.02 & 9.95 &  10.31 &  13.90 &  983 & 31.69 & 0.99 & 2.59 & 1.1496 & 1.2948 \\
$\mu_< = 1400 ^{\dagger,\ddagger}$ & 1.99 & 9.94 &  -- -- &  13.90 &  983 & 30.23 & 0.99 & 2.54 & 1.1489 & 1.2982 \\
\hline
DD2F--VBAG(a) & 1.90 & 11.19 &   -- -- &  12.48 &  438 & 22.66 & 0.35 & 3.30 & 1.2026 & 1.3578 \\
DD2F--VBAG(b) & 2.10 & 11.79 &  12.51 &  12.76 &  552 & 18.77 & 0.35 & 4.44 & 1.2218 & 1.3608 \\
MITBAG(139) & 2.05 & 11.67 &  12.23 &  12.46 & 562 & 19.64 & 0.34 & 4.60 & 1.2468 & 1.3729 \\
MITBAG(145) & 2.01 & 11.70 &  11.91 &  13.02 &  557 & 20.56 & 0.34 & 3.84 & 1.2037 & 1.3533 \\
\hline
DD2F-RDF 1.1 & 2.13 & 10.12 & 10.93 &  12.42 &  425 & 29.20 & 0.85 & 3.19 & 1.1758 & 1.2897 \\ 
DD2Fev-RDF 1.2  & 2.15 & 10.90 &  11.56 &  12.43 &  424 & 23.53 & 0.44 & 3.86 & 1.1981 & 1.3104 \\ 
DD2F-RDF 1.3 & 2.02 & 10.26 &  10.71 &  12.41 &  426 & 28.89 & 0.65 & 3.03 & 1.1787 & 1.2982 \\ 
DD2F-RDF 1.4  & 2.02 & 10.29 &  10.73 &  12.41 &  426 & 26.48 & 0.64 & 3.03 & 1.1766 & 1.3117 \\ 
DD2F-RDF 1.5  &  2.03 & 10.24 &  10.73 &  12.41 &  426 & 28.89 & 0.65 & 3.05 & 1.1789 & 1.3000 \\ 
DD2F-RDF 1.6$^{\ \ddagger}$  & 2.00 & 9.99 &  10.31 &  12.42 & 426 & 29.56 & 0.65 & 2.82 & 1.1704 & 1.3190 \\ 
DD2F-RDF 1.7  &  2.11 & 10.71 &  11.42 &  12.42 &  426 & 24.38 & 0.64 & 3.54 & 1.1898 & 1.3102 \\ 
\hline
DDQuark$\eta_{\rm V}$08 $\rho_{\rm tr}=2.5 ^{\ \dagger,\ddagger}$&   1.99 & 13.65 &   -- -- &   14.48 &  893 & 11.73 & 0.07 & 4.59 & 1.2112 & 1.3611 \\
DDQuark$\eta_{\rm V}$08 $\rho_{\rm tr}=3.0 ^{\ \dagger,\ddagger}$&   2.08 & 13.91 &  14.05 & 14.46 &  893 & 15.41 & 0.16 & 5.35 & 1.2244 & 1.3544 \\
\hline
DD2F--CSS(a)   &  2.21 & 10.70 &  11.37 & 11.23 &  254 & 23.61 & 0.60 & 4.39 & 1.2305 & 1.3242 \\
DD2F--CSS(b)   &  2.16 & 10.85 &  11.76 &  12.37 &  433 & 23.61 & 0.60 & 3.93 & 1.1994 & 1.3386 \\
\hline
\end{tabular}
\begin{flushleft}
{$^\dagger$:~Excluded due to the tidal deformability constraint \citep[see][and references therein]{TheLIGOScientific:2018}} \\
{$^\ddagger$:~Excluded due to the NICER constraints \citep[][]{Riley:2019,Miller:2019,Riley2021ApJ,Miller2021ApJ}}

\end{flushleft}
\label{tab:EoS-properties_hybrid_T0}
\end{table*}

\begin{table*}
\centering
\caption{The same as Table~\ref{tab:EoS-properties_hybrid_T0} but for the considered hybrid EoS at $s=3~k_{\rm B}$ and a constant lepton fraction $Y_{\rm L}=0.3$.}
\begin{tabular}{l c c c c c c c c}
\hline
EoS & $M_{\rm max}$ & $R_{\rm max}$ & $\left.\rho_{\rm central}\right\vert_{M_{\rm max}}$  & $\left.c^2_{s\rm ,central}\right\vert_{M_{\rm max}}$ & $\left.I^*\right\vert_{M^*_{\rm max}}$ &$M^*_{\rm max}/M_{\rm max}$ & $R^*_{\rm max}/R_{\rm max}$   \\ 
 &  $[$M$_\odot]$ & $[$km$]$ & $[10^{14}\,\,\,$g~cm$^{-3}]$ & $[$c$^2]$ &$[10^{45}\,\,\,$g~cm$^{2}]$& & \\
\hline
DD2F--VBAG(a) & 1.97 & 17.71 & 11.08 & 0.01 & 4.18 & 1.0984 & 1.3718 \\
DD2F--VBAG(b) & 1.92 & 13.16 & 22.02 & 0.36 & 2.56 & 1.0893 & 1.3480 \\
MITBAG(139) & 2.25 & 14.36 & 16.55 & 0.35 & 4.31 & 1.1152 & 1.3398\\
MITBAG(145) & 2.18 & 14.48 & 17.43 & 0.37 & 3.86 & 1.1021 & 1.3457\\
DD2F-RDF 1.1 & 2.12 & 11.17 & 28.58 & 0.84 & 2.60 & 1.0969 & 1.2916 \\
DD2Fev-RDF 1.2 & 2.17 & 12.17 & 22.99 & 0.44 & 3.11 & 1.1035 & 1.3146  \\
DD2F-RDF 1.3 & 2.03 & 11.51 & 29.18 & 0.62 & 2.43 & 1.0919 & 1.3156 \\
DD2F-RDF 1.4 & 2.02 & 11.59 & 28.29 & 0.62 & 2.42 & 1.0900 & 1.3218\\
DD2F-RDF 1.5 & 2.04 & 11.34 & 26.78 & 0.62 & 2.47 & 1.0931 & 1.3325 \\
DD2F-RDF 1.6 & 2.01 & 11.19 & 29.45 & 0.78 & 2.28 & 1.0899 & 1.3147\\
DD2F-RDF 1.7 & 2.12 & 12.10 & 24.53 & 0.50 & 2.84 & 1.0974 & 1.3056 \\
DD2-RDF 1.8 & 2.09 & 11.37 & 26.33 & 0.61 & 2.66 & 1.1035 & 1.2977 \\
DD2-RDF 1.9 & 2.21 & 12.09 & 22.83 & 0.44 & 3.31 & 1.1147 & 1.3003\\
\hline
\end{tabular}
\label{tab:EoS-properties_hybrid_s3}
\end{table*}

\subsubsection{Non-relativistic semiphenomenological models}

Our collection of non-relativistic EoS includes the microscopic EoS (APR)
of \citet{AP:1997} and \citet{APR:1998}. It is based on variational theory and an Argonne class of
two-body potentials plus semiphenomenological Urbana class of three-body forces. 

The remaining non-relativistic models are based on mean-field theory and Skyrme functional. The LS models of \cite{Lattimer:1991}, which in addition to the description of bulk nuclear matter incorporate a compressible liquid-drop model for nuclei, are distinguished by their value of nuclear compressibility $K$ and we use $K=220$~MeV and $K=375$~MeV models that are denoted as LS220 and LS375, respectively. 

Finally, we use also two models (BSk20 and BSk21) from the Brussel-Montreal class of models. They are likewise based on a Skyrme functional. These models have been applied to nuclei (and the model parameters were tuned) using the Hartree-Fock-Bogoliubov theory of finite nuclei \citep[see][]{Pearson2018,Goriely:2010}.  

\subsubsection{Finite-temperature equations of state}

Some of the models and their EoS were extended to finite temperatures. We employ in our collection the finite-temperature extensions of DD2, DD2F, DD2esym, LS220, LS375, NL3, SFHo/x, TM1, TMA, IUFSU, and BHB$\Lambda\phi$ models. These have already been used extensively in dynamical studies of binary neutron star mergers and core-collapse supernovae. As a representative value for the (averaged) entropy per particle for the finite-temperature stars, we chose $s=3~k_{\rm B}$, which is motivated by core-collapse supernova studies~\citep[for a comprehensive review about the role of the EoS in simulations of core-collapse supernovae, see][and references therein]{Fischer:2017}.

 In the case of $T=0$, we assume the usual condition of $\beta$-equilibrium, i.e. zero neutrino chemical potential $\mu_{\nu_e}=0$, which resembles the case of equal electron lepton abundance and electron fraction, $Y_{\rm L}\equiv Y_e$. Contrary, in the case of finite entropy, which corresponds to temperatures on the order of several tens of MeV at the interior of these protoneutron stars, this assumption cannot be justified. In appendix~\ref{sec:appendixB}, we provide a comparison of these cases at the example of the DD2 RMF hadronic EoS. It results in a moderate enhancement of the radii and maximum masses. Hence, for the determination of the universal relations for the maximum masses and corresponding radii, we select a representative and constant value of the electron lepton fraction of $Y_{\rm L}=0.3$ \citep[][]{Huedepohl:2010,Fischer:2009af,Fischer:2020PhRvC101}. Note that we omit a finite muon lepton fraction since it has been demonstrated recently in radiation-hydrodynamics simulations of core-collapse supernovae, featuring six species neutrino transport including muons and associated muonic weak interaction, that the abundance of both muons and muon (anti)neutrinos is generally much lower than those of the electrons and electron (anti)neutrinos leaving a negligible impact on the structure and stability of hot neutron stars \citep[cf.][]{Bollig:2017,Fischer:2020PhRvC101,Fischer:2021PhRvD}.

\subsection{Global parameters}
Next, we turn to some (selected) properties of the models listed above. Table~\ref{tab:EoS-properties_hadron_T0} tabulates the following global parameters of 
the configurations: the maximum static and Keplerian masses, corresponding radii and the radius $R_{1.4}$ of a $M = 1.4~M_{\odot}$ star together with the corresponding tidal deformability $\Lambda_{1.4}$. Table~\ref{tab:EoS-properties_hadron_s3} provides the same quantities for hot configurations.  In comparison to the zero-temperature cases, one observes systematically larger radii and higher maximum masses.  The mass-radius relations for the cold and $\beta$-equilibrium EoS are shown in Fig.~\ref{fig:MR_hadronic_cold}. The mass-radius relations for hot configurations at fixed $s = 3~k_{\rm B}$ and constant electron lepton fraction of $Y_{\rm L}=0.3$ is shown in Fig.~\ref{fig:MR_hadronic_hot}. These panels include both static and maximally rapidly rotating configurations. Note that the radii of the rapidly rotating configuration correspond to the equatorial radii. Clearly, they exhibit both the well-known enhancement of the maximum mass of a sequence when rotation is included and the increase in the circumferential radius at the equator due to the centrifugal force for rotating stars.

In closing our review of hadronic models, we would like to stress that all models are subject to the maximum-mass constraint and radius constraints, specifically on the tidal deformability from the GW170817 event~\citep{TheLIGOScientific:2017qsa,Lattimer:2018,TheLIGOScientific:2018} and on the radii of canonical and $2M_{\odot}$ stars from NICER experiment~\citep{Riley2021ApJ,Miller2021ApJ}.
 This leads to a narrowing of the original set of EoS.
Especially, we point out that several hadronic models
predict tidal deformabilities that are incompatible with the constraint on this quantity deduced from the analysis of GW170817 event at the 90\% credible level \citep[][]{TheLIGOScientific:2018}. 
 Those EoS are marked in Table~\ref{tab:EoS-hadron} with a $\dagger$-symbol. Models that are excluded due to the radius constraints obtained from the analysis of NICER data \citep[][]{Riley:2019,Miller:2019,Riley2021ApJ,Miller2021ApJ}
are marked with a $\ddagger$-symbol.
Furthermore, the sound speed for several models
at the maximum mass exceeds the speed of light, see Table~\ref{tab:EoS-hadron}, marked by a $\diamond$-symbol, which implies that the EoS based on these models become acausal on the stable branch of the respective stellar sequence.
 Ab initio, non-relativistic nuclear EoS based on potential models such as the Brueckner-Hartree-Fock calculations \citep[cf.][]{Burgio:2010ek,Wei:2021veo} will be investigated in more detail in a future study \citep[see also][and references therein]{Lu:2019mza}.


\subsection{Hybrid models}
\label{sec:EoS_hybrid}
The set of $T=0$ hybrid EoS used in this study is given in Table~\ref{tab:EoS-properties_hybrid_T0}.  The models are distinguished by the approaches by which the EoS of the quark phase was derived; in any case, the models considered here are phenomenological to a different extent, ranging from Lagrangian-based ones to schematic parametrizations. The hadronic EoS to which the quark phases are matched either via Maxwell construction or mixed-phase construction are also indicated in Table~\ref{tab:EoS-properties_hybrid_T0}.  Below we briefly review the models pertaining to the quark matter. \\

\begin{figure}
\includegraphics[width=0.475\textwidth]{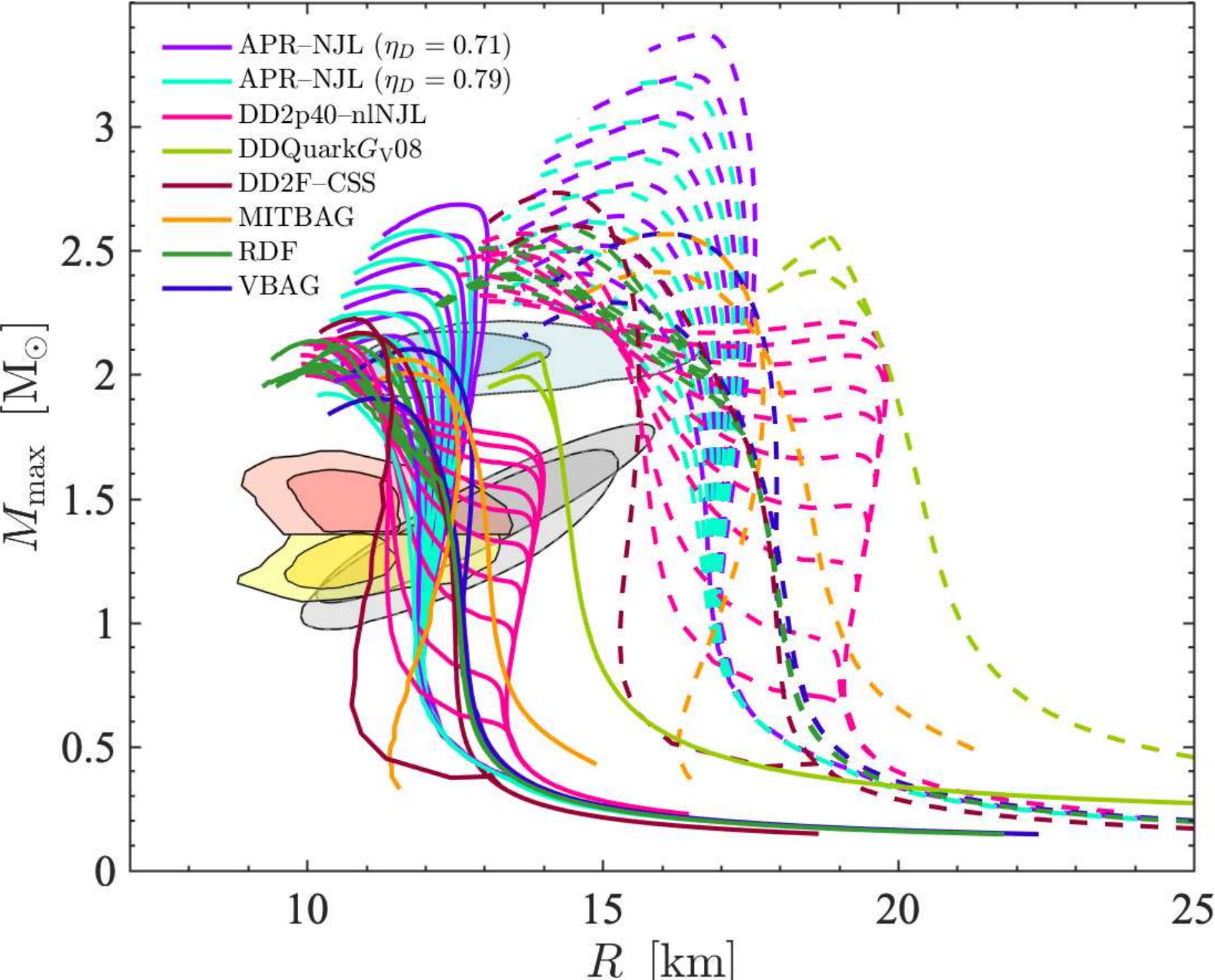}
\caption{(colour online) The same as Fig~\ref{fig:MR_hadronic_cold} but for the cold hybrid EoS.}
\label{fig:MR_hybrid_cold}
\end{figure}

\begin{figure}
\includegraphics[width=0.475\textwidth]{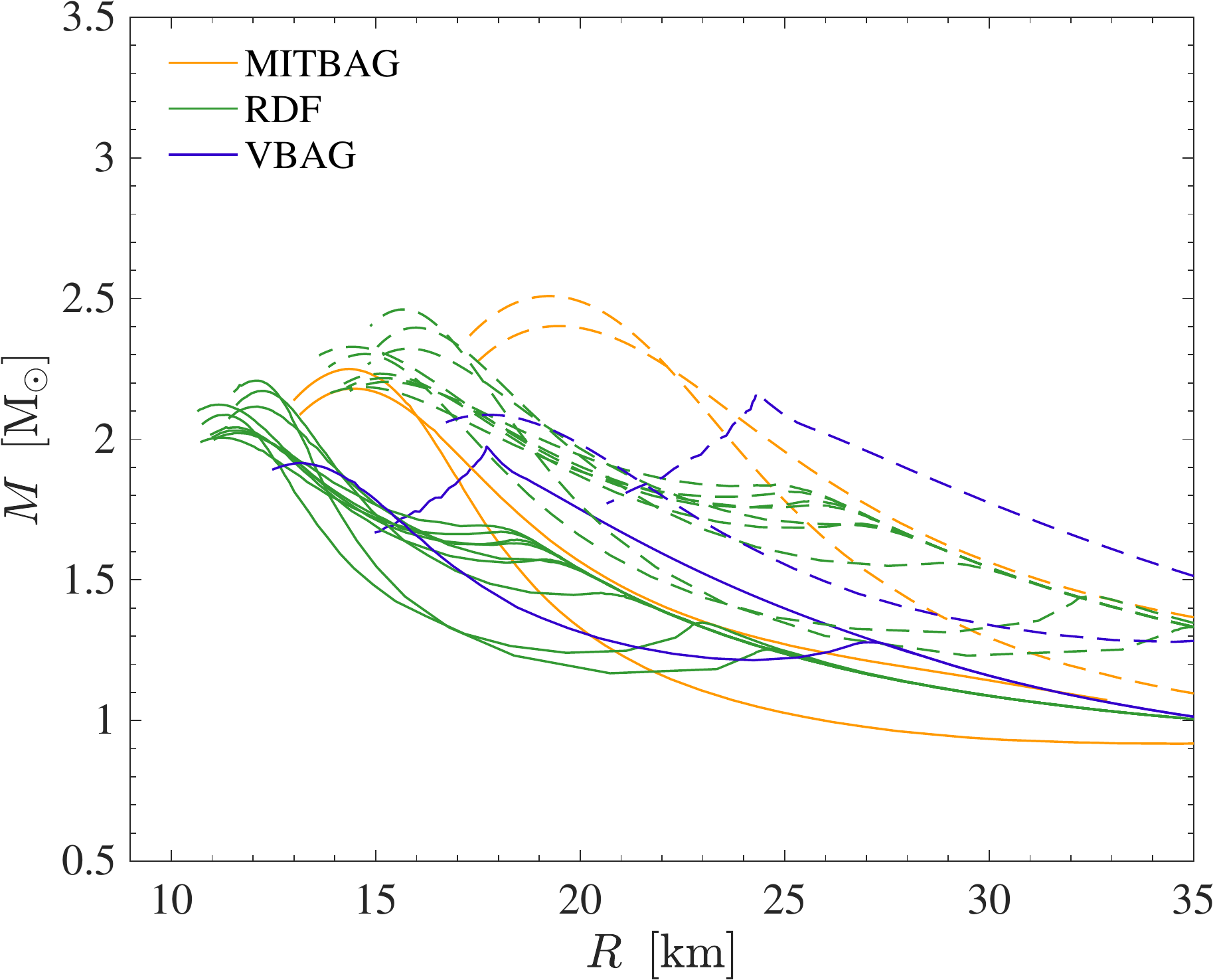}
\caption{(colour online) The same as Fig~\ref{fig:MR_hadronic_hot} but for the hot hybrid EoS.}
\label{fig:MR_hybrid_hot}
\end{figure}

\subsubsection{Bag models}
This class of hybrid EoS solves the non-interacting Fermi-gas integrals for the particle densities $n$, pressure $P^{\rm FG}$, energy density $\epsilon^{\rm FG}$ and entropy $S^{\rm FG}$, for arbitrary degeneracy. In the case of two quark flavours, one has to consider two chemical potentials correspondingly, $\mu_u$ and $\mu_d$. In the case of strange matter, one has a third chemical potential for the strange quarks, which is assumed to be equal to that of the down quarks due to weak equilibrium, $\mu_s=\mu_d$, which is considered the main channel for the production and equilibration of strangeness in astrophysical systems. The final EoS is then shifted by a constant bag pressure $B$, as follows: $P=P^{\rm FG}-B$ and $\epsilon=\epsilon^{\rm FG}+B$, which ensures the Euler equation to be fulfilled $P+\epsilon = S T - \mu_i\,n_i$. Here, we employ two bag model EoS for three-flavour quark matter, one with a bag constant of $B^{1/4}=139$~ MeV and the second with $B^{1/4}=145$~ MeV, henceforth denoted as MITBAG(139) and MITBAG(145),  respectively, both with corrections due to the strong coupling constant of $\alpha_S=0.7$~\citep[further details can be found in][and references therein]{Fischer:2011}. These hybrid MITBAG EoS feature a phase transition based on Gibb's construction, from the RMF model of \citet{Hempel:2010}, with the TM1 parametrization \citep[for further details, see][]{Sagert:2009,Fischer:2011}.

An extension to this ``classical'', thermodynamic bag model is the vector-interaction-enhanced bag model (VBAG), which was derived in \citet{Klaehn:2015} from the QCD Dyson-Schwinger equations truncated by assuming a gluon propagator constant in momentum space~\citep[effectively imposing only quark--quark contact interactions; cf.][]{Cierniak:2017dxr,Cierniak:2018}. It can be related to the NJL models for deconfined quarks. VBAG differs from ordinary thermodynamic bag models by the inclusion of effective vector repulsion. This is realized by shifting the chemical potential $\mu$ by a density-dependant factor $\mu^* = \mu - K_V \cdot n$. The parameter $K_V$ is related to the coupling constant of vector interactions in NJL models. This shift is necessary to achieve a quark matter EoS capable of supporting a stable neutron star with mass $\gtrsim$2~M$_\odot$~\citep[cf.][]{Sagert:2009}. All other calculations of the VBAG model EoS follow closely those of the thermodynamic bag models, i.e. solving Fermi-gas integrals for arbitrary degeneracy, however, defined with respect to $\mu^{*}$. The resulting thermodynamic quantities are shifted to match those of the  NJL model in the chirally symmetric phase, hence this shift is labelled as the chiral bag constant $B_\chi$ and the chiral chemical potential is defined as $P_{(VBAG)}(\mu_\chi)=0$. The advantage of VBAG over NJL is that no gap equations have to be solved. To avoid inconsistencies in the description of chiral physics between VBAG and an independent low-density hadronic EoS, another shift is introduced to VBAG, the deconfinement bag $B_{dc}$, defined as the hadron EoS pressure at $\mu_\chi$. This ensures that chiral symmetry restoration and deconfinement coincide and all chiral effects are present exclusively on the quark side. $B_{dc}$ is a function of temperature and isospin asymmetry, aspects of which are explored in detail in \citet{Klaehn:2017}. Here we employ the DD2F-VBAG hybrid EoS for the following two-parameter sets, with chiral bag constant $B_{\chi}^{1/4}=165$~MeV and vector coupling constant $K_{V}^{1/2}=10^{-3}$~MeV$^{-1}$ denoted as DD2F--VBAG(a), and the second one with chiral bag constant $B_{\chi}^{1/4}=130$~MeV and vector coupling constant $K_{V}^{1/2}=2\times 10^{-3}$~MeV$^{-1}$, denoted as DD2F--VBAG(b) \citep[for more details, also about the definition of the parameters, see][]{Cierniak:2018}.

\subsubsection{String-flip models}
This class of quark matter EoS models is based on the relativistic density functional (RDF) approach in the mean-field approximation, where an effective quark-quark interaction potential $U$ is expanded around the expectation values of the bilinear forms in the quark fields $q$ and $\bar{q}$ that define the scalar and vector densities $n_{\rm S}=\langle \bar{q}q\rangle$ and $n_{\rm V}=\langle\bar{q}\gamma_0 q\rangle$, respectively.
The thermodynamic potential is given by the following expression:
\begin{equation}
\Omega = \Omega^{\rm quasi} + U(n_{\rm V},n_{\rm S}) - n_{\rm V} \Sigma_{\rm V} - n_{\rm S} \Sigma_{\rm S}~,
\end{equation}
with scalar and vector self-energies, $\Sigma_{\rm S}$ and $\Sigma_{\rm V}$, respectively. Further details can be found in \citet{Bastian:2018}. The density functional takes the following explicit form: $U(n_{\rm V},n_{\rm S})=D(n_{\rm V})n_{\rm S}^{2/3} + a n_{\rm V}^2 + b n_{\rm V}^4/(1+c n_{\rm V}^2)$, which has been related to an effective Lagrangian, $\mathcal{L}_{\rm eff}=\mathcal{L}_{\rm free}-U$, in \citet{Kaltenborn:2017}. The first term stems from a confining interaction between quarks that is saturated within nearest neighbours \citep[this is the string-flip concept of][]{Ropke:1986qs}
and contains a string tension parameter $D(n_{\rm V})= D_0 \Phi(n_{\rm V})$ with a medium modification through an excluded volume approach, $\Phi(n_{\rm V})=\exp\{-\alpha(n_{\rm V}-n_0)^2\}$.
This first term gives rise to divergent masses with decreasing density and provides a mechanism for confinement, depending on the parameters $D_0$, $\alpha$, and $n_0$. The second term of $U$ is the usual linear vector repulsion well known from NJL-like models, with the vector coupling constant $a$. The third term of $U$ is interpreted as higher order repulsive vector interaction, depending on the parameter $b$, which can be motivated by NJL-like models that include eight-quark interactions according to \citet{Benic:2015}. Note that the denominator has been introduced in \citet{Kaltenborn:2017} to ensure that the speed of sound never exceeds the speed of light, however, at the cost of an additional parameter $c$.

The model was extended to finite temperature and arbitrary isospin asymmetry, featuring a isovector $\rho$-meson interaction equivalent term in \citet{Fischer:2018} and \citet{Bastian:2021}. In this study, we consider all available nine parametrizations denoted by RDF~1.1--1.9, which cover a  wide range of onset densities for quark matter \citep[details, including the values of the parameters used, can be found in][]{Bastian:2021}. All of these aforementioned hadron-quark matter hybrid EoS feature a first-order phase transition. For all of them, a standard Maxwell construction was applied for the phase transition construction. 

\subsubsection{NJL Lagrangian based models}
The class of NJL Lagrangian-based models is constructed using four-fermion interaction terms in the Lagrangian with local (constant) couplings while gluons are integrated out from the theory. 
NJL models are well suited for describing chiral symmetry breaking and restoration in the matter, but lack confining features. The original NJL Lagrangian included scalar and pseudo-scalar interaction terms characterized by a common coupling constant $G_S$ that is fitted to the vacuum properties of the pion in this model~\citep{Klevansky:1992qe,Hatsuda:1994pi,Buballa:2003qv}. %
Subsequently, the model was extended to include ad hoc pairing interaction (or diquark term) that accounts for Cooper pairing amongst quarks with leading pairing channels being two-flavour pairing in $ud$ quark matter (2SC phase) and three-flavour pairing in $uds$ quark matter (CFL phase).
The first self-consistent solution for the mass and diquark pairing gaps in the three-flavour case was performed by~\citet{Blaschke2005} and \citet{Ruester2005}.  
It was subsequently used for studies of hybrid neutron star phenomenology, e.g. in \citet{Bonanno2012} and \citet{Klaehn2013}.
The corresponding diquark coupling $G_D$ is often measured in terms of scalar coupling $G_S$ via the (dimensionless) parameter $\eta_D = G_D/G_S$. A further extension that was introduced to allow for massive compact stars is the inclusion in the Lagrangian the vector interaction characterized by coupling constant $G_V$, which is frequently parametrized in terms of the parameter $\eta_V = G_V/G_S$ and 
serve for stiffening of the EoS of quark matter to describe massive hybrid stars \citep[cf.][and references therein]{Klahn:2006iw}.

Here we consider some example cases of NJL Lagrangian-based models. First, we include a model derived in \citet{Ayvazyan:2013}, which corresponds to their set-up with $\eta_{\rm V}= 0.8$ (hereafter DDQuark$\eta_{\rm V}0.8$) model; their EoS of nucleonic matter corresponds to the density-dependent relativistic functionals with a parametrization similar (but not identical) to the DD2 models described above and it is matched to the quark EoS via standard Maxwell construction. This model has a low-density 2SC pairing phase and a high-density CFL pairing phase. Secondly, we use two more hybrid EoS that are based on the matching of the non-relativistic nucleonic EoS to a non-local quark NJL EoS of~\citet{Ayriyan:2021}, hereafter denoted as nlNJL. In this case, the nucleonic EoS is soft (as opposed to the previous case), whereas the stiffness of the quark EoS is varied in a range of values $0.06\le \eta_V \le 0.20$.
 In this case, there is only a single 2SC paired phase of quark matter and in contrast to the previous case, an interpolation between quark and nucleonic phases is used to match these EoS as described in \citet{Baym:2018} and \citet{Ayriyan:2021prr}. 
 Here we implement two sets of the nlNJL models for two selected values of the diquark coupling, $\eta_D=0.71$ and $\eta_D=0.79$, as representative cases, for both of which the vector coupling constant is varied systematically over a large range of allowed parameters $0.06\le \eta_V \le 0.20$. The corresponding, systematic increase of the hybrid star's maximum masses and radii are given in Table~\ref{tab:EoS-properties_hybrid_T0}. Thirdly, we use the so-called DD2p40-nlNJL EoS, which implements a transition from the class of DD2ev models to the non-local chiral NJL model as described in \citet{Alvarez:2019}. 
 Here the suffix "p40" denotes an excluded volume parameter ${\rm v}=4$~fm$^3$ \citep[for details, see][]{Typel:2016}, which results in substantial stiffening of the EoS. In addition, the quark matter EoS features a transition from a lower density parametrization to a higher density one using a smooth interpolation. For these models, a Maxwell critical point ($\mu_c, P_c$) construction between hadronic and quark matter has been applied. The effect of geometrical structures at the hadron-quark interface has been taken into account with an EoS replacement around the phase transition that is an interpolation that slightly deviates from the Maxwell case, i.e. around the Maxwell critical point by adding some extra pressure $\Delta P=0.02P_c$, as described in \citet{Abgaryan:2018}. All these DD2p40-nlNJL EoS models share the same $\Delta P$ value, whereas the chemical potential parameter $\mu_<$ of the switching function $f_<(\mu)$ for quark matter varies between $\mu_<=950$~MeV and $\mu_<=1400$~MeV \citep[see][and  Table~\ref{tab:EoS-properties_hybrid_T0}]{Blaschke:2020qqj}. Note that the $\mu_<$ parameter controls the onset density for the phase transition construction through the switching function. Its value is very close to $\mu_c$ where the effect of the switching interpolation is more pronounced than at lower chemical potentials.

\subsubsection{Constant speed of sound models}
The constant speed of sound (CSS) parametrizations are motivated by the results found in the NJL Lagrangian-based calculations of \citet{Bonanno2012} and \citet{Klaehn2013}.  They allow for a flexible parametrization of the quark EoS with the help of a few parameters, specifically the speed of sound in the quark phase $c_s$ and the value of pressure (or energy density) at the phase transition~\citep[for details, see][]{Zdunik2013,Alford:2013,Alford:2017qgh,Lijj_2021}. Our models based on CSS EoS for quark matter were obtained by applying a Maxwell construction to match
them to the DD2F nucleonic EoS.  The EoS is calculated for two different parameters, both of which employ a speed of sound of $c_s^2=0.6$ and the same pressure slope parameter of 140~MeV~fm$^{-3}$ but different bag constants of 130~MeV~fm$^{-3}$, denoted as DD2F--CSS(a), and 142~MeV~fm$^{-3}$ denoted as DD2F--CSS(b).
The latter exhibits the masquerade phenomenon 
\citep{Alford:2004pf}
in which the mass-radius relation differs only marginally from the reference hadronic DD2F EoS.

\subsubsection{Finite temperature equations of state and global parameters}
Presently, only a few finite temperatures hadron-quark hybrid EoS models are available for studies of hot compact stars. Table~\ref{tab:EoS-properties_hybrid_s3} lists our selection of EoS that includes besides the simplistic MITBAG models, the RDF EoS collection, and the two VBAG models introduced above. It also lists selected global properties, including maximum masses of non-rotating and maximally rotating configurations, of the hybrid stars for $s=3~k_{\rm B}$ and $Y_{\rm L}=0.3$. In Fig.~\ref{fig:MR_hybrid_hot} we show the mass-radius relations for corresponding hot hybrid stars. The $T=0$ and $\beta$-equilibrium counterparts are shown in Fig.~\ref{fig:MR_hybrid_cold}. Finite entropy MITBAG models have been studied by \citet{Hempel:2016} systematically, also for various selected and constant values of the lepton fraction. Similarly as reported in \citet{Hempel:2016}, we observe the appearance of a second maximum in the mass-radius relations, which is associated with the onset of the hadron-quark phase transition, for all hybrid models shown in Fig.~\ref{fig:MR_hybrid_hot}, for the example value of $s=3~k_{\rm B}$ and a constant $Y_{\rm L}=0.3$. 

The origin of this is the different thermal behaviours of hadronic and quark matter phases, e.g. the increase of the density jump between hadronic and quark matter phases with increasing temperature, partly due to the first-order phase transition constructions.

Particularly interesting is the appearance of an unstable branch at the transition between the hadronic and the quark matter phases for the hot cases, known as twin-phenomenon, which is absent for the $T=0$ cases. This is particularly pronounced for some of the RDF EoS parametrizations, which feature an early onset of quark matter. In the case of higher entropies per particle, e.g. $s>5~k_{\rm B}$ the second maximum becomes the global maximum. This is also the case for the VBAG(b) EoS here (blue lines in Fig.~\ref{fig:MR_hybrid_hot}). In addition, the finite entropy RDF EoS have been discussed in \citet{Fischer:2018} and \citet{Fischer:2021}, however, only for the two RDF~1.1 and DRF~1.2 EoS, respectively.

\begin{figure*}
\includegraphics[width=0.475\textwidth]{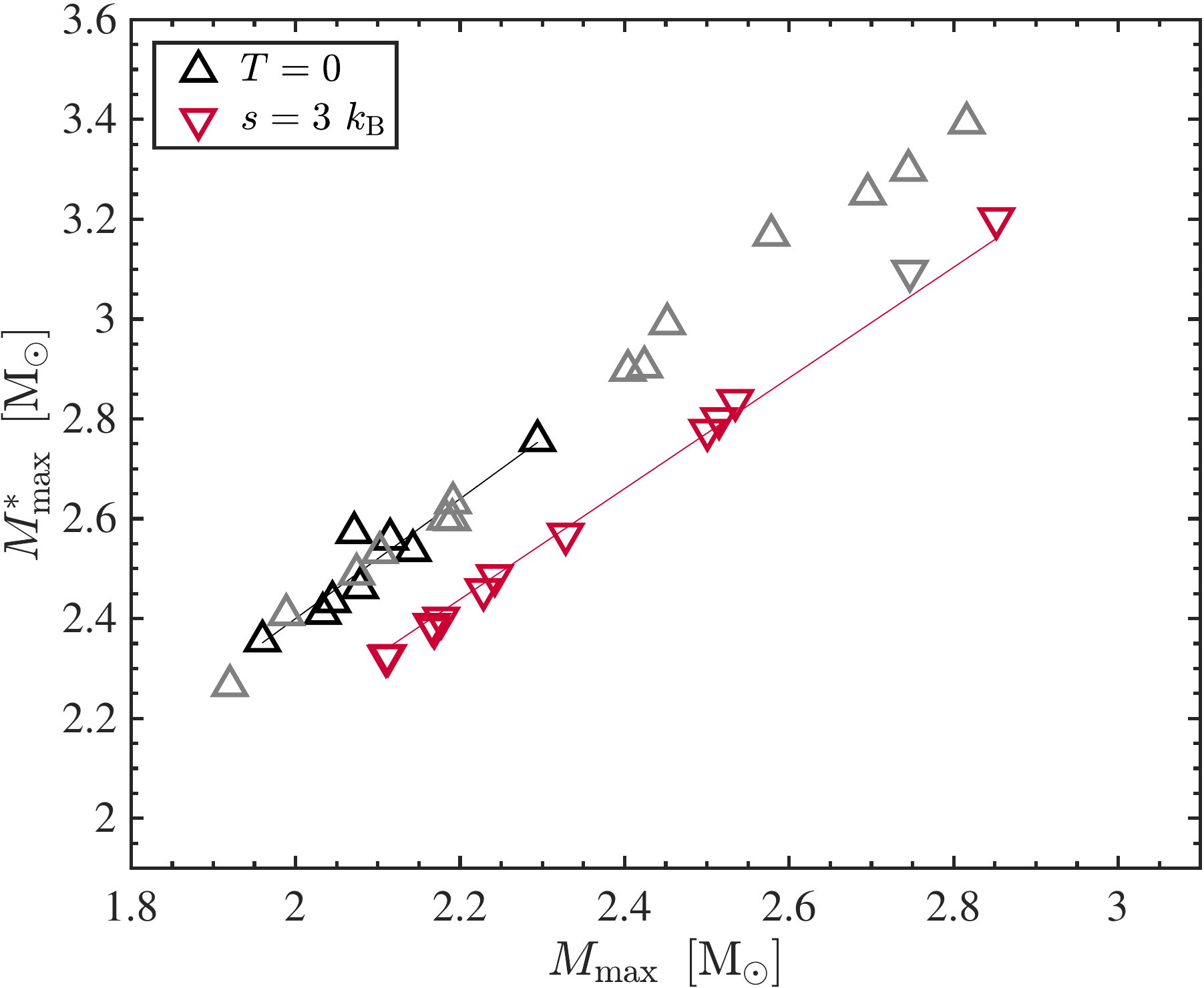}
\hfill
\includegraphics[width=0.475\textwidth]{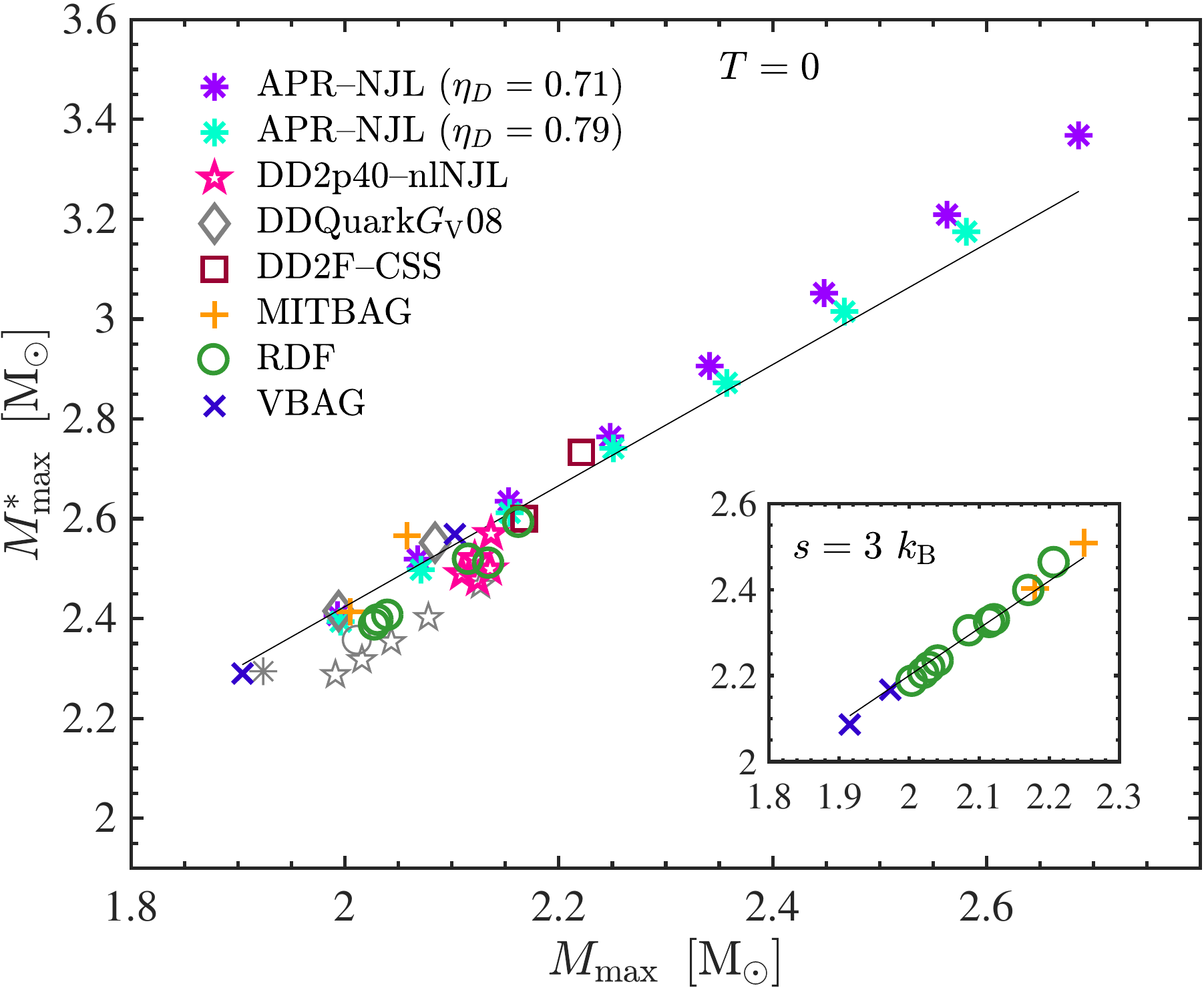}
\\
\includegraphics[width=0.475\textwidth]{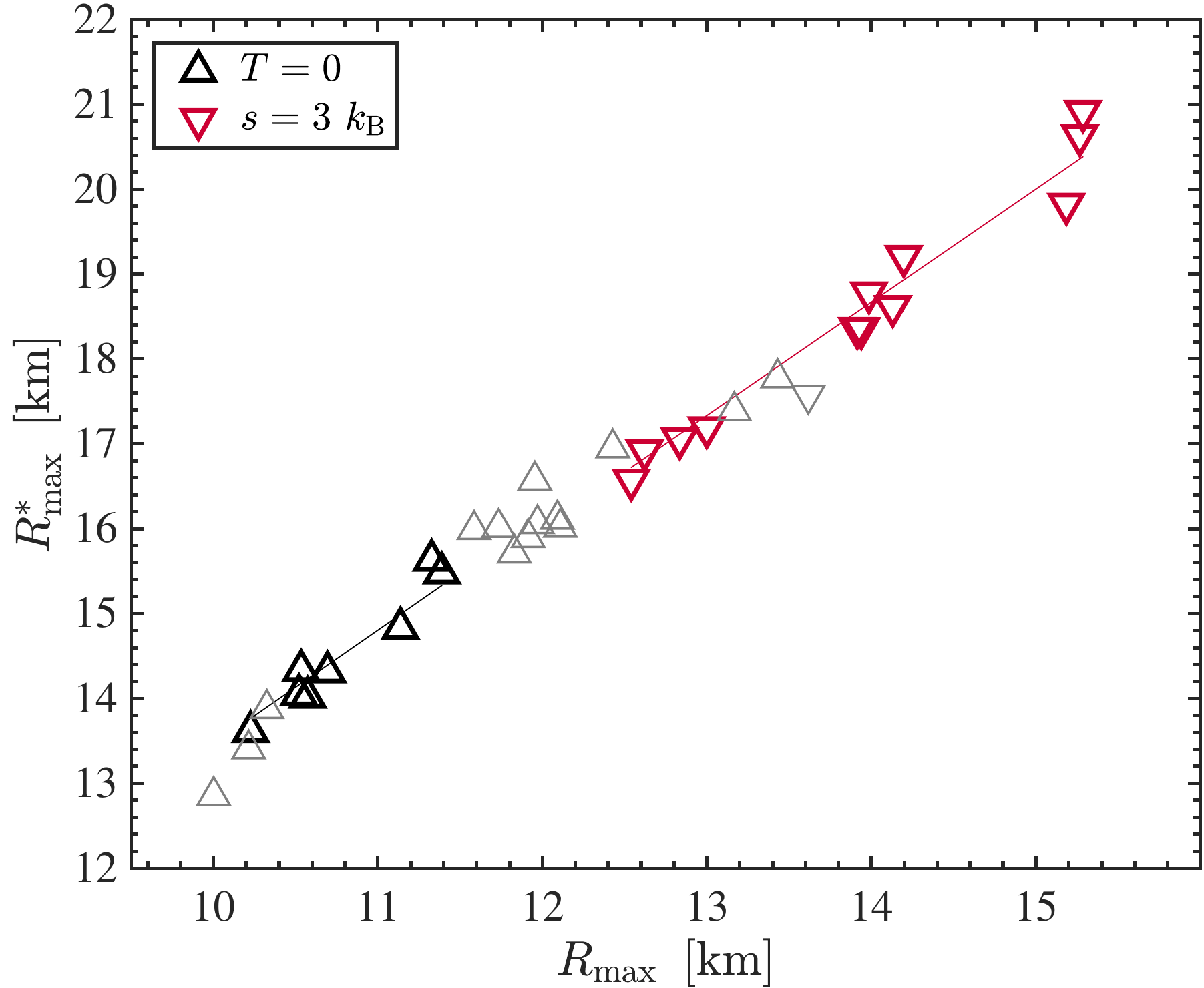}
\hfill
\includegraphics[width=0.475\textwidth]{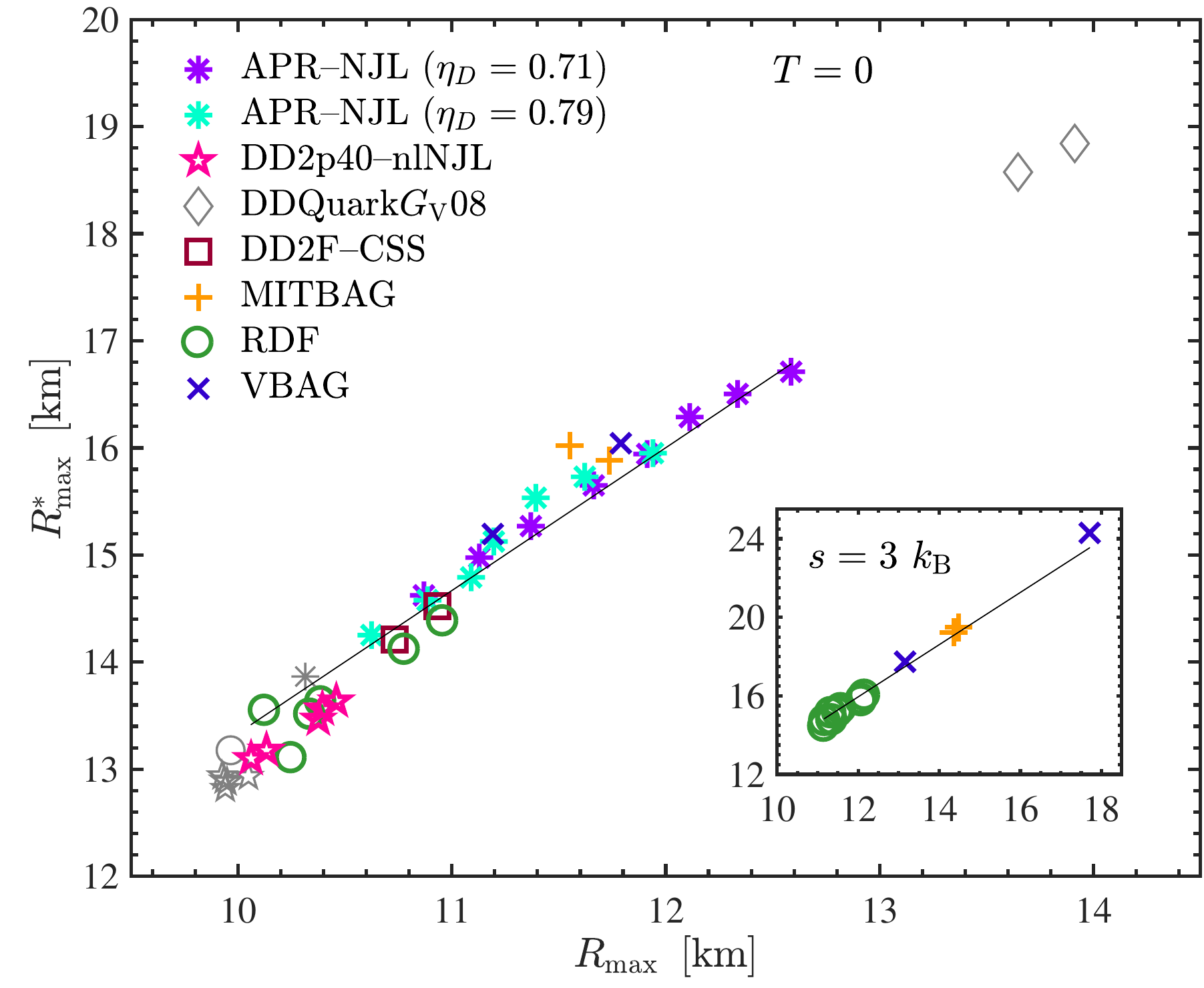}
\\
\includegraphics[width=0.475\textwidth]{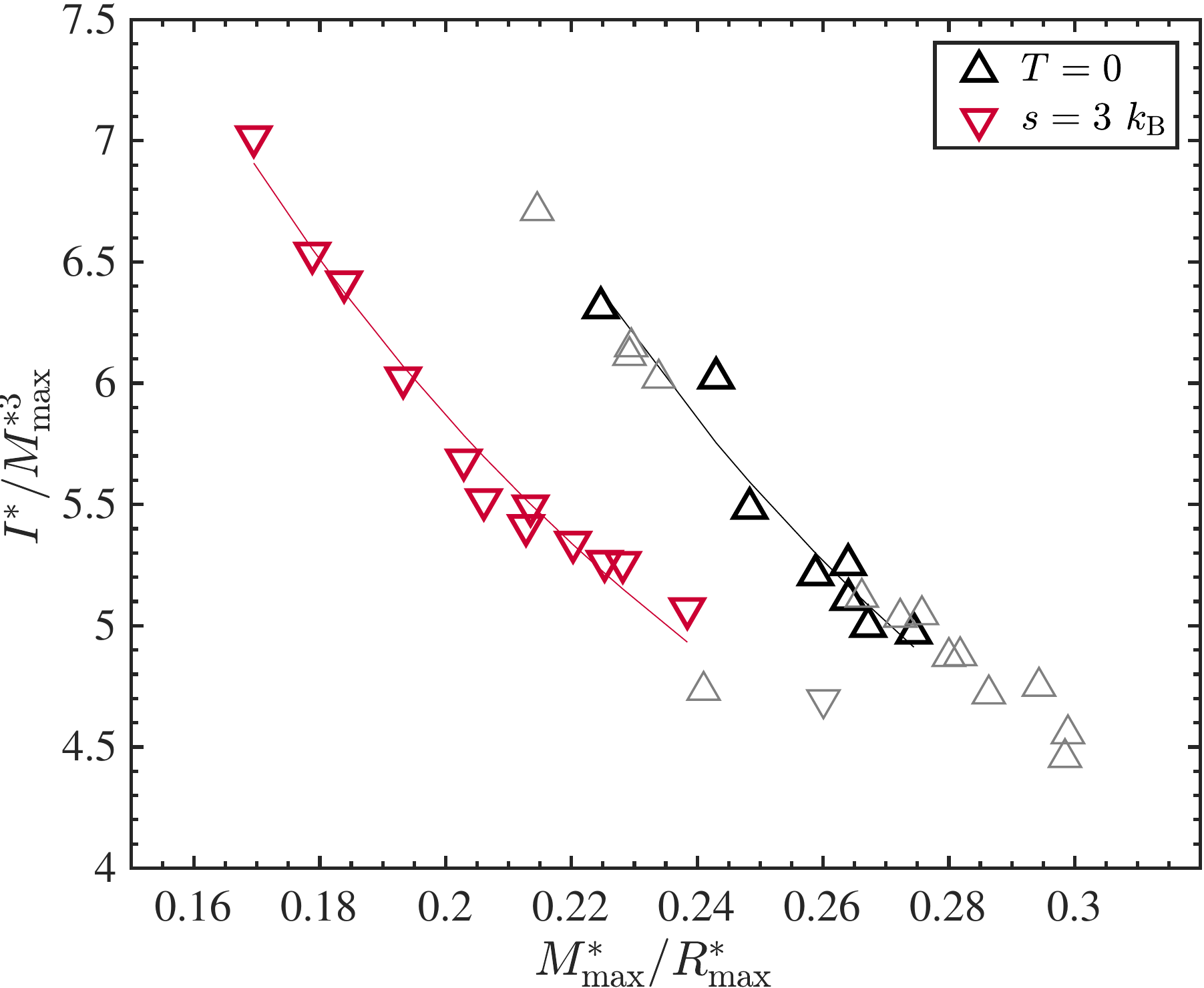}
\hfill
\includegraphics[width=0.475\textwidth]{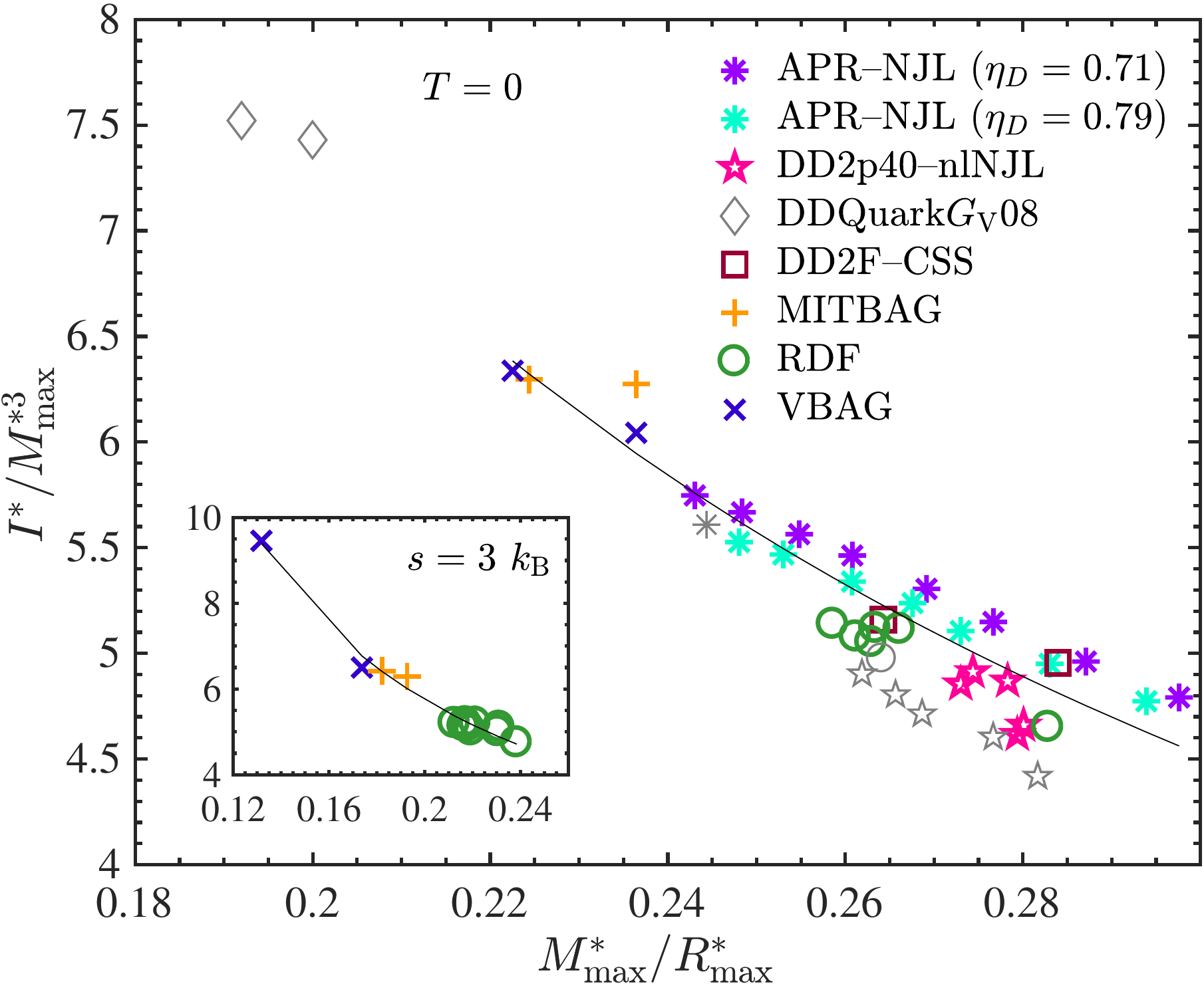}
\caption{(colour online) Relations for rapidly rotating compact stars, maximum masses, $M^*_{\rm max}$ and $M_{\rm max}$ (top panels), the corresponding radii, $R^*_{\rm max}$ and $R_{\rm max}$ (middle panels), and the normalized moment of inertia $I^*/M_{\rm max}^3$, given in geometric units, with respect to the compactness $M^*_{\rm max}/R^*_{\rm max}$ (bottom panels), for the $T=0$ and $s=3~k_{\rm B}$ EoS, distinguishing hadronic model (left-hand panels) and hybrid models (right-hand panels). The straight lines represent the universal relations~\eqref{eq:universal_M} and \eqref{eq:universal_R}, with the corresponding parameters $C_M$ and $C_R$ given in Table~\ref{tab:universal}, as well as relation~\eqref{eq:universal_I} with the coefficients $a_1$ and $a_2$ given in Table~\ref{tab:universal_I}.
 The grey markers belong to the EoS sets that are excluded from the analysis due to the constraints on the tidal deformability deduced from GW180817, together with radius constraints due to the NICER observations as well as the causality constraint. 
}
\label{fig:universal}
\end{figure*}

\section{Universal relations for rapidly rotating stars}
\label{sec:results}
In this section, we extend the studies of universal relations for isentropic finite-temperature compact stars of \cite{Khadkikar2021} to a comprehensive sample of hadron-quark matter hybrid models. We implemented the hadronic and hybrid EoS, introduced in Secs.~\ref{sec:EoS_hadronic} and \ref{sec:EoS_hybrid}, into the {\tt{RNS}} code to obtain the global parameters of non-rotating (Tolman-Oppenheimer-Volkov--TOV) and maximally rotating (Keplerian) stars. Having computed the relevant quantities we now evaluate the universal relations for the maximum masses and corresponding radii,
\begin{eqnarray}
M_{\rm max}^* &=& C_M\,M_{\rm max}~, \label{eq:universal_M} \\
R_{\rm max}^* &=& C_R\,R_{\rm max}~, \label{eq:universal_R}
\end{eqnarray}
where $M_{\rm max}$ and $M^*_{\rm max}$ are the TOV and Keplerian masses of configurations, respectively, $R_{\rm max}$ and $R^*_{\rm max}$ are their corresponding radii. The universality of relations~\eqref{eq:universal_M} and \eqref{eq:universal_R} implies that the coefficients $C_M$ and $C_R$ of the linear relation between the masses and between the radii of maximally rapidly rotating and non-rotating configurations are independent of the EoS. These relations were first obtained for zero-temperature and $\beta$-equilibrium matter. As shown in \cite{Khadkikar2021} they also hold for finite temperature matter at fixed (not very large) entropy per particle and constant lepton number fractions. Here, we test these relations of the representative value of entropy per particle $s=3~k_{\rm B}$ and a constant electron lepton fraction of $Y_{\rm L}=0.3$ and for the $T=0$ case assuming $\beta$-equilibrium. 

\begin{table}
\centering
\caption{Coefficients for the universal relations \eqref{eq:universal_M} and \eqref{eq:universal_R} including the maximum deviations.} 
\begin{tabular}{l c c c}
\hline
Condition & EoS type & $C_M$ & $C_R$ \\
\hline
$T=0$, $\beta$-eq. & hadronic &1.2000 $\pm 0.0160$ &1.3460 $\pm0.0160 $ \\
$s=3~k_{\rm B}$, $Y_{\rm L}=0.3$ & hadronic &1.1085 $\pm 0.0055 $ &1.3335 $\pm 0.0125 $ \\
$T=0$, $\beta$-eq. & hybrid   &1.2120 $\pm 0.0090$ & 1.3335 $\pm 0.0085 $ \\
$s=3~k_{\rm B}$, $Y_{\rm L}=0.3$ & hybrid   & $1.0995\pm 0.0055$ & $1.3290\pm 0.0160$ \\
\hline
\end{tabular}
\label{tab:universal}
\end{table}

The complete sample of hadronic mass-radius relations is shown in Figs.~\ref{fig:MR_hadronic_cold} and \ref{fig:MR_hadronic_hot}, for the cold and hot neutron star configurations respectively, including the non-rotating (solid lines) and the rapidly rotating setup (dashed lines), corresponding to the set of Table~\ref{tab:EoS-properties_hadron_T0}. We confirm the previous findings for purely hadronic models, i.e. systematically higher maximum masses and radii for the rapidly rotating stars, both, for cold and hot configurations. The universal relations for the maximum mass and the corresponding radii are shown in Fig.~\ref{fig:universal} in the top and bottom panels, respectively. The case of cold stars is shown by black triangles and the case $s=3~k_{\rm B}$ is shown by red triangles (upside-down). The linear dependencies for masses and radii, i.e. the coefficients $C_M$ and $C_R$ of the universal relations \eqref{eq:universal_M} and \eqref{eq:universal_R}, are listed in Table~\ref{tab:universal}. They correspond to the solid lines in Figs.~\ref{fig:universal} (top and middle panels), confirming the previous findings for both hot and cold models quantitatively. The error bars correspond to the maximum standard deviation from a linear regression without an intercept term.

It is interesting to note that the $s=3~k_{\rm B}$ maximum masses are systematically shifted to higher values for the non-rotating cases, in comparison to the $T=0$ cases, whereas the maximum masses are systematically shifted to lower values for the rapidly rotating cases. On the other hand, the $s=3~k_{\rm B}$ and $Y_{\rm L}=0.3$ configurations have systematically larger radii for both rotating and non-rotating models. The values we find for $C_M$ and $C_R$, for the $T=0$ cases, are in quantitative agreement of more than $98.5\%$ with \citet{Khadkikar2021}, i.e. they find $C_M=1.2187$ and $C_R=1.3587$, respectively, and the results obtained for the finite entropy of  $s=3~k_{\rm B}$ are   $C_M=1.0887$ and $C_R=1.3506$. For the latter, an equally tiny deviation is found in comparison to the results of \citet{Khadkikar2021}, even though they consider a constant value of $Y_e=0.4$. 

The universal relations for the hybrid models, for which the coefficients of the universal relations are given in Table~\ref{tab:universal} (bottom panel), are shown in Fig.~\ref{fig:universal} (top and middle panels), for the masses and radii, respectively. The corresponding mass-radius relations for these EoS are shown in the upper and lower
panels of Fig.~\ref{fig:MR_hybrid_cold} and \ref{fig:MR_hybrid_hot}. While we find an overall quantitative agreement with the universal relations for maximum masses and radii, see the coefficients $C_M$ and $C_R$ in Table~\ref{tab:universal}, the values of the coefficients in the hybrid cases are somewhat different than for the hadronic cases for $T=0$. This can be attributed to the systematic deviation of some of the extreme parametrizations of the $\eta$NJL hybrid EoS class, for which the largest maximum masses and radii are obtained for the highest values of the vector coupling constants resulting in a speed of sound close to the speed of light. Note that the different quark matter EoS parameters for $\eta$NJL provide systematically additional pressure with increasing density due to the increasing vector coupling constants $\eta_V$ and $\mu_<$ (see therefore Table~\ref{tab:EoS-properties_hybrid_T0}). These are reflected in the systematically varying global properties of the hybrid star configurations, i.e. the increasing maximum masses and radii from about $M_{\rm max}^{(*)}=2.0(2.4)-2.7(3.4)$~M$_\odot$ and $R_{\rm max}^{(*)}=11(13)-12.5(16.5)$~km for $\eta$NJL with $\eta_D=0.71$ and $\eta_V=0.06-0.20$ (values in the  parenthesis correspond to the rapidly rotating ones). The same systematics is obtained for $\eta_D=0.79$, as well as for nlNJL and RDF (see the right-hand panels in Fig.~\ref{fig:universal}). 

For the hot hybrid models, we find systematically lower maximum masses for both rotating and non-rotating configurations, which is contrary to what was found for their cold counterparts. An exceptional case is the VBAG(b) EoS, for which the maximum mass corresponds to the onset of the phase transition for both rotating and non-rotating configurations (see the lines in Fig.~\ref{fig:MR_hybrid_hot}). The lower maximum mass and corresponding radius range for the universal relations of the hot EoS, shown in   Fig.~\ref{fig:universal} right-hand side top and middle panels, are due to the limited set of hot hybrid models. Nevertheless, this demonstrates the validity of the universal relations~\eqref{eq:universal_M} and \eqref{eq:universal_R} for hot hybrid models, within the maximum deviation given in Table~\ref{tab:universal} for the coefficients $C_M$ and $C_R$. 

\begin{table}
\centering
\caption{Coefficients for the universal relation \eqref{eq:universal_I} including the maximum deviations.}
\begin{tabular}{l c c c}
\hline
Condition & EoS type & $a_1$ & $a_2$ \\
\hline
$T=0$, $\beta$-eq. & hadronic &0.9584 $\pm0.4745$ &0.1069 $\pm0.1196 $ \\
$s=3~k_{\rm B}$, $Y_{\rm L}=0.3$ & hadronic & 1.1885$\pm 0.1275$ & -0.0030 $\pm 0.0253 $ \\
$T=0$, $\beta$-eq. & hybrid   &1.1710 $\pm 0.2020$ &0.0555 $\pm 0.0525$ \\
$s=3~k_{\rm B}$, $Y_{\rm L}=0.3$ & hybrid   & $0.9843\pm 0.0966$ & $0.0333\pm 0.0177$ \\
\hline
\end{tabular}
\label{tab:universal_I}
\end{table}

The universal relation for the normalized moment of inertia, 
\begin{eqnarray}
\frac{I^*}{M_{\rm max}^{*3}} &=& a_1 \left(\frac{M^*_{\rm max}}{R^*_{\rm max}}\right)^{-1} + a_2 \left(\frac{M^*_{\rm max}}{R^*_{\rm max}}\right)^{-2}~, \label{eq:universal_I}
\end{eqnarray}
with the corresponding value of the moment of inertia, $I^*$, is evaluated at the Keplerian limit. It implies the quadratic dependence of the inverse of the compactness, $M/R$, with the two coefficients, $a_1$ and $a_2$, for the linear and quadratic terms, respectively. The average values for $a_1$ and $a_2$ obtained for all EoS under investigation here are listed in Table~\ref{tab:universal_I}. The corresponding values obtained for the universal relations are shown in Fig.~\ref{fig:universal} (bottom panel), for the hadronic EoS (left-hand panel) and for the hybrid EoS (right-hand panel), at both conditions ($T=0$, $\beta$-equilibrium) and ($s=3~k_{\rm B}$, $Y_{\rm L}=0.3$).
  The grey markers belong to the EoS sets that are excluded from the analysis due to the constraints on the tidal deformability deduced from GW180817, as well as radius constraints due to the NICER observations. Acausal EOS sets have been excluded too.
 Note that the moment of inertia shown here are given in geometric units, as commonly used in the literature.  The values we find for the coefficients $a_1$ and $a_2$ are in quantitative agreement with the hadronic set explored in \citet{Khadkikar2021}; for the $T=0$ case they find $a_1=0.9398$ and $a_2=0.1246$, respectively, and the results obtained for the finite entropy of $s=3~k_{\rm B}$ are $a_1=1.1777$ and $a_2=-0.0109$. For the latter, however, there is a better agreement with our results, within about 95.5\%, despite the different assumption of the constant value of $Y_e=0.4$ considered in \citet{Khadkikar2021}. 

\section{Summary}
\label{sec:summary}

This paper investigates the universal relations for rapidly rotating compact stars, in particular those which are at the Keplerian limit. We confirm the previously obtained universal relations for the maximum masses and the corresponding radii and the moment of inertia for a comprehensive set of hadronic models, evaluated for zero-temperature and $\beta$-equilibrium as well as at finite entropy per particle EoS for a fixed value of $s=3~k_{\rm B}$ and a constant electron lepton fraction of $Y_{\rm L}=0.3$. The results we found, which also qualitatively agree with Appendix~\ref{sec:appendixB} before exclusions of EOS sets due to the tidal deformability constraints deduced from GW170817 and NICER measurements for intermediate-mass and high-mass pulsars, are in quantitative agreement with the literature, namely the maximum masses and corresponding radii for rapidly rotating stars are enhanced compared to their non-rotating counterparts by about a constant factor, with $M^*_{\rm max}/M_{\rm max}= 1.2000$ and $R^*_{\rm max}/R_{\rm max}=1.3460$ for $T=0$ and $M^*_{\rm max}/M_{\rm masx}= 1.1085$ and $R^*_{\rm max}/R_{\rm max}=1.3335$ for $s=3~k_{\rm B}$. We also show that the normalized moments of inertia at the Keplerian limit and the corresponding compactness follow a quadratic behaviour, with constant coefficients of $a_{\rm 1}= 0.9584$ and $a_{\rm 2}= 0.1069$ for $T=0$ and $a_{\rm 1}= 1.1885$ and $a_{\rm 2}= -0.0030$ for the hot EoS. Considering the radius constraints derived in \citet{Lattimer:2018}, $R_{\rm 1.4}=9.8-13.2$~km deduced from GW170817 measurements allowing in addition for the binary neutron star system's asymmetry, would result in a similarly restricted set of cold hadronic and hybrid models as the ones listed in Tables~\ref{tab:EoS-hadron} and \ref{tab:EoS-hybrid}.

 In addition, we extended the analysis to include a set of zero-temperature hadron-quark hybrid EoS. Also here we confirm the universal relations for the maximum masses and corresponding radii, with  $M^*_{\rm max}/M_{\rm max}= 1.2120$ and $R^*_{\rm max}/R_{\rm max}=1.3335$, as well as the quadratic relation for the moments of inertia and their corresponding compactness with the coefficients $a_1= 1.1710$ and $a_2= 0.0555$. Note that we find different values for the coefficients than for the hadronic models. This is due to EoS sets that have extreme values of parameters in our collection of hybrid models; these EoS are close to the limit of causality at the maximum mass configurations. We stress here that the underlying systematics cannot reflect an EoS dependence, which would imply the violation of the universal relations. 

Furthermore, we complemented the analysis above with a representative set of hot hybrid EoS, for the same constant value of the entropy per particle of $s=3~k_{\rm B}$ and a constant electron lepton fraction of $Y_{\rm L}=0.3$ as was adopted for hadronic stars. We find that the hot hybrid configurations obey the universal relations for maximum masses and corresponding radii, with $M^*_{\rm max}/M_{\rm max}= 1.0995$ and $R^*_{\rm max}/R_{\rm max}=1.3290$, as well as the quadratic relation for the moments of inertia and their corresponding compactness with the coefficients $a_1= 0.9843$ and $a_2= 0.0333$.

Note that all our hybrid EoS feature a quark core that is surrounded by a hadronic crust, i.e. in this analysis we do not consider the possibility of absolutely stable strange quark matter and stable strange stars. Other universal relations, such as the ''$I$-Love-$Q$'' relations, which involve the moment of inertia $I$, tidal deformability $\Lambda$ and the quadrupole moment $Q$, remain to be explored in a future study. Moreover, the set of finite entropy hybrid models considered here is not exhaustive and should be extended in the future.

\section*{Acknowledgements}
We would like to thank Andreas Bauswein for providing the data for the cold hadronic reference EoS used in this study.  The authors acknowledge support by the Polish National Science Centre (NCN) under grants No. 2019/33/B/ST9/03059 (N.K.L, M.C. and D.B.B.), 2020/37/B/ST9/00691 (T.F., N.K.L) and 2020/37/B/ST9/01937 (A.S.) A.S.  acknowledges the DFG Grant No. SE 1836/5-1. We also acknowledge the support of the PHAROS COST Action CA16214. This project has received funding from the European Union’s Horizon 2020 research and innovation programme under grant agreement No 824093. The numerical calculations were performed at the Wroc\l{}aw Centre for Scientific Computing and Networking (WCSS).

\section*{Data Availability}
The EoS sets and the {\tt{RNS}} simulation results are available based upon a reasonable request to the authors. The EoS used in this study are available at the compOSE database under the following link:
http://compose.obspm.fr \citep[][]{Oertel:2017RvMP}.
%



\bibliographystyle{mnras}
\bibliography{rotation} 

\begin{thebibliography}{}
\makeatletter
\relax
\def\mn@urlcharsother{\let\do\@makeother \do\$\do\&\do\#\do\^\do\_\do\%\do\~}
\def\mn@doi{\begingroup\mn@urlcharsother \@ifnextchar [ {\mn@doi@}
  {\mn@doi@[]}}
\def\mn@doi@[#1]#2{\def\@tempa{#1}\ifx\@tempa\@empty \href
  {http://dx.doi.org/#2} {doi:#2}\else \href {http://dx.doi.org/#2} {#1}\fi
  \endgroup}
\def\mn@eprint#1#2{\mn@eprint@#1:#2::\@nil}
\def\mn@eprint@arXiv#1{\href {http://arxiv.org/abs/#1} {{\tt arXiv:#1}}}
\def\mn@eprint@dblp#1{\href {http://dblp.uni-trier.de/rec/bibtex/#1.xml}
  {dblp:#1}}
\def\mn@eprint@#1:#2:#3:#4\@nil{\def\@tempa {#1}\def\@tempb {#2}\def\@tempc
  {#3}\ifx \@tempc \@empty \let \@tempc \@tempb \let \@tempb \@tempa \fi \ifx
  \@tempb \@empty \def\@tempb {arXiv}\fi \@ifundefined
  {mn@eprint@\@tempb}{\@tempb:\@tempc}{\expandafter \expandafter \csname
  mn@eprint@\@tempb\endcsname \expandafter{\@tempc}}}

\bibitem[\protect\citeauthoryear{Abbott et~al.}{Abbott
  et~al.}{2017}]{TheLIGOScientific:2017qsa}
Abbott B.~P.,  et~al., 2017, \mn@doi [Phys. Rev. Lett.]
  {10.1103/PhysRevLett.119.161101}, 119, 161101

\bibitem[\protect\citeauthoryear{Abbott et~al.}{Abbott
  et~al.}{2018}]{TheLIGOScientific:2018}
Abbott B.~P.,  et~al., 2018, \mn@doi [\prl] {10.1103/PhysRevLett.121.161101},
  \href {https://ui.adsabs.harvard.edu/abs/2018PhRvL.121p1101A} {121, 161101}

\bibitem[\protect\citeauthoryear{{Abgaryan}, {Alvarez-Castillo}, {Ayriyan},
  {Blaschke}  \& {Grigorian}}{{Abgaryan} et~al.}{2018}]{Abgaryan:2018}
{Abgaryan} V.,  {Alvarez-Castillo} D.,  {Ayriyan} A.,  {Blaschke} D.,
  {Grigorian} H.,  2018, \mn@doi [Universe] {10.3390/universe4090094}, \href
  {https://ui.adsabs.harvard.edu/abs/2018Univ....4...94A} {4, 94}

\bibitem[\protect\citeauthoryear{{Akmal} \& {Pandharipande}}{{Akmal} \&
  {Pandharipande}}{1997}]{AP:1997}
{Akmal} A.,  {Pandharipande} V.~R.,  1997, \mn@doi [\prc]
  {10.1103/PhysRevC.56.2261}, \href
  {https://ui.adsabs.harvard.edu/abs/1997PhRvC..56.2261A} {56, 2261}

\bibitem[\protect\citeauthoryear{{Akmal}, {Pandharipande}  \&
  {Ravenhall}}{{Akmal} et~al.}{1998}]{APR:1998}
{Akmal} A.,  {Pandharipande} V.~R.,   {Ravenhall} D.~G.,  1998, \mn@doi [\prc]
  {10.1103/PhysRevC.58.1804}, \href
  {https://ui.adsabs.harvard.edu/abs/1998PhRvC..58.1804A} {58, 1804}

\bibitem[\protect\citeauthoryear{Alford \& Sedrakian}{Alford \&
  Sedrakian}{2017}]{Alford:2017qgh}
Alford M.~G.,  Sedrakian A.,  2017, \mn@doi [Phys. Rev. Lett.]
  {10.1103/PhysRevLett.119.161104}, 119, 161104

\bibitem[\protect\citeauthoryear{Alford, Braby, Paris  \& Reddy}{Alford
  et~al.}{2005}]{Alford:2004pf}
Alford M.,  Braby M.,  Paris M.~W.,   Reddy S.,  2005, \mn@doi [Astrophys. J.]
  {10.1086/430902}, 629, 969

\bibitem[\protect\citeauthoryear{{Alford}, {Han}  \& {Prakash}}{{Alford}
  et~al.}{2013}]{Alford:2013}
{Alford} M.~G.,  {Han} S.,   {Prakash} M.,  2013, \mn@doi [\prd]
  {10.1103/PhysRevD.88.083013}, \href
  {https://ui.adsabs.harvard.edu/abs/2013PhRvD..88h3013A} {88, 083013}

\bibitem[\protect\citeauthoryear{Alvarez-Castillo \& Blaschke}{Alvarez-Castillo
  \& Blaschke}{2017}]{Alvarez-Castillo:2017qki}
Alvarez-Castillo D.~E.,  Blaschke D.~B.,  2017, \mn@doi [Phys. Rev. C]
  {10.1103/PhysRevC.96.045809}, 96, 045809

\bibitem[\protect\citeauthoryear{Alvarez-Castillo, Ayriyan, Benic, Blaschke,
  Grigorian  \& Typel}{Alvarez-Castillo
  et~al.}{2016}]{Alvarez-Castillo:2016oln}
Alvarez-Castillo D.,  Ayriyan A.,  Benic S.,  Blaschke D.,  Grigorian H.,
  Typel S.,  2016, \mn@doi [Eur. Phys. J. A] {10.1140/epja/i2016-16069-2}, 52,
  69

\bibitem[\protect\citeauthoryear{{Alvarez-Castillo}, {Blaschke}, {Grunfeld}  \&
  {Pagura}}{{Alvarez-Castillo} et~al.}{2019}]{Alvarez:2019}
{Alvarez-Castillo} D.~E.,  {Blaschke} D.~B.,  {Grunfeld} A.~G.,   {Pagura}
  V.~P.,  2019, \mn@doi [\prd] {10.1103/PhysRevD.99.063010}, \href
  {https://ui.adsabs.harvard.edu/abs/2019PhRvD..99f3010A} {99, 063010}

\bibitem[\protect\citeauthoryear{{Annala}, {Gorda}, {Kurkela},
  {N{\"a}ttil{\"a}}  \& {Vuorinen}}{{Annala} et~al.}{2020}]{Kurkela:2020NatPh}
{Annala} E.,  {Gorda} T.,  {Kurkela} A.,  {N{\"a}ttil{\"a}} J.,   {Vuorinen}
  A.,  2020, \mn@doi [Nature Physics] {10.1038/s41567-020-0914-9}, \href
  {https://ui.adsabs.harvard.edu/abs/2020NatPh..16..907A} {16, 907}

\bibitem[\protect\citeauthoryear{{Ayriyan}, {Blaschke}, {Grunfeld},
  {Alvarez-Castillo}, {Grigorian}  \& {Abgaryan}}{{Ayriyan}
  et~al.}{2021a}]{Ayriyan:2021}
{Ayriyan} A.,  {Blaschke} D.,  {Grunfeld} A.~G.,  {Alvarez-Castillo} D.,
  {Grigorian} H.,   {Abgaryan} V.,  2021a, arXiv e-prints, \href
  {https://ui.adsabs.harvard.edu/abs/2021arXiv210213485A} {p.
  arXiv:2102.13485v1}

\bibitem[\protect\citeauthoryear{Ayriyan, Blaschke, Grunfeld, Alvarez-Castillo,
  Grigorian  \& Abgaryan}{Ayriyan et~al.}{2021b}]{Ayriyan:2021prr}
Ayriyan A.,  Blaschke D.,  Grunfeld A.~G.,  Alvarez-Castillo D.,  Grigorian H.,
    Abgaryan V.,  2021b, \mn@doi [Eur. Phys. J. A]
  {10.1140/epja/s10050-021-00619-0}, 57, 318

\bibitem[\protect\citeauthoryear{{Ayvazyan}, {Colucci}, {Rischke}  \&
  {Sedrakian}}{{Ayvazyan} et~al.}{2013}]{Ayvazyan:2013}
{Ayvazyan} N.~S.,  {Colucci} G.,  {Rischke} D.~H.,   {Sedrakian} A.,  2013,
  \mn@doi [\aap] {10.1051/0004-6361/201322484}, \href
  {https://ui.adsabs.harvard.edu/abs/2013A&A...559A.118A} {559, A118}

\bibitem[\protect\citeauthoryear{Baiotti}{Baiotti}{2019}]{Baiotti2019}
Baiotti L.,  2019, \mn@doi [Prog. Part. Nucl. Phys.]
  {10.1016/j.ppnp.2019.103714}, 109, 103714

\bibitem[\protect\citeauthoryear{{Banik}, {Hempel}  \& {Bandyopadhyay}}{{Banik}
  et~al.}{2014}]{Banik:2014}
{Banik} S.,  {Hempel} M.,   {Bandyopadhyay} D.,  2014, \mn@doi [\apjs]
  {10.1088/0067-0049/214/2/22}, \href
  {https://ui.adsabs.harvard.edu/abs/2014ApJS..214...22B} {214, 22}

\bibitem[\protect\citeauthoryear{{Bastian}}{{Bastian}}{2021}]{Bastian:2021}
{Bastian} N.-U.~F.,  2021, \mn@doi [\prd] {10.1103/PhysRevD.103.023001}, \href
  {https://ui.adsabs.harvard.edu/abs/2021PhRvD.103b3001B} {103, 023001}

\bibitem[\protect\citeauthoryear{{Bastian}, {Blaschke}, {Fischer}  \&
  {R{\"o}pke}}{{Bastian} et~al.}{2018}]{Bastian:2018}
{Bastian} N.-U.,  {Blaschke} D.,  {Fischer} T.,   {R{\"o}pke} G.,  2018,
  \mn@doi [Universe] {10.3390/universe4060067}, \href
  {https://ui.adsabs.harvard.edu/abs/2018Univ....4...67B} {4, 67}

\bibitem[\protect\citeauthoryear{{Bauswein}, {Baumgarte}  \&
  {Janka}}{{Bauswein} et~al.}{2013}]{Bauswein:2013}
{Bauswein} A.,  {Baumgarte} T.~W.,   {Janka} H.~T.,  2013, \mn@doi [\prl]
  {10.1103/PhysRevLett.111.131101}, \href
  {https://ui.adsabs.harvard.edu/abs/2013PhRvL.111m1101B} {111, 131101}

\bibitem[\protect\citeauthoryear{Bauswein, Just, Janka  \&
  Stergioulas}{Bauswein et~al.}{2017}]{Bauswein:2017vtn}
Bauswein A.,  Just O.,  Janka H.-T.,   Stergioulas N.,  2017, \mn@doi
  [Astrophys. J. Lett.] {10.3847/2041-8213/aa9994}, 850, L34

\bibitem[\protect\citeauthoryear{{Bauswein}, {Bastian}, {Blaschke},
  {Chatziioannou}, {Clark}, {Fischer}  \& {Oertel}}{{Bauswein}
  et~al.}{2019}]{Bauswein:2019}
{Bauswein} A.,  {Bastian} N.-U.~F.,  {Blaschke} D.~B.,  {Chatziioannou} K.,
  {Clark} J.~A.,  {Fischer} T.,   {Oertel} M.,  2019, \mn@doi [\prl]
  {10.1103/PhysRevLett.122.061102}, \href
  {https://ui.adsabs.harvard.edu/abs/2019PhRvL.122f1102B} {122, 061102}

\bibitem[\protect\citeauthoryear{{Baym}, {Hatsuda}, {Kojo}, {Powell}, {Song}
  \& {Takatsuka}}{{Baym} et~al.}{2018}]{Baym:2018}
{Baym} G.,  {Hatsuda} T.,  {Kojo} T.,  {Powell} P.~D.,  {Song} Y.,
  {Takatsuka} T.,  2018, \mn@doi [Reports on Progress in Physics]
  {10.1088/1361-6633/aaae14}, \href
  {https://ui.adsabs.harvard.edu/abs/2018RPPh...81e6902B} {81, 056902}

\bibitem[\protect\citeauthoryear{{Bazavov} et~al.,}{{Bazavov}
  et~al.}{2014}]{Bazavov:2014}
{Bazavov} A.,  et~al., 2014, \mn@doi [\prd] {10.1103/PhysRevD.90.094503}, \href
  {https://ui.adsabs.harvard.edu/abs/2014PhRvD..90i4503B} {90, 094503}

\bibitem[\protect\citeauthoryear{{Bazavov} et~al.,}{{Bazavov}
  et~al.}{2019}]{Bazavov:2019}
{Bazavov} A.,  et~al., 2019, \mn@doi [Physics Letters B]
  {10.1016/j.physletb.2019.05.013}, \href
  {https://ui.adsabs.harvard.edu/abs/2019PhLB..795...15B} {795, 15}

\bibitem[\protect\citeauthoryear{{Beni{\'c}}, {Blaschke}, {Alvarez-Castillo},
  {Fischer}  \& {Typel}}{{Beni{\'c}} et~al.}{2015}]{Benic:2015}
{Beni{\'c}} S.,  {Blaschke} D.,  {Alvarez-Castillo} D.~E.,  {Fischer} T.,
  {Typel} S.,  2015, \mn@doi [\aap] {10.1051/0004-6361/201425318}, \href
  {https://ui.adsabs.harvard.edu/abs/2015A&A...577A..40B} {577, A40}

\bibitem[\protect\citeauthoryear{{Blaschke}, {Fredriksson}, {Grigorian},
  {{\"O}zta{\c{s}}}  \& {Sandin}}{{Blaschke} et~al.}{2005}]{Blaschke2005}
{Blaschke} D.,  {Fredriksson} S.,  {Grigorian} H.,  {{\"O}zta{\c{s}}} A.~M.,
  {Sandin} F.,  2005, \mn@doi [\prd] {10.1103/PhysRevD.72.065020}, \href
  {https://ui.adsabs.harvard.edu/abs/2005PhRvD..72f5020B} {72, 065020}

\bibitem[\protect\citeauthoryear{Blaschke, Ayriyan, Alvarez-Castillo  \&
  Grigorian}{Blaschke et~al.}{2020}]{Blaschke:2020qqj}
Blaschke D.,  Ayriyan A.,  Alvarez-Castillo D.~E.,   Grigorian H.,  2020,
  \mn@doi [Universe] {10.3390/universe6060081}, 6, 81

\bibitem[\protect\citeauthoryear{{Bollig}, {Janka}, {Lohs},
  {Mart{\'\i}nez-Pinedo}, {Horowitz}  \& {Melson}}{{Bollig}
  et~al.}{2017}]{Bollig:2017}
{Bollig} R.,  {Janka} H.~T.,  {Lohs} A.,  {Mart{\'\i}nez-Pinedo} G.,
  {Horowitz} C.~J.,   {Melson} T.,  2017, \mn@doi [\prl]
  {10.1103/PhysRevLett.119.242702}, \href
  {https://ui.adsabs.harvard.edu/abs/2017PhRvL.119x2702B} {119, 242702}

\bibitem[\protect\citeauthoryear{{Bonanno} \& {Sedrakian}}{{Bonanno} \&
  {Sedrakian}}{2012}]{Bonanno2012}
{Bonanno} L.,  {Sedrakian} A.,  2012, \mn@doi [\aap]
  {10.1051/0004-6361/201117832}, \href
  {https://ui.adsabs.harvard.edu/abs/2012A&A...539A..16B} {539, A16}

\bibitem[\protect\citeauthoryear{{Bonazzola}, {Gourgoulhon}, {Salgado}  \&
  {Marck}}{{Bonazzola} et~al.}{1993}]{Bonazzola1993}
{Bonazzola} S.,  {Gourgoulhon} E.,  {Salgado} M.,   {Marck} J.~A.,  1993, \aap,
  \href {https://ui.adsabs.harvard.edu/abs/1993A&A...278..421B} {278, 421}

\bibitem[\protect\citeauthoryear{{Bors{\'a}nyi}, {Fodor}, {Hoelbling}, {Katz},
  {Krieg}  \& {Szab{\'o}}}{{Bors{\'a}nyi} et~al.}{2014}]{Borsanyi:2014}
{Bors{\'a}nyi} S.,  {Fodor} Z.,  {Hoelbling} C.,  {Katz} S.~D.,  {Krieg} S.,
  {Szab{\'o}} K.~K.,  2014, \mn@doi [Physics Letters B]
  {10.1016/j.physletb.2014.01.007}, \href
  {https://ui.adsabs.harvard.edu/abs/2014PhLB..730...99B} {730, 99}

\bibitem[\protect\citeauthoryear{Bozzola, Stergioulas  \& Bauswein}{Bozzola
  et~al.}{2018}]{Bozzola:2017qbu}
Bozzola G.,  Stergioulas N.,   Bauswein A.,  2018, \mn@doi [Mon. Not. Roy.
  Astron. Soc.] {10.1093/mnras/stx3002}, 474, 3557

\bibitem[\protect\citeauthoryear{Bozzola, Espino, Lewin  \&
  Paschalidis}{Bozzola et~al.}{2019}]{Bozzola:2019tit}
Bozzola G.,  Espino P.~L.,  Lewin C.~D.,   Paschalidis V.,  2019, \mn@doi [Eur.
  Phys. J. A] {10.1140/epja/i2019-12831-2}, 55, 149

\bibitem[\protect\citeauthoryear{Breu \& Rezzolla}{Breu \&
  Rezzolla}{2016}]{Breu:2016ufb}
Breu C.,  Rezzolla L.,  2016, \mn@doi [Mon. Not. Roy. Astron. Soc.]
  {10.1093/mnras/stw575}, 459, 646

\bibitem[\protect\citeauthoryear{Buballa}{Buballa}{2005}]{Buballa:2003qv}
Buballa M.,  2005, \mn@doi [Phys. Rept.] {10.1016/j.physrep.2004.11.004}, 407,
  205

\bibitem[\protect\citeauthoryear{Burgio \& Schulze}{Burgio \&
  Schulze}{2010}]{Burgio:2010ek}
Burgio G.~F.,  Schulze H.~J.,  2010, \mn@doi [Astron. Astrophys.]
  {10.1051/0004-6361/201014308}, 518, A17

\bibitem[\protect\citeauthoryear{Chandrasekhar}{Chandrasekhar}{1970}]{Chandrasekhar:1970pjp}
Chandrasekhar S.,  1970, \mn@doi [Phys. Rev. Lett.]
  {10.1103/PhysRevLett.24.611}, 24, 611

\bibitem[\protect\citeauthoryear{{Chatziioannou}}{{Chatziioannou}}{2020}]{Chatziioannou2020GReGr}
{Chatziioannou} K.,  2020, \mn@doi [General Relativity and Gravitation]
  {10.1007/s10714-020-02754-3}, \href
  {https://ui.adsabs.harvard.edu/abs/2020GReGr..52..109C} {52, 109}

\bibitem[\protect\citeauthoryear{Cierniak \& Kl\"ahn}{Cierniak \&
  Kl\"ahn}{2017}]{Cierniak:2017dxr}
Cierniak M.,  Kl\"ahn T.,  2017, \mn@doi [Acta Phys. Polon. Supp.]
  {10.5506/APhysPolBSupp.10.811}, 10, 811

\bibitem[\protect\citeauthoryear{{Cierniak}, {Kl{\"a}hn}, {Fischer}  \&
  {Bastian}}{{Cierniak} et~al.}{2018}]{Cierniak:2018}
{Cierniak} M.,  {Kl{\"a}hn} T.,  {Fischer} T.,   {Bastian} N.-U.,  2018,
  \mn@doi [Universe] {10.3390/universe4020030}, \href
  {https://ui.adsabs.harvard.edu/abs/2018Univ....4...30C} {4, 30}

\bibitem[\protect\citeauthoryear{{Cook}, {Shapiro}  \& {Teukolsky}}{{Cook}
  et~al.}{1994}]{Cook:1994}
{Cook} G.~B.,  {Shapiro} S.~L.,   {Teukolsky} S.~A.,  1994, \mn@doi [\apj]
  {10.1086/173721}, \href
  {https://ui.adsabs.harvard.edu/abs/1994ApJ...422..227C} {422, 227}

\bibitem[\protect\citeauthoryear{{Cromartie} et~al.,}{{Cromartie}
  et~al.}{2020}]{Cromartie:2020NatAs}
{Cromartie} H.~T.,  et~al., 2020, \mn@doi [Nature Astronomy]
  {10.1038/s41550-019-0880-2}, \href
  {https://ui.adsabs.harvard.edu/abs/2020NatAs...4...72C} {4, 72}

\bibitem[\protect\citeauthoryear{{Danielewicz}, {Lacey}  \&
  {Lynch}}{{Danielewicz} et~al.}{2002}]{Danielewicz:2002Sci}
{Danielewicz} P.,  {Lacey} R.,   {Lynch} W.~G.,  2002, \mn@doi [Science]
  {10.1126/science.1078070}, \href
  {https://ui.adsabs.harvard.edu/abs/2002Sci...298.1592D} {298, 1592}

\bibitem[\protect\citeauthoryear{{De}, {Finstad}, {Lattimer}, {Brown}, {Berger}
   \& {Biwer}}{{De} et~al.}{2018}]{Lattimer:2018}
{De} S.,  {Finstad} D.,  {Lattimer} J.~M.,  {Brown} D.~A.,  {Berger} E.,
  {Biwer} C.~M.,  2018, \mn@doi [\prl] {10.1103/PhysRevLett.121.091102}, \href
  {https://ui.adsabs.harvard.edu/abs/2018PhRvL.121i1102D} {121, 091102}

\bibitem[\protect\citeauthoryear{{Doneva}, {Yazadjiev}, {Staykov}  \&
  {Kokkotas}}{{Doneva} et~al.}{2014}]{Doneva2014PhRvD}
{Doneva} D.~D.,  {Yazadjiev} S.~S.,  {Staykov} K.~V.,   {Kokkotas} K.~D.,
  2014, \mn@doi [\prd] {10.1103/PhysRevD.90.104021}, \href
  {https://ui.adsabs.harvard.edu/abs/2014PhRvD..90j4021D} {90, 104021}

\bibitem[\protect\citeauthoryear{{Essick}, {Landry}  \& {Holz}}{{Essick}
  et~al.}{2020}]{Essick2020}
{Essick} R.,  {Landry} P.,   {Holz} D.~E.,  2020, \mn@doi [\prd]
  {10.1103/PhysRevD.101.063007}, \href
  {https://ui.adsabs.harvard.edu/abs/2020PhRvD.101f3007E} {101, 063007}

\bibitem[\protect\citeauthoryear{Farhi \& Jaffe}{Farhi \&
  Jaffe}{1984}]{Farhi:1984qu}
Farhi E.,  Jaffe R.,  1984, \mn@doi [Phys.Rev.] {10.1103/PhysRevD.30.2379},
  D30, 2379

\bibitem[\protect\citeauthoryear{{Ferreira}, {Pereira}  \&
  {Provid{\^e}ncia}}{{Ferreira} et~al.}{2021}]{Ferreira2021}
{Ferreira} M.,  {Pereira} R.~C.,   {Provid{\^e}ncia} C.,  2021, \mn@doi [\prd]
  {10.1103/PhysRevD.103.123020}, \href
  {https://ui.adsabs.harvard.edu/abs/2021PhRvD.103l3020F} {103, 123020}

\bibitem[\protect\citeauthoryear{{Fischer}}{{Fischer}}{2021}]{Fischer:2021}
{Fischer} T.,  2021, \mn@doi [Eur.\ Phys.\ J.\ A]
  {10.1140/epja/s10050-021-00571-z}, \href
  {https://ui.adsabs.harvard.edu/abs/2021arXiv210800196F} {57, 270}

\bibitem[\protect\citeauthoryear{{Fischer}, {Whitehouse}, {Mezzacappa},
  {Thielemann}  \& {Liebend{\"o}rfer}}{{Fischer} et~al.}{2010}]{Fischer:2009af}
{Fischer} T.,  {Whitehouse} S.~C.,  {Mezzacappa} A.,  {Thielemann} F.~K.,
  {Liebend{\"o}rfer} M.,  2010, \mn@doi [\aap] {10.1051/0004-6361/200913106},
  \href {https://ui.adsabs.harvard.edu/abs/2010A&A...517A..80F} {517, A80}

\bibitem[\protect\citeauthoryear{{Fischer} et~al.,}{{Fischer}
  et~al.}{2011}]{Fischer:2011}
{Fischer} T.,  et~al., 2011, \mn@doi [\apjs] {10.1088/0067-0049/194/2/39},
  \href {https://ui.adsabs.harvard.edu/abs/2011ApJS..194...39F} {194, 39}

\bibitem[\protect\citeauthoryear{{Fischer}, {Hempel}, {Sagert}, {Suwa}  \&
  {Schaffner-Bielich}}{{Fischer} et~al.}{2014}]{Fischer:2014}
{Fischer} T.,  {Hempel} M.,  {Sagert} I.,  {Suwa} Y.,   {Schaffner-Bielich} J.,
   2014, \mn@doi [European Physical Journal A] {10.1140/epja/i2014-14046-5},
  \href {https://ui.adsabs.harvard.edu/abs/2014EPJA...50...46F} {50, 46}

\bibitem[\protect\citeauthoryear{{Fischer} et~al.,}{{Fischer}
  et~al.}{2017}]{Fischer:2017}
{Fischer} T.,  et~al., 2017, \mn@doi [\pasa] {10.1017/pasa.2017.63}, \href
  {https://ui.adsabs.harvard.edu/abs/2017PASA...34...67F} {34, e067}

\bibitem[\protect\citeauthoryear{{Fischer} et~al.,}{{Fischer}
  et~al.}{2018}]{Fischer:2018}
{Fischer} T.,  et~al., 2018, \mn@doi [Nature Astronomy]
  {10.1038/s41550-018-0583-0}, \href
  {https://ui.adsabs.harvard.edu/abs/2018NatAs...2..980F} {2, 980}

\bibitem[\protect\citeauthoryear{{Fischer}, {Guo}, {Dzhioev},
  {Mart{\'\i}nez-Pinedo}, {Wu}, {Lohs}  \& {Qian}}{{Fischer}
  et~al.}{2020}]{Fischer:2020PhRvC101}
{Fischer} T.,  {Guo} G.,  {Dzhioev} A.~A.,  {Mart{\'\i}nez-Pinedo} G.,  {Wu}
  M.-R.,  {Lohs} A.,   {Qian} Y.-Z.,  2020, \mn@doi [\prc]
  {10.1103/PhysRevC.101.025804}, \href
  {https://ui.adsabs.harvard.edu/abs/2020PhRvC.101b5804F} {101, 025804}

\bibitem[\protect\citeauthoryear{{Fischer}, {Carenza}, {Fore}, {Giannotti},
  {Mirizzi}  \& {Reddy}}{{Fischer} et~al.}{2021}]{Fischer:2021PhRvD}
{Fischer} T.,  {Carenza} P.,  {Fore} B.,  {Giannotti} M.,  {Mirizzi} A.,
  {Reddy} S.,  2021, \mn@doi [\prd] {10.1103/PhysRevD.104.103012}, \href
  {https://ui.adsabs.harvard.edu/abs/2021PhRvD.104j3012F} {104, 103012}

\bibitem[\protect\citeauthoryear{{Fonseca} et~al.,}{{Fonseca}
  et~al.}{2016}]{Fonseca:2016}
{Fonseca} E.,  et~al., 2016, \mn@doi [\apj] {10.3847/0004-637X/832/2/167},
  \href {https://ui.adsabs.harvard.edu/abs/2016ApJ...832..167F} {832, 167}

\bibitem[\protect\citeauthoryear{{Friedman} \& {Schutz}}{{Friedman} \&
  {Schutz}}{1975}]{1975ApJ...199L.157F}
{Friedman} J.~L.,  {Schutz} B.~F.,  1975, \mn@doi [\apjl] {10.1086/181872},
  \href {https://ui.adsabs.harvard.edu/abs/1975ApJ...199L.157F} {199, L157}

\bibitem[\protect\citeauthoryear{{Friedman} \& {Schutz}}{{Friedman} \&
  {Schutz}}{1978}]{1978ApJ...221L..99F}
{Friedman} J.~L.,  {Schutz} B.~F.,  1978, \apjl, \href
  {https://ui.adsabs.harvard.edu/abs/1978ApJ...221L..99F} {221, L99}

\bibitem[\protect\citeauthoryear{{Friedman}, {Ipser}  \& {Parker}}{{Friedman}
  et~al.}{1989}]{Friedman:1989}
{Friedman} J.~L.,  {Ipser} J.~R.,   {Parker} L.,  1989, \mn@doi [\prl]
  {10.1103/PhysRevLett.62.3015}, \href
  {https://ui.adsabs.harvard.edu/abs/1989PhRvL..62.3015F} {62, 3015}

\bibitem[\protect\citeauthoryear{{Glendenning}, {Pei}  \&
  {Weber}}{{Glendenning} et~al.}{1997}]{Glendenning1997}
{Glendenning} N.~K.,  {Pei} S.,   {Weber} F.,  1997, \mn@doi [\prl]
  {10.1103/PhysRevLett.79.1603}, \href
  {https://ui.adsabs.harvard.edu/abs/1997PhRvL..79.1603G} {79, 1603}

\bibitem[\protect\citeauthoryear{{Goriely}, {Chamel}  \& {Pearson}}{{Goriely}
  et~al.}{2010}]{Goriely:2010}
{Goriely} S.,  {Chamel} N.,   {Pearson} J.~M.,  2010, \mn@doi [\prc]
  {10.1103/PhysRevC.82.035804}, \href
  {https://ui.adsabs.harvard.edu/abs/2010PhRvC..82c5804G} {82, 035804}

\bibitem[\protect\citeauthoryear{{Haensel} \& {Zdunik}}{{Haensel} \&
  {Zdunik}}{1989}]{HaenselZdunik:1989}
{Haensel} P.,  {Zdunik} J.~L.,  1989, \mn@doi [\nat] {10.1038/340617a0}, \href
  {https://ui.adsabs.harvard.edu/abs/1989Natur.340..617H} {340, 617}

\bibitem[\protect\citeauthoryear{{Haensel}, {Salgado}  \&
  {Bonazzola}}{{Haensel} et~al.}{1995}]{Haensel:1995}
{Haensel} P.,  {Salgado} M.,   {Bonazzola} S.,  1995, \aap, \href
  {https://ui.adsabs.harvard.edu/abs/1995A&A...296..745H} {296, 745}

\bibitem[\protect\citeauthoryear{{Haensel}, {Zdunik}, {Bejger}  \&
  {Lattimer}}{{Haensel} et~al.}{2009}]{Haensel:2009}
{Haensel} P.,  {Zdunik} J.~L.,  {Bejger} M.,   {Lattimer} J.~M.,  2009, \mn@doi
  [\aap] {10.1051/0004-6361/200811605}, \href
  {https://ui.adsabs.harvard.edu/abs/2009A&A...502..605H} {502, 605}

\bibitem[\protect\citeauthoryear{{Haensel}, {Bejger}, {Fortin}  \&
  {Zdunik}}{{Haensel} et~al.}{2016}]{Haensel2016EPJA}
{Haensel} P.,  {Bejger} M.,  {Fortin} M.,   {Zdunik} L.,  2016, \mn@doi
  [European Physical Journal A] {10.1140/epja/i2016-16059-4}, \href
  {https://ui.adsabs.harvard.edu/abs/2016EPJA...52...59H} {52, 59}

\bibitem[\protect\citeauthoryear{{Haskell}, {Ciolfi}, {Pannarale}  \&
  {Rezzolla}}{{Haskell} et~al.}{2014}]{Haskell2014MNRAS}
{Haskell} B.,  {Ciolfi} R.,  {Pannarale} F.,   {Rezzolla} L.,  2014, \mn@doi
  [\mnras] {10.1093/mnrasl/slt161}, \href
  {https://ui.adsabs.harvard.edu/abs/2014MNRAS.438L..71H} {438, L71}

\bibitem[\protect\citeauthoryear{Hatsuda \& Kunihiro}{Hatsuda \&
  Kunihiro}{1994}]{Hatsuda:1994pi}
Hatsuda T.,  Kunihiro T.,  1994, \mn@doi [Phys. Rept.]
  {10.1016/0370-1573(94)90022-1}, 247, 221

\bibitem[\protect\citeauthoryear{{Hempel} \& {Schaffner-Bielich}}{{Hempel} \&
  {Schaffner-Bielich}}{2010}]{Hempel:2010}
{Hempel} M.,  {Schaffner-Bielich} J.,  2010, \mn@doi [\nphysa]
  {10.1016/j.nuclphysa.2010.02.010}, \href
  {https://ui.adsabs.harvard.edu/abs/2010NuPhA.837..210H} {837, 210}

\bibitem[\protect\citeauthoryear{{Hempel}, {Heinimann}, {Yudin}, {Iosilevskiy},
  {Liebend{\"o}rfer}  \& {Thielemann}}{{Hempel} et~al.}{2016}]{Hempel:2016}
{Hempel} M.,  {Heinimann} O.,  {Yudin} A.,  {Iosilevskiy} I.,
  {Liebend{\"o}rfer} M.,   {Thielemann} F.-K.,  2016, \mn@doi [\prd]
  {10.1103/PhysRevD.94.103001}, \href
  {https://ui.adsabs.harvard.edu/abs/2016PhRvD..94j3001H} {94, 103001}

\bibitem[\protect\citeauthoryear{{Hotokezaka}, {Kyutoku}, {Sekiguchi}  \&
  {Shibata}}{{Hotokezaka} et~al.}{2016}]{Shibata:2016PhRvD}
{Hotokezaka} K.,  {Kyutoku} K.,  {Sekiguchi} Y.-i.,   {Shibata} M.,  2016,
  \mn@doi [\prd] {10.1103/PhysRevD.93.064082}, \href
  {https://ui.adsabs.harvard.edu/abs/2016PhRvD..93f4082H} {93, 064082}

\bibitem[\protect\citeauthoryear{{H{\"u}depohl}, {M{\"u}ller}, {Janka}, {Marek}
   \& {Raffelt}}{{H{\"u}depohl} et~al.}{2010}]{Huedepohl:2010}
{H{\"u}depohl} L.,  {M{\"u}ller} B.,  {Janka} H.~T.,  {Marek} A.,   {Raffelt}
  G.~G.,  2010, \mn@doi [\prl] {10.1103/PhysRevLett.104.251101}, \href
  {https://ui.adsabs.harvard.edu/abs/2010PhRvL.104y1101H} {104, 251101}

\bibitem[\protect\citeauthoryear{{Janka}, {Langanke}, {Marek},
  {Mart{\'\i}nez-Pinedo}  \& {M{\"u}ller}}{{Janka}
  et~al.}{2007}]{Janka:2007PhR}
{Janka} H.~T.,  {Langanke} K.,  {Marek} A.,  {Mart{\'\i}nez-Pinedo} G.,
  {M{\"u}ller} B.,  2007, \mn@doi [\physrep] {10.1016/j.physrep.2007.02.002},
  \href {https://ui.adsabs.harvard.edu/abs/2007PhR...442...38J} {442, 38}

\bibitem[\protect\citeauthoryear{{Kaltenborn}, {Bastian}  \&
  {Blaschke}}{{Kaltenborn} et~al.}{2017}]{Kaltenborn:2017}
{Kaltenborn} M. A.~R.,  {Bastian} N.-U.~F.,   {Blaschke} D.~B.,  2017, \mn@doi
  [\prd] {10.1103/PhysRevD.96.056024}, \href
  {https://ui.adsabs.harvard.edu/abs/2017PhRvD..96e6024K} {96, 056024}

\bibitem[\protect\citeauthoryear{{Khadkikar}, {Raduta}, {Oertel}  \&
  {Sedrakian}}{{Khadkikar} et~al.}{2021}]{Khadkikar2021}
{Khadkikar} S.,  {Raduta} A.~R.,  {Oertel} M.,   {Sedrakian} A.,  2021, \mn@doi
  [\prc] {10.1103/PhysRevC.103.055811}, \href
  {https://ui.adsabs.harvard.edu/abs/2021PhRvC.103e5811K} {103, 055811}

\bibitem[\protect\citeauthoryear{{Kl{\"a}hn} \& {Fischer}}{{Kl{\"a}hn} \&
  {Fischer}}{2015}]{Klaehn:2015}
{Kl{\"a}hn} T.,  {Fischer} T.,  2015, \mn@doi [\apj]
  {10.1088/0004-637X/810/2/134}, \href
  {https://ui.adsabs.harvard.edu/abs/2015ApJ...810..134K} {810, 134}

\bibitem[\protect\citeauthoryear{Kl{\"a}hn, Blaschke, Sandin, Fuchs, Faessler,
  Grigorian, R{\"o}pke  \& Trumper}{Kl{\"a}hn et~al.}{2007}]{Klahn:2006iw}
Kl{\"a}hn T.,  Blaschke D.,  Sandin F.,  Fuchs C.,  Faessler A.,  Grigorian H.,
   R{\"o}pke G.,   Trumper J.,  2007, \mn@doi [Phys. Lett. B]
  {10.1016/j.physletb.2007.08.048}, 654, 170

\bibitem[\protect\citeauthoryear{{Kl{\"a}hn}, {{\L}astowiecki}  \&
  {Blaschke}}{{Kl{\"a}hn} et~al.}{2013}]{Klaehn2013}
{Kl{\"a}hn} T.,  {{\L}astowiecki} R.,   {Blaschke} D.,  2013, \mn@doi [\prd]
  {10.1103/PhysRevD.88.085001}, \href
  {https://ui.adsabs.harvard.edu/abs/2013PhRvD..88h5001K} {88, 085001}

\bibitem[\protect\citeauthoryear{{Kl{\"a}hn}, {Fischer}  \&
  {Hempel}}{{Kl{\"a}hn} et~al.}{2017}]{Klaehn:2017}
{Kl{\"a}hn} T.,  {Fischer} T.,   {Hempel} M.,  2017, \mn@doi [\apj]
  {10.3847/1538-4357/836/1/89}, \href
  {https://ui.adsabs.harvard.edu/abs/2017ApJ...836...89K} {836, 89}

\bibitem[\protect\citeauthoryear{Klevansky}{Klevansky}{1992}]{Klevansky:1992qe}
Klevansky S.~P.,  1992, \mn@doi [Rev. Mod. Phys.] {10.1103/RevModPhys.64.649},
  64, 649

\bibitem[\protect\citeauthoryear{{Koliogiannis} \&
  {Moustakidis}}{{Koliogiannis} \& {Moustakidis}}{2020}]{Koliogiannis:2020}
{Koliogiannis} P.~S.,  {Moustakidis} C.~C.,  2020, \mn@doi [\prc]
  {10.1103/PhysRevC.101.015805}, \href
  {https://ui.adsabs.harvard.edu/abs/2020PhRvC.101a5805K} {101, 015805}

\bibitem[\protect\citeauthoryear{{Komatsu}, {Eriguchi}  \& {Hachisu}}{{Komatsu}
  et~al.}{1989}]{Komatsu:1989}
{Komatsu} H.,  {Eriguchi} Y.,   {Hachisu} I.,  1989, \mn@doi [\mnras]
  {10.1093/mnras/237.2.355}, \href
  {https://ui.adsabs.harvard.edu/abs/1989MNRAS.237..355K} {237, 355}

\bibitem[\protect\citeauthoryear{Kramer et~al.,}{Kramer
  et~al.}{2021}]{Kramer:2021}
Kramer M.,  et~al., 2021, \mn@doi [Phys. Rev. X] {10.1103/PhysRevX.11.041050},
  11, 041050

\bibitem[\protect\citeauthoryear{Kr\"uger \& Kokkotas}{Kr\"uger \&
  Kokkotas}{2020}]{Kruger:2019zuz}
Kr\"uger C.~J.,  Kokkotas K.~D.,  2020, \mn@doi [Phys. Rev. Lett.]
  {10.1103/PhysRevLett.125.111106}, 125, 111106

\bibitem[\protect\citeauthoryear{{Kumar} \& {Landry}}{{Kumar} \&
  {Landry}}{2019}]{Kumar2019PhRvD}
{Kumar} B.,  {Landry} P.,  2019, \mn@doi [\prd] {10.1103/PhysRevD.99.123026},
  \href {https://ui.adsabs.harvard.edu/abs/2019PhRvD..99l3026K} {99, 123026}

\bibitem[\protect\citeauthoryear{{Kuroda}, {Fischer}, {Takiwaki}  \&
  {Kotake}}{{Kuroda} et~al.}{2022}]{Kuroda:2022}
{Kuroda} T.,  {Fischer} T.,  {Takiwaki} T.,   {Kotake} K.,  2022, \mn@doi
  [\apj] {10.3847/1538-4357/ac31a8}, \href
  {https://ui.adsabs.harvard.edu/abs/2022ApJ...924...38K} {924, 38}

\bibitem[\protect\citeauthoryear{{Lalazissis}, {K{\"o}nig}  \&
  {Ring}}{{Lalazissis} et~al.}{1997}]{Lalazissis:1997}
{Lalazissis} G.~A.,  {K{\"o}nig} J.,   {Ring} P.,  1997, \mn@doi [\prc]
  {10.1103/PhysRevC.55.540}, \href
  {https://ui.adsabs.harvard.edu/abs/1997PhRvC..55..540L} {55, 540}

\bibitem[\protect\citeauthoryear{{Lasota}, {Haensel}  \& {Abramowicz}}{{Lasota}
  et~al.}{1996}]{Haensel:1996}
{Lasota} J.-P.,  {Haensel} P.,   {Abramowicz} M.~A.,  1996, \mn@doi [\apj]
  {10.1086/176650}, \href
  {https://ui.adsabs.harvard.edu/abs/1996ApJ...456..300L} {456, 300}

\bibitem[\protect\citeauthoryear{{Lattimer} \& {Schutz}}{{Lattimer} \&
  {Schutz}}{2005}]{LattimerSchutz:2005}
{Lattimer} J.~M.,  {Schutz} B.~F.,  2005, \mn@doi [\apj] {10.1086/431543},
  \href {https://ui.adsabs.harvard.edu/abs/2005ApJ...629..979L} {629, 979}

\bibitem[\protect\citeauthoryear{{Lattimer} \& {Swesty}}{{Lattimer} \&
  {Swesty}}{1991}]{Lattimer:1991}
{Lattimer} J.~M.,  {Swesty} D.~F.,  1991, \mn@doi [\nphysa]
  {10.1016/0375-9474(91)90452-C}, \href
  {https://ui.adsabs.harvard.edu/abs/1991NuPhA.535..331L} {535, 331}

\bibitem[\protect\citeauthoryear{{Lenka}, {Char}  \& {Banik}}{{Lenka}
  et~al.}{2017}]{Lenka2017}
{Lenka} S.~S.,  {Char} P.,   {Banik} S.,  2017, \mn@doi [International Journal
  of Modern Physics D] {10.1142/S0218271817501279}, \href
  {https://ui.adsabs.harvard.edu/abs/2017IJMPD..2650127L} {26, 1750127}

\bibitem[\protect\citeauthoryear{Li, Sedrakian  \& Alford}{Li
  et~al.}{2020}]{Lijj_2020a}
Li J.~J.,  Sedrakian A.,   Alford M.,  2020, \mn@doi [Phys. Rev. D]
  {10.1103/PhysRevD.101.063022}, 101, 063022

\bibitem[\protect\citeauthoryear{Li, Sedrakian  \& Alford}{Li
  et~al.}{2021}]{Lijj_2021}
Li J.~J.,  Sedrakian A.,   Alford M.,  2021, \mn@doi [Phys. Rev. D]
  {10.1103/PhysRevD.104.L121302}, 104, L121302

\bibitem[\protect\citeauthoryear{Lu, Li, Burgio, Figura  \& Schulze}{Lu
  et~al.}{2019}]{Lu:2019mza}
Lu J.-J.,  Li Z.-H.,  Burgio G.~F.,  Figura A.,   Schulze H.~J.,  2019, \mn@doi
  [Phys. Rev. C] {10.1103/PhysRevC.100.054335}, 100, 054335

\bibitem[\protect\citeauthoryear{{Majumder}, {Yagi}  \& {Yunes}}{{Majumder}
  et~al.}{2015}]{Majumder2015PhRvD}
{Majumder} B.,  {Yagi} K.,   {Yunes} N.,  2015, \mn@doi [\prd]
  {10.1103/PhysRevD.92.024020}, \href
  {https://ui.adsabs.harvard.edu/abs/2015PhRvD..92b4020M} {92, 024020}

\bibitem[\protect\citeauthoryear{{Manoharan}, {Kr{\"u}ger}  \&
  {Kokkotas}}{{Manoharan} et~al.}{2021}]{Manoharan2021}
{Manoharan} P.,  {Kr{\"u}ger} C.~J.,   {Kokkotas} K.~D.,  2021, \mn@doi [\prd]
  {10.1103/PhysRevD.104.023005}, \href
  {https://ui.adsabs.harvard.edu/abs/2021PhRvD.104b3005M} {104, 023005}

\bibitem[\protect\citeauthoryear{{Mart{\'\i}nez-Pinedo}, {Fischer}, {Lohs}  \&
  {Huther}}{{Mart{\'\i}nez-Pinedo} et~al.}{2012}]{MartinezPinedo:2012}
{Mart{\'\i}nez-Pinedo} G.,  {Fischer} T.,  {Lohs} A.,   {Huther} L.,  2012,
  \mn@doi [\prl] {10.1103/PhysRevLett.109.251104}, \href
  {https://ui.adsabs.harvard.edu/abs/2012PhRvL.109y1104M} {109, 251104}

\bibitem[\protect\citeauthoryear{{Miller} et~al.,}{{Miller}
  et~al.}{2019}]{Miller:2019}
{Miller} M.~C.,  et~al., 2019, \mn@doi [\apjl] {10.3847/2041-8213/ab50c5},
  \href {https://ui.adsabs.harvard.edu/abs/2019ApJ...887L..24M} {887, L24}

\bibitem[\protect\citeauthoryear{{Miller} et~al.,}{{Miller}
  et~al.}{2021}]{Miller2021ApJ}
{Miller} M.~C.,  et~al., 2021, \mn@doi [\apjl] {10.3847/2041-8213/ac089b},
  \href {https://ui.adsabs.harvard.edu/abs/2021ApJ...918L..28M} {918, L28}

\bibitem[\protect\citeauthoryear{{Monta{\~n}a}, {Tol{\'o}s}, {Hanauske}  \&
  {Rezzolla}}{{Monta{\~n}a} et~al.}{2019}]{Montana2019}
{Monta{\~n}a} G.,  {Tol{\'o}s} L.,  {Hanauske} M.,   {Rezzolla} L.,  2019,
  \mn@doi [\prd] {10.1103/PhysRevD.99.103009}, \href
  {https://ui.adsabs.harvard.edu/abs/2019PhRvD..99j3009M} {99, 103009}

\bibitem[\protect\citeauthoryear{{Most}, {Papenfort}, {Dexheimer}, {Hanauske},
  {Schramm}, {St{\"o}cker}  \& {Rezzolla}}{{Most}
  et~al.}{2019}]{Most:2019PhRvL}
{Most} E.~R.,  {Papenfort} L.~J.,  {Dexheimer} V.,  {Hanauske} M.,  {Schramm}
  S.,  {St{\"o}cker} H.,   {Rezzolla} L.,  2019, \mn@doi [\prl]
  {10.1103/PhysRevLett.122.061101}, \href
  {https://ui.adsabs.harvard.edu/abs/2019PhRvL.122f1101M} {122, 061101}

\bibitem[\protect\citeauthoryear{{Nambu} \& {Jona-Lasinio}}{{Nambu} \&
  {Jona-Lasinio}}{1961}]{NJL:1961}
{Nambu} Y.,  {Jona-Lasinio} G.,  1961, \mn@doi [Physical Review]
  {10.1103/PhysRev.122.345}, \href
  {https://ui.adsabs.harvard.edu/abs/1961PhRv..122..345N} {122, 345}

\bibitem[\protect\citeauthoryear{{Nozawa}, {Stergioulas}, {Gourgoulhon}  \&
  {Eriguchi}}{{Nozawa} et~al.}{1998}]{Nozawa1998}
{Nozawa} T.,  {Stergioulas} N.,  {Gourgoulhon} E.,   {Eriguchi} Y.,  1998,
  \mn@doi [\aaps] {10.1051/aas:1998304}, \href
  {https://ui.adsabs.harvard.edu/abs/1998A&AS..132..431N} {132, 431}

\bibitem[\protect\citeauthoryear{{Oertel}, {Hempel}, {Kl{\"a}hn}  \&
  {Typel}}{{Oertel} et~al.}{2017}]{Oertel:2017RvMP}
{Oertel} M.,  {Hempel} M.,  {Kl{\"a}hn} T.,   {Typel} S.,  2017, \mn@doi
  [Reviews of Modern Physics] {10.1103/RevModPhys.89.015007}, \href
  {https://ui.adsabs.harvard.edu/abs/2017RvMP...89a5007O} {89, 015007}

\bibitem[\protect\citeauthoryear{{Otto}, {Oertel}  \& {Schaefer}}{{Otto}
  et~al.}{2020}]{Otto2020}
{Otto} K.,  {Oertel} M.,   {Schaefer} B.-J.,  2020, \mn@doi [European Physical
  Journal Special Topics] {10.1140/epjst/e2020-000155-y}, \href
  {https://ui.adsabs.harvard.edu/abs/2020EPJST.229.3629O} {229, 3629}

\bibitem[\protect\citeauthoryear{{Pappas}, {Doneva}, {Sotiriou}, {Yazadjiev}
  \& {Kokkotas}}{{Pappas} et~al.}{2019}]{Pappas2019}
{Pappas} G.,  {Doneva} D.~D.,  {Sotiriou} T.~P.,  {Yazadjiev} S.~S.,
  {Kokkotas} K.~D.,  2019, \mn@doi [\prd] {10.1103/PhysRevD.99.104014}, \href
  {https://ui.adsabs.harvard.edu/abs/2019PhRvD..99j4014P} {99, 104014}

\bibitem[\protect\citeauthoryear{Paschalidis, Yagi, Alvarez-Castillo, Blaschke
  \& Sedrakian}{Paschalidis et~al.}{2018}]{Paschalidis:2017qmb}
Paschalidis V.,  Yagi K.,  Alvarez-Castillo D.,  Blaschke D.~B.,   Sedrakian
  A.,  2018, \mn@doi [Phys. Rev. D] {10.1103/PhysRevD.97.084038}, 97, 084038

\bibitem[\protect\citeauthoryear{{Pearson}, {Chamel}, {Potekhin}, {Fantina},
  {Ducoin}, {Dutta}  \& {Goriely}}{{Pearson} et~al.}{2018}]{Pearson2018}
{Pearson} J.~M.,  {Chamel} N.,  {Potekhin} A.~Y.,  {Fantina} A.~F.,  {Ducoin}
  C.,  {Dutta} A.~K.,   {Goriely} S.,  2018, \mn@doi [\mnras]
  {10.1093/mnras/sty2413}, \href
  {https://ui.adsabs.harvard.edu/abs/2018MNRAS.481.2994P} {481, 2994}

\bibitem[\protect\citeauthoryear{{Philipsen}}{{Philipsen}}{2021}]{Philipsen2021}
{Philipsen} O.,  2021, \mn@doi [Indian Journal of Physics]
  {10.1007/s12648-021-02164-4}, \href
  {https://ui.adsabs.harvard.edu/abs/2021InJPh..95.1599P} {95, 1599}

\bibitem[\protect\citeauthoryear{{Popchev}, {Staykov}, {Doneva}  \&
  {Yazadjiev}}{{Popchev} et~al.}{2019}]{Popchev2019EPJC}
{Popchev} D.,  {Staykov} K.~V.,  {Doneva} D.~D.,   {Yazadjiev} S.~S.,  2019,
  \mn@doi [European Physical Journal C] {10.1140/epjc/s10052-019-6691-x}, \href
  {https://ui.adsabs.harvard.edu/abs/2019EPJC...79..178P} {79, 178}

\bibitem[\protect\citeauthoryear{{Prakash} et~al.,}{{Prakash}
  et~al.}{2021}]{Prakash:2021}
{Prakash} A.,  et~al., 2021, arXiv e-prints, \href
  {https://ui.adsabs.harvard.edu/abs/2021arXiv210607885P} {p. arXiv:2106.07885}

\bibitem[\protect\citeauthoryear{{Raaijmakers} et~al.,}{{Raaijmakers}
  et~al.}{2019}]{Raaijmakers:2019}
{Raaijmakers} G.,  et~al., 2019, \mn@doi [\apjl] {10.3847/2041-8213/ab451a},
  \href {https://ui.adsabs.harvard.edu/abs/2019ApJ...887L..22R} {887, L22}

\bibitem[\protect\citeauthoryear{{Raduta}, {Oertel}  \& {Sedrakian}}{{Raduta}
  et~al.}{2020}]{Raduta2020MNRAS}
{Raduta} A.~R.,  {Oertel} M.,   {Sedrakian} A.,  2020, \mn@doi [\mnras]
  {10.1093/mnras/staa2491}, \href
  {https://ui.adsabs.harvard.edu/abs/2020MNRAS.499..914R} {499, 914}

\bibitem[\protect\citeauthoryear{{Ravenhall} \& {Pethick}}{{Ravenhall} \&
  {Pethick}}{1994}]{RavenhallPethick:1994}
{Ravenhall} D.~G.,  {Pethick} C.~J.,  1994, \mn@doi [\apj] {10.1086/173935},
  \href {https://ui.adsabs.harvard.edu/abs/1994ApJ...424..846R} {424, 846}

\bibitem[\protect\citeauthoryear{Rezzolla, Most  \& Weih}{Rezzolla
  et~al.}{2018}]{Rezzolla:2017aly}
Rezzolla L.,  Most E.~R.,   Weih L.~R.,  2018, \mn@doi [Astrophys. J. Lett.]
  {10.3847/2041-8213/aaa401}, 852, L25

\bibitem[\protect\citeauthoryear{{Riahi}, {Kalantari}  \& {Rueda}}{{Riahi}
  et~al.}{2019}]{Riahi2019}
{Riahi} R.,  {Kalantari} S.~Z.,   {Rueda} J.~A.,  2019, \mn@doi [\prd]
  {10.1103/PhysRevD.99.043004}, \href
  {https://ui.adsabs.harvard.edu/abs/2019PhRvD..99d3004R} {99, 043004}

\bibitem[\protect\citeauthoryear{{Riley} et~al.,}{{Riley}
  et~al.}{2019}]{Riley:2019}
{Riley} T.~E.,  et~al., 2019, \mn@doi [\apjl] {10.3847/2041-8213/ab481c}, \href
  {https://ui.adsabs.harvard.edu/abs/2019ApJ...887L..21R} {887, L21}

\bibitem[\protect\citeauthoryear{{Riley} et~al.,}{{Riley}
  et~al.}{2021}]{Riley2021ApJ}
{Riley} T.~E.,  et~al., 2021, \mn@doi [\apjl] {10.3847/2041-8213/ac0a81}, \href
  {https://ui.adsabs.harvard.edu/abs/2021ApJ...918L..27R} {918, L27}

\bibitem[\protect\citeauthoryear{{Roberts}, {Reddy}  \& {Shen}}{{Roberts}
  et~al.}{2012}]{Roberts:2012}
{Roberts} L.~F.,  {Reddy} S.,   {Shen} G.,  2012, \mn@doi [\prc]
  {10.1103/PhysRevC.86.065803}, \href
  {https://ui.adsabs.harvard.edu/abs/2012PhRvC..86f5803R} {86, 065803}

\bibitem[\protect\citeauthoryear{{Roca-Maza} \& {Piekarewicz}}{{Roca-Maza} \&
  {Piekarewicz}}{2008}]{RocaMaza:2008}
{Roca-Maza} X.,  {Piekarewicz} J.,  2008, \mn@doi [\prc]
  {10.1103/PhysRevC.78.025807}, \href
  {https://ui.adsabs.harvard.edu/abs/2008PhRvC..78b5807R} {78, 025807}

\bibitem[\protect\citeauthoryear{R{\"o}pke, Blaschke  \& Schulz}{R{\"o}pke
  et~al.}{1986}]{Ropke:1986qs}
R{\"o}pke G.,  Blaschke D.,   Schulz H.,  1986, \mn@doi [Phys. Rev. D]
  {10.1103/PhysRevD.34.3499}, 34, 3499

\bibitem[\protect\citeauthoryear{{Rosswog}}{{Rosswog}}{2015}]{Rosswog_15}
{Rosswog} S.,  2015, \mn@doi [Int. J. Mod. Phys. D]
  {10.1142/S0218271815300128}, \href
  {http://cdsads.u-strasbg.fr/abs/2015IJMPD..2430012R} {24, 30012}

\bibitem[\protect\citeauthoryear{{R{\"u}ster}, {Werth}, {Buballa}, {Shovkovy}
  \& {Rischke}}{{R{\"u}ster} et~al.}{2005}]{Ruester2005}
{R{\"u}ster} S.~B.,  {Werth} V.,  {Buballa} M.,  {Shovkovy} I.~A.,   {Rischke}
  D.~H.,  2005, \mn@doi [\prd] {10.1103/PhysRevD.72.034004}, \href
  {https://ui.adsabs.harvard.edu/abs/2005PhRvD..72c4004R} {72, 034004}

\bibitem[\protect\citeauthoryear{{Sagert}, {Fischer}, {Hempel}, {Pagliara},
  {Schaffner-Bielich}, {Mezzacappa}, {Thielemann}  \&
  {Liebend{\"o}rfer}}{{Sagert} et~al.}{2009}]{Sagert:2009}
{Sagert} I.,  {Fischer} T.,  {Hempel} M.,  {Pagliara} G.,  {Schaffner-Bielich}
  J.,  {Mezzacappa} A.,  {Thielemann} F.~K.,   {Liebend{\"o}rfer} M.,  2009,
  \mn@doi [\prl] {10.1103/PhysRevLett.102.081101}, \href
  {https://ui.adsabs.harvard.edu/abs/2009PhRvL.102h1101S} {102, 081101}

\bibitem[\protect\citeauthoryear{{Shapiro}, {Teukolsky}  \&
  {Wasserman}}{{Shapiro} et~al.}{1989}]{Shapiro:1989}
{Shapiro} S.~L.,  {Teukolsky} S.~A.,   {Wasserman} I.,  1989, \mn@doi [\nat]
  {10.1038/340451a0}, \href
  {https://ui.adsabs.harvard.edu/abs/1989Natur.340..451S} {340, 451}

\bibitem[\protect\citeauthoryear{{Shen}, {Horowitz}  \& {Teige}}{{Shen}
  et~al.}{2011}]{Shen:2011}
{Shen} G.,  {Horowitz} C.~J.,   {Teige} S.,  2011, \mn@doi [\prc]
  {10.1103/PhysRevC.83.035802}, \href
  {https://ui.adsabs.harvard.edu/abs/2011PhRvC..83c5802S} {83, 035802}

\bibitem[\protect\citeauthoryear{Shibata, Fujibayashi, Hotokezaka, Kiuchi,
  Kyutoku, Sekiguchi  \& Tanaka}{Shibata et~al.}{2017}]{Shibata:2017xdx}
Shibata M.,  Fujibayashi S.,  Hotokezaka K.,  Kiuchi K.,  Kyutoku K.,
  Sekiguchi Y.,   Tanaka M.,  2017, \mn@doi [Phys. Rev. D]
  {10.1103/PhysRevD.96.123012}, 96, 123012

\bibitem[\protect\citeauthoryear{Shibata, Zhou, Kiuchi  \& Fujibayashi}{Shibata
  et~al.}{2019}]{Shibata:2019ctb}
Shibata M.,  Zhou E.,  Kiuchi K.,   Fujibayashi S.,  2019, \mn@doi [Phys. Rev.
  D] {10.1103/PhysRevD.100.023015}, 100, 023015

\bibitem[\protect\citeauthoryear{{Sieniawska}, {Turcza{\'n}ski}, {Bejger}  \&
  {Zdunik}}{{Sieniawska} et~al.}{2019}]{Sieniawska2019}
{Sieniawska} M.,  {Turcza{\'n}ski} W.,  {Bejger} M.,   {Zdunik} J.~L.,  2019,
  \mn@doi [\aap] {10.1051/0004-6361/201833969}, \href
  {https://ui.adsabs.harvard.edu/abs/2019A&A...622A.174S} {622, A174}

\bibitem[\protect\citeauthoryear{{Silva}, {Sotani}  \& {Berti}}{{Silva}
  et~al.}{2016}]{Silva2016MNRAS}
{Silva} H.~O.,  {Sotani} H.,   {Berti} E.,  2016, \mn@doi [\mnras]
  {10.1093/mnras/stw969}, \href
  {https://ui.adsabs.harvard.edu/abs/2016MNRAS.459.4378S} {459, 4378}

\bibitem[\protect\citeauthoryear{{Sotani}, {Nakazato}, {Iida}  \&
  {Oyamatsu}}{{Sotani} et~al.}{2013}]{Sotani2013MNRAS}
{Sotani} H.,  {Nakazato} K.,  {Iida} K.,   {Oyamatsu} K.,  2013, \mn@doi
  [\mnras] {10.1093/mnras/stt1152}, \href
  {https://ui.adsabs.harvard.edu/abs/2013MNRAS.434.2060S} {434, 2060}

\bibitem[\protect\citeauthoryear{{Steiner}, {Lattimer}  \& {Brown}}{{Steiner}
  et~al.}{2010}]{Steiner:2010}
{Steiner} A.~W.,  {Lattimer} J.~M.,   {Brown} E.~F.,  2010, \mn@doi [\apj]
  {10.1088/0004-637X/722/1/33}, \href
  {https://ui.adsabs.harvard.edu/abs/2010ApJ...722...33S} {722, 33}

\bibitem[\protect\citeauthoryear{{Steiner}, {Hempel}  \& {Fischer}}{{Steiner}
  et~al.}{2013}]{Steiner:2013}
{Steiner} A.~W.,  {Hempel} M.,   {Fischer} T.,  2013, \mn@doi [\apj]
  {10.1088/0004-637X/774/1/17}, \href
  {https://ui.adsabs.harvard.edu/abs/2013ApJ...774...17S} {774, 17}

\bibitem[\protect\citeauthoryear{{Steiner}, {Lattimer}  \& {Brown}}{{Steiner}
  et~al.}{2016}]{Steiner2016EPJA}
{Steiner} A.~W.,  {Lattimer} J.~M.,   {Brown} E.~F.,  2016, \mn@doi [European
  Physical Journal A] {10.1140/epja/i2016-16018-1}, \href
  {https://ui.adsabs.harvard.edu/abs/2016EPJA...52...18S} {52, 18}

\bibitem[\protect\citeauthoryear{{Stergioulas} \& {Friedman}}{{Stergioulas} \&
  {Friedman}}{1995}]{Stergioulas:1995}
{Stergioulas} N.,  {Friedman} J.~L.,  1995, \mn@doi [\apj] {10.1086/175605},
  \href {https://ui.adsabs.harvard.edu/abs/1995ApJ...444..306S} {444, 306}

\bibitem[\protect\citeauthoryear{{Sugahara} \& {Toki}}{{Sugahara} \&
  {Toki}}{1994}]{Sugahara:1994}
{Sugahara} Y.,  {Toki} H.,  1994, \mn@doi [\nphysa]
  {10.1016/0375-9474(94)90923-7}, \href
  {https://ui.adsabs.harvard.edu/abs/1994NuPhA.579..557S} {579, 557}

\bibitem[\protect\citeauthoryear{{Suleiman}, {Fortin}, {Zdunik}  \&
  {Haensel}}{{Suleiman} et~al.}{2021}]{Suleiman2021PhRvC}
{Suleiman} L.,  {Fortin} M.,  {Zdunik} J.~L.,   {Haensel} P.,  2021, \mn@doi
  [\prc] {10.1103/PhysRevC.104.015801}, \href
  {https://ui.adsabs.harvard.edu/abs/2021PhRvC.104a5801S} {104, 015801}

\bibitem[\protect\citeauthoryear{{Toki}, {Hirata}, {Sugahara}, {Sumiyoshi}  \&
  {Tanihata}}{{Toki} et~al.}{1995}]{Toki:1995}
{Toki} H.,  {Hirata} D.,  {Sugahara} Y.,  {Sumiyoshi} K.,   {Tanihata} I.,
  1995, \mn@doi [\nphysa] {10.1016/0375-9474(95)00161-S}, \href
  {https://ui.adsabs.harvard.edu/abs/1995NuPhA.588..357T} {588, 357}

\bibitem[\protect\citeauthoryear{{Typel}}{{Typel}}{2014}]{Typel:2014b}
{Typel} S.,  2014, \mn@doi [\prc] {10.1103/PhysRevC.89.064321}, \href
  {https://ui.adsabs.harvard.edu/abs/2014PhRvC..89f4321T} {89, 064321}

\bibitem[\protect\citeauthoryear{{Typel}}{{Typel}}{2016}]{Typel:2016}
{Typel} S.,  2016, \mn@doi [European Physical Journal A]
  {10.1140/epja/i2016-16016-3}, \href
  {https://ui.adsabs.harvard.edu/abs/2016EPJA...52...16T} {52, 16}

\bibitem[\protect\citeauthoryear{{Typel}, {R{\"o}pke}, {Kl{\"a}hn}, {Blaschke}
  \& {Wolter}}{{Typel} et~al.}{2010}]{Typel:2010}
{Typel} S.,  {R{\"o}pke} G.,  {Kl{\"a}hn} T.,  {Blaschke} D.,   {Wolter} H.~H.,
   2010, \mn@doi [\prc] {10.1103/PhysRevC.81.015803}, \href
  {https://ui.adsabs.harvard.edu/abs/2010PhRvC..81a5803T} {81, 015803}

\bibitem[\protect\citeauthoryear{{Typel}, {Wolter}, {R{\"o}pke}  \&
  {Blaschke}}{{Typel} et~al.}{2014}]{Typel:2014a}
{Typel} S.,  {Wolter} H.~H.,  {R{\"o}pke} G.,   {Blaschke} D.,  2014, \mn@doi
  [European Physical Journal A] {10.1140/epja/i2014-14017-x}, \href
  {https://ui.adsabs.harvard.edu/abs/2014EPJA...50...17T} {50, 17}

\bibitem[\protect\citeauthoryear{{Wei}, {Figura}, {Burgio}, {Chen}  \&
  {Schulze}}{{Wei} et~al.}{2019}]{Wei2019JPhG}
{Wei} J.~B.,  {Figura} A.,  {Burgio} G.~F.,  {Chen} H.,   {Schulze} H.~J.,
  2019, \mn@doi [Journal of Physics G Nuclear Physics]
  {10.1088/1361-6471/aaf95c}, \href
  {https://ui.adsabs.harvard.edu/abs/2019JPhG...46c4001W} {46, 034001}

\bibitem[\protect\citeauthoryear{Wei, Burgio, Raduta  \& Schulze}{Wei
  et~al.}{2021}]{Wei:2021veo}
Wei J.-B.,  Burgio G.~F.,  Raduta A.~R.,   Schulze H.~J.,  2021, \mn@doi [Phys.
  Rev. C] {10.1103/PhysRevC.104.065806}, 104, 065806

\bibitem[\protect\citeauthoryear{{Yagi} \& {Stepniczka}}{{Yagi} \&
  {Stepniczka}}{2021}]{Yagi2021PhRvD}
{Yagi} K.,  {Stepniczka} M.,  2021, \mn@doi [\prd]
  {10.1103/PhysRevD.104.044017}, \href
  {https://ui.adsabs.harvard.edu/abs/2021PhRvD.104d4017Y} {104, 044017}

\bibitem[\protect\citeauthoryear{{Yagi} \& {Yunes}}{{Yagi} \&
  {Yunes}}{2013}]{Yagi2013Sci}
{Yagi} K.,  {Yunes} N.,  2013, \mn@doi [Science] {10.1126/science.1236462},
  \href {https://ui.adsabs.harvard.edu/abs/2013Sci...341..365Y} {341, 365}

\bibitem[\protect\citeauthoryear{{Yagi} \& {Yunes}}{{Yagi} \&
  {Yunes}}{2017}]{Yagi2017x}
{Yagi} K.,  {Yunes} N.,  2017, \mn@doi [\physrep]
  {10.1016/j.physrep.2017.03.002}, \href
  {https://ui.adsabs.harvard.edu/abs/2017PhR...681....1Y} {681, 1}

\bibitem[\protect\citeauthoryear{{Zdunik} \& {Haensel}}{{Zdunik} \&
  {Haensel}}{2013}]{Zdunik2013}
{Zdunik} J.~L.,  {Haensel} P.,  2013, \mn@doi [\aap]
  {10.1051/0004-6361/201220697}, \href
  {https://ui.adsabs.harvard.edu/abs/2013A&A...551A..61Z} {551, A61}

\bibitem[\protect\citeauthoryear{{Zha}, {O'Connor}, {Chu}, {Lin}  \&
  {Couch}}{{Zha} et~al.}{2020}]{OConnor:2020}
{Zha} S.,  {O'Connor} E.~P.,  {Chu} M.-c.,  {Lin} L.-M.,   {Couch} S.~M.,
  2020, \mn@doi [\prl] {10.1103/PhysRevLett.125.051102}, \href
  {https://ui.adsabs.harvard.edu/abs/2020PhRvL.125e1102Z} {125, 051102}

\makeatother
\end{thebibliography}
\appendix

\section{Role of finite neutrino abundance}
\label{sec:appendixA}

In the case of finite temperatures the neutrinos, which are produced from a variety of weak reactions, are no longer freely streaming, above around $T\simeq 0.5-1$~MeV. Frequent re-absorption via the Urca processes, $\nu_e+n\rightleftarrows p + e^-$ and $\bar\nu_e+p\rightleftarrows n + e^+$, as well as re-scattering, e.g. on electrons and nucleons leads to the establishment of a finite neutrino abundance that is associated with the neutrino equilibrium chemical potential, $\mu_{\nu_e}=\mu_e-\hat\mu$ with the charged chemical potential defined as the difference between neutron and proton chemical potentials, $\hat\mu=\mu_n-\mu_p$. A situation like this is usually realized at the interior of protoneutron stars, which result from a core-collapse supernova explosion \citep[for a review, cf.][and references therein]{Janka:2007PhR}. After the supernova explosion onset, protoneutron stars are still hot and lepton rich, and hence they deleptonize, i.e. the diffusion of neutrinos of all flavours, lasting for several tens of seconds and the later Kelvin-Helmholtz cooling phase \citep[][]{Huedepohl:2010,Fischer:2009af}. Particular focus is thereby on the treatment of the weak interactions consistent with the nuclear EoS, which determines details of the deleptonization, e.g. the magnitude of the neutrino fluxes and spectra and their evolution \citep[][]{MartinezPinedo:2012,Roberts:2012,Fischer:2020PhRvC101}. 

\begin{figure}
\includegraphics[width=0.475\textwidth]{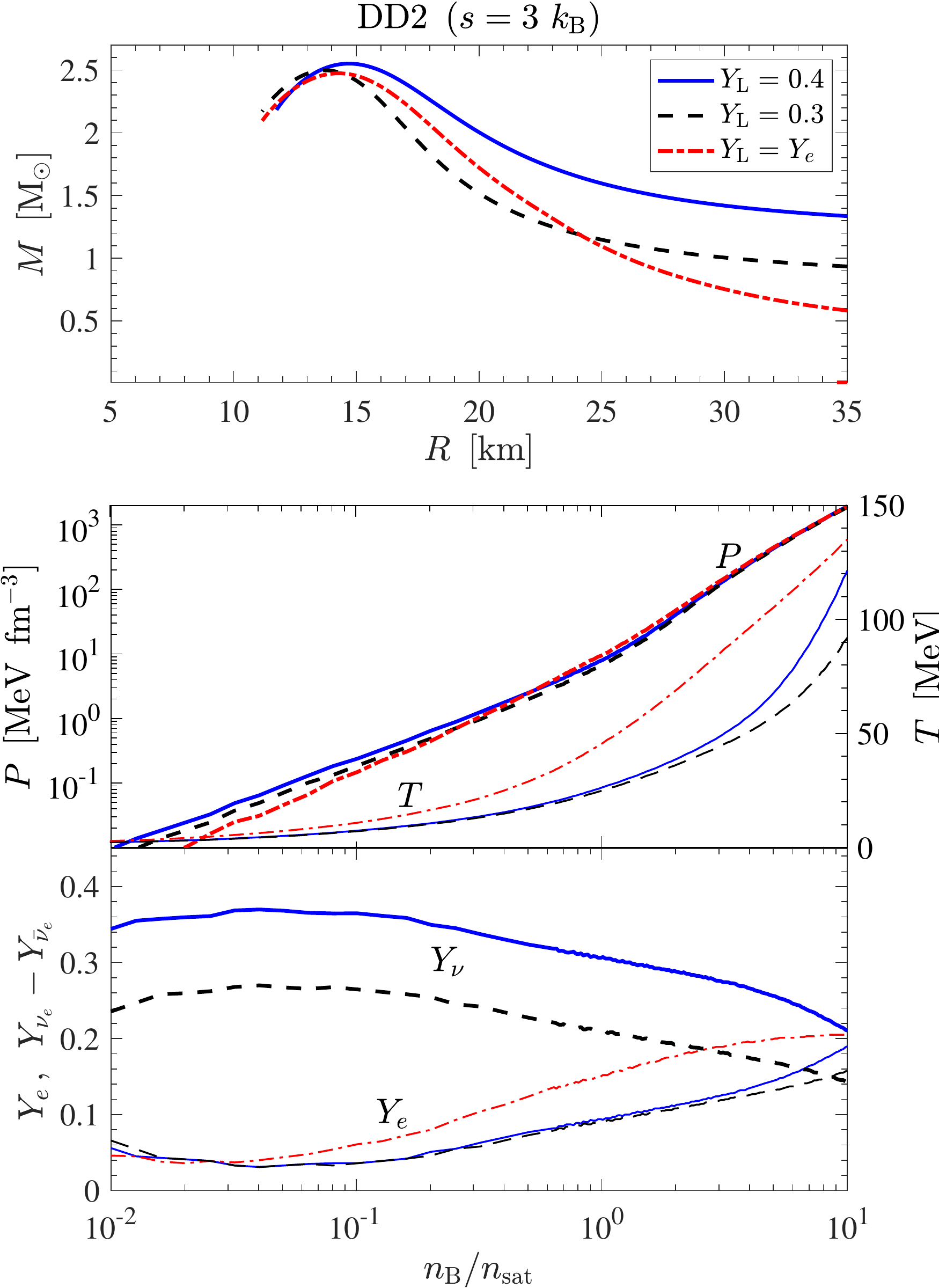}
\caption{(colour online) Comparison of the EoS (middle and bottom panels), pressure $P$ (thick lines), temperature $T$ (thin lines) and electron as well as net neutrino abundances $Y_e$ (thin lines)  and $Y_{\nu_e}-Y_{\bar\nu_e}$ (thick lines), respectively, and the corresponding mass-radius relations (top panel) for the hadronic DD2 RMF EoS, evaluated at constant entropy per baryon of $s=3~k_{\rm B}$ and different values of the electron lepton fraction of $Y_{\rm L}=Y_e$ (red dash-dotted lines), $Y_{\rm L}=0.3$ (black dashed lines), and $Y_{\rm L}=0.4$ (blue solid lines).}
\label{fig:appendix}
\end{figure}

\begin{figure*}
\includegraphics[width=0.475\textwidth]{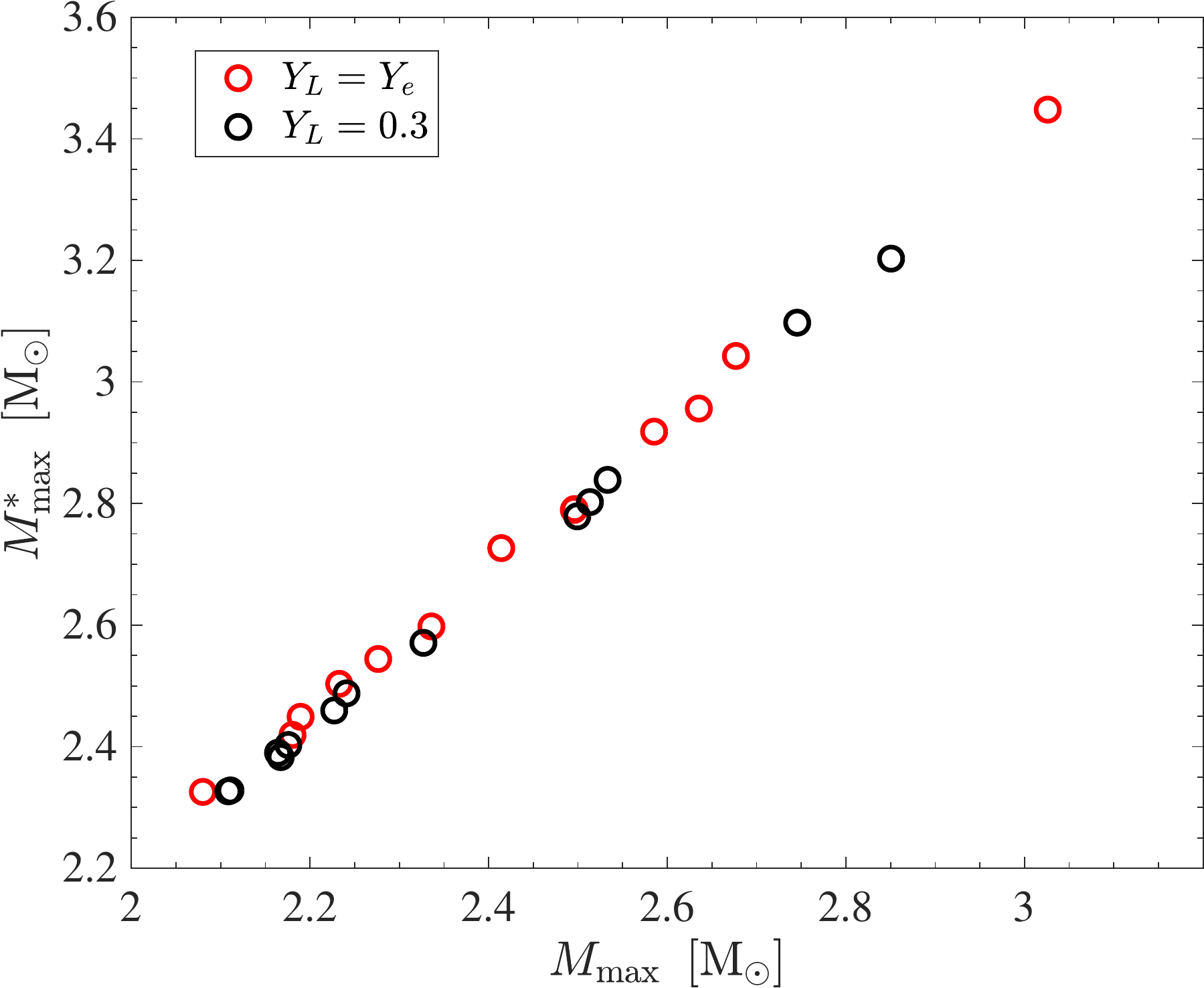}
\includegraphics[width=0.475\textwidth]{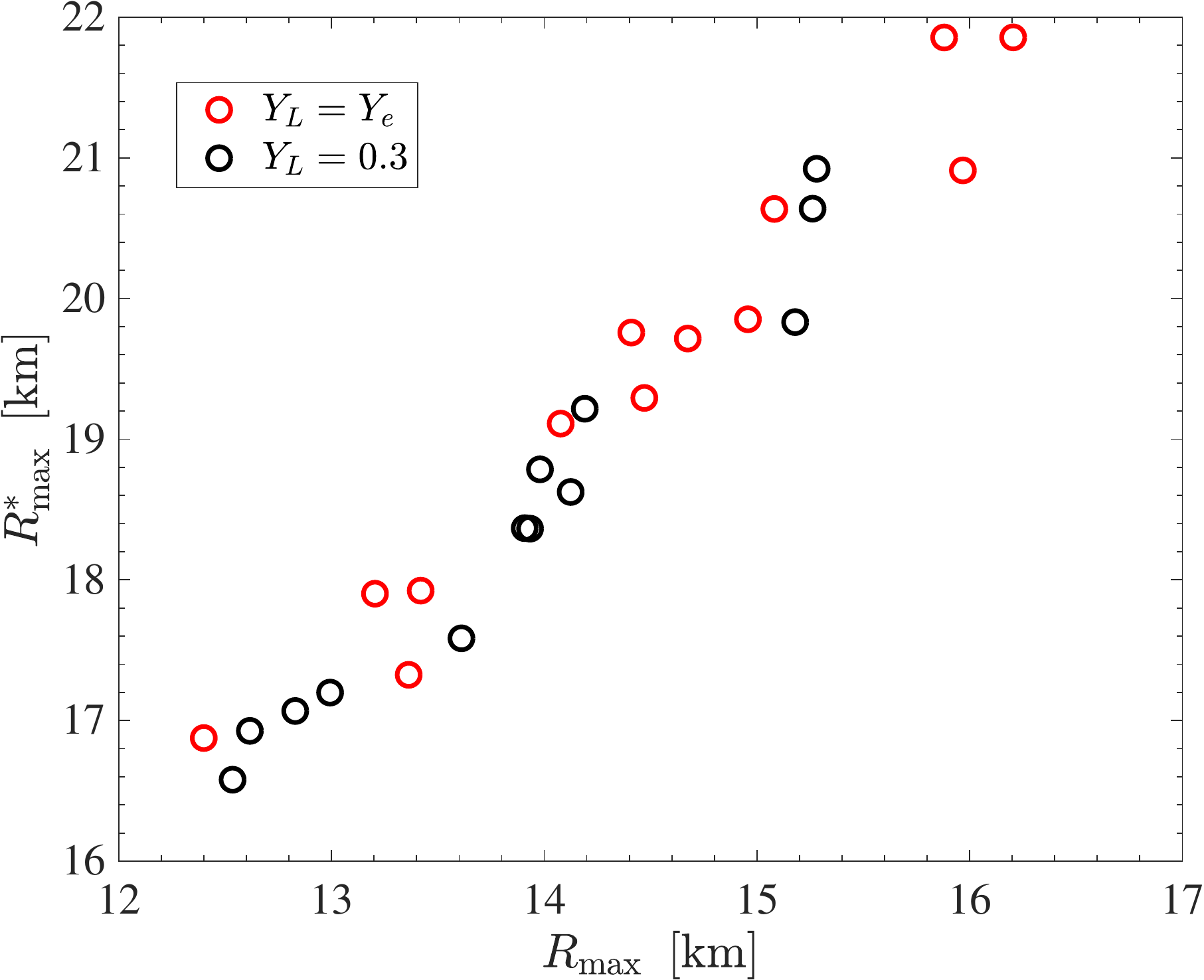}
\caption{(colour online) Comparison of the relations for rapidly rotating compact stars, maximum masses, $M^*_{\rm max}$ and $M_{\rm max}$ (left-hand panel), and the corresponding radii, $R^*_{\rm max}$ and $R_{\rm max}$ (right-hand panel) evaluated at constant entropy per baryon of $s=3~k_{\rm B}$ and different values of the electron lepton fraction of $Y_{\rm L}=Y_e$ (red circles), $Y_{\rm L}=0.3$ (black circles) ).}
\label{fig:appendix_universal_hadronic_MR}
\end{figure*}

\begin{figure*}
\includegraphics[width=0.475\textwidth]{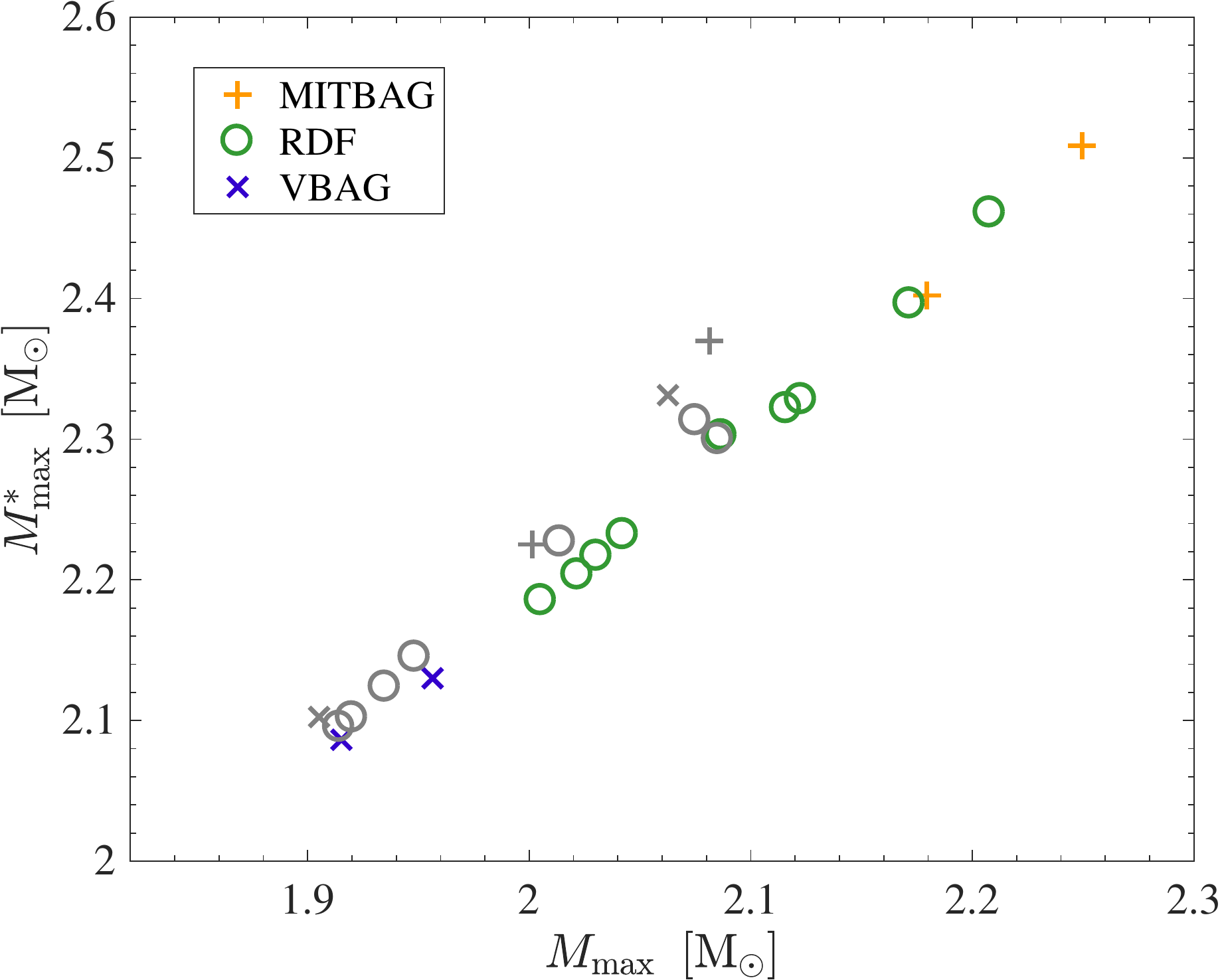}
\includegraphics[width=0.46\textwidth]{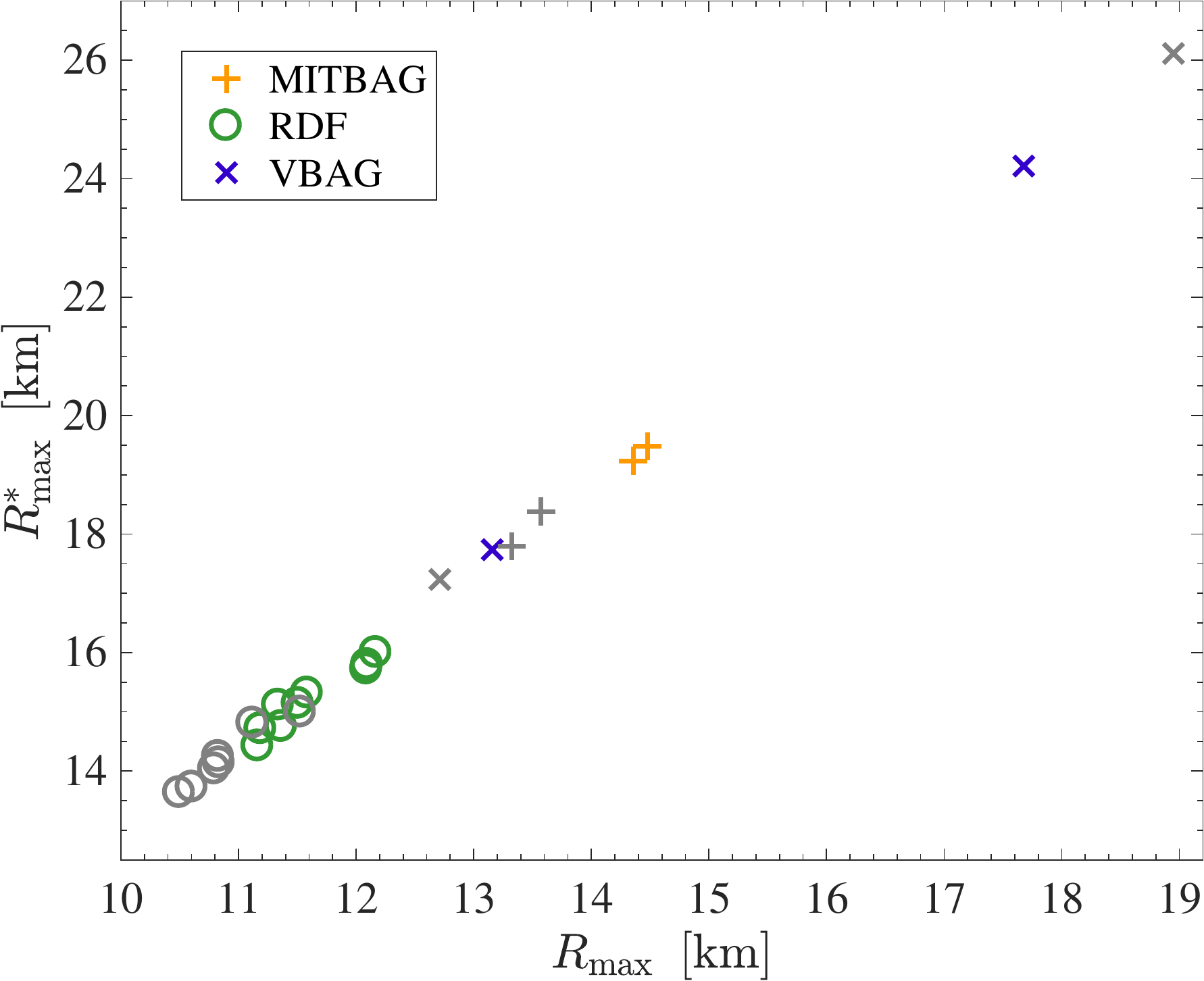}
\caption{Same as Fig.~\ref{fig:appendix_universal_hadronic_MR} but for the considered hybrid EoS at $s=3~k_{\rm B}$. The coloured markers refer to $Y_{\rm L}=0.3$ and gray markers correspond to $Y_{\rm L}=Y_e$.}
\label{fig:appendix_universal_hybrid_MR}
\end{figure*}

These studies indicate that the electron lepton number can be very different from that of the electron abundance. Hence, we decide to consider the finite entropy per particle EoS for a finite and constant electron lepton number, $Y_{\rm L} = Y_e + Y_{\nu_e} - Y_{\bar\nu_e}$. Note that the electron abundance is related with the electron vector density, $n_e=\rho Y_e/m_{\rm B}$ with baryon mass $m_{\rm B}$, and hence represents the net number of negatively charged leptons, $Y_e:=Y_{e^-}-Y_{e^+}$. The associated results are shown in Fig.~\ref{fig:appendix}. In comparison to the usual case, $Y_{\rm L}=Y_e$, i.e. zero neutrino abundance (red dashed lines), the finite neutrino abundances (bottom panel) have a negligible contribution to the pressure and energy density in the relevant density range. However, their contribution to the entropy is substantial, which, in turn, requires a lower temperature for the same entropy of $s=3~k_{\rm B}$ (thin lines in the middle panel show the temperatures). The related feedback to the temperature dependence of the nuclear EoS is then responsible for the slightly higher pressure (thick lines in the middle panel). The magnitude of this effect is proportional to the value assumed for the lepton fraction; here, we explore the two representative cases $Y_{\rm L}=0.3$ and $Y_{\rm L}=0.4$. Consequently, the mass-radius relations (top panel) show somewhat larger radii, due to the larger impact seen at densities around and slightly above nuclear saturation density. At higher densities, when the neutrino abundances decrease, the impact is somewhat smaller. The reduction of the temperature is smaller and, moreover, the EoS is less sensitive to thermal effects. Hence, the impact, more precisely the enhancement of the maximum masses for the cases of a finite neutrino abundance, is $\left.M_{\rm max}\right\vert_{Y_{\rm L}=Y_e}= 2.46$~M$_\odot$, $\left.M_{\rm max}\right\vert_{Y_{\rm L}=0.3}= 2.51$~M$_\odot$ and $\left.M_{\rm max}\right\vert_{Y_{\rm L}=0.4}= 2.57$~M$_\odot$.

\begin{table}
\centering
\caption{Coefficients for the universal relations \eqref{eq:universal_M} and \eqref{eq:universal_R} including the maximum deviations, assuming $Y_{\rm L}=Y_{\rm e}$ }
\begin{tabular}{l c c c}
\hline
Condition & EoS type & $C_M$ & $C_R$ \\
\hline
$s=3~k_{\rm B}$, $Y_{\rm L}= Y_{\rm e}$ & hadronic & $1.1220\pm 0.0060$ & $1.3440\pm 0.0150 $ \\
$s=3~k_{\rm B}$, $Y_{\rm L}=Y_{\rm e}$ & hybrid   & $1.1095\pm 0.0095$ & $1.3345\pm 0.0215$ \\
\hline
\end{tabular}
\label{tab:old_universal}
\end{table}

We remark here that we find qualitatively the same behaviour for all hadronic and hybrid finite entropy EoS explored in this study. Therefore, to quantify the differences between the two setups, $Y_{\rm L}=Y_e$ and fixed $Y_{\rm L}=0.3$, Figs.~\ref{fig:appendix_universal_hadronic_MR} and \ref{fig:appendix_universal_hybrid_MR} compare the results obtained for the universal relations for mass (left-hand panels) and radius (right-hand panels), for both hadronic and hybrid models, respectively. The corresponding coefficients, $C_M$ and $C_R$, of the universal relations for $Y_{\rm L}=Y_e$ are given in Table~\ref{tab:old_universal}. With this setup, the coefficients are generally slightly larger, by about one percent for both hadronic and hybrid EoS, in comparison to the more physical case of a constant lepton fraction. The discrepancy might become larger when compared with a higher lepton fraction as well as for higher values of the entropy. 

\section{Universal relations without constraints}
\label{sec:appendixB}
  In this appendix the universal relations for maximum mass and radius at the maximum mass and the moment of inertia are computed based on the complete set of EOS, taking into account the maximum neutron star mass constraints but otherwise ignoring the constraints derived for $\Lambda_{1.4}$ and the radii of 1.4 and 2.0~M$_\odot$ neutron stars due to GW170817 and NICER respectively. These results are shown in Tables~\ref{tab:appB_universal_MR} and \ref{tab:appB_universal_I}. From these, it becomes evident that the same universal relations as discussed in sec.~\ref{sec:results} apply, with error margins of a similar order of magnitude.

\begin{table}
\centering
\caption{Coefficients for the universal relations \eqref{eq:universal_M} and \eqref{eq:universal_R} including the maximum deviations, without taking into account the constraints for radius and tidal deformability. }
\begin{tabular}{l c c c}
\hline
Condition & EoS type & $C_M$ & $C_R$ \\
\hline
$T=0$, $\beta$-eq. & hadronic & $1.2025\pm 0.0065$ & $1.3415\pm 0.0105$ \\
$s=3~k_{\rm B}$, $Y_{\rm L}=0.3$ & hadronic & $1.1105\pm 0.0055$ & $1.3300\pm 0.0130 $ \\
$T=0$, $\beta$-eq. & hybrid   & $1.2045\pm 0.0095$ & $1.3315\pm 0.0075$ \\
$s=3~k_{\rm B}$, $Y_{\rm L}=0.3$ & hybrid   & $1.0995\pm 0.0055$ & $1.3290\pm 0.0160$ \\
\hline
\end{tabular}
\label{tab:appB_universal_MR}
\end{table}

\begin{table}
\centering
\caption{Coefficients for the universal relation \eqref{eq:universal_I} including the maximum deviations, without taking into account the constraints for radius and tidal deformability.}
\begin{tabular}{l c c c}
\hline
Condition & EoS type & $a_1$ & $a_2$ \\
\hline
$T=0$, $\beta$-eq. & hadronic & $1.1345\pm 0.1175$ & $0.0638\pm 0.0297$ \\
$s=3~k_{\rm B}$, $Y_{\rm L}=0.3$ & hadronic & $1.2265\pm 0.1175$ & $-0.0103\pm 0.0236$ \\
$T=0$, $\beta$-eq. & hybrid   & $1.0840\pm 0.1444$ & $0.0743\pm 0.0365$ \\
$s=3~k_{\rm B}$, $Y_{\rm L}=0.3$ & hybrid   & $0.9843\pm 0.0966$ & $0.0333\pm 0.0177$ \\
\hline
\end{tabular}
\label{tab:appB_universal_I}
\end{table}
\bsp	
\label{lastpage}
\end{document}